\documentclass[journal]{IEEEtran}

\usepackage[font=footnotesize]{caption}
\usepackage{amsmath,graphicx,setspace,hyperref,algpseudocode,amssymb,tabularx,bm}
\usepackage{siunitx}
\usepackage{comment}
\usepackage{lipsum} 
\usepackage{enumitem,kantlipsum}
\usepackage[algoruled,vlined,linesnumbered]{algorithm2e}
\usepackage{mathtools, cuted}
\usepackage[font=footnotesize,labelformat=simple]{subcaption}
\usepackage{xcolor}

\usepackage{makecell}
\usepackage{multirow}
\usepackage{cite}
\usepackage{float}
\usepackage{multicol}

\makeatletter
\newcommand{\multiline}[1]{%
  \begin{tabularx}{\dimexpr\linewidth-\ALG@thistlm}[t]{@{}X@{}}
    #1
  \end{tabularx}
}
\makeatother

\DeclareMathOperator*{\argmin}{arg\,min}

\usepackage[colorinlistoftodos]{todonotes}

\usepackage{stackengine}

\SetKwInput{KwInput}{Input}                
\SetKwInput{KwOutput}{Output}              

\begin{document}
\bstctlcite{IEEEexample:BSTcontrol}

\title{A Fast Automatic Method for Deconvoluting Macro X-ray Fluorescence Data Collected from Easel Paintings}

\author{Su~Yan,~\IEEEmembership{Student~Member,~IEEE,}
        Jun-Jie~Huang,~\IEEEmembership{Member,~IEEE,}
        Herman Verinaz-Jadan,~\IEEEmembership{Student~Member,~IEEE,}
        Nathan~Daly,
        Catherine~Higgitt,
        and~Pier~Luigi~Dragotti,~\IEEEmembership{Fellow,~IEEE}
\thanks{This work is in part supported by EPSRC grant EP/R032785/1. Su Yan is supported by China Scholarship Council (CSC) scholarship. Herman Verinaz-Jadan was in part supported by SENESCYT.}
}


\maketitle

\begin{abstract}
Macro X-ray Fluorescence (MA-XRF) scanning is increasingly widely used by researchers in heritage science to analyse easel paintings as one of a suite of non-invasive imaging techniques. The task of processing the resulting MA-XRF datacube generated in order to produce individual chemical element maps is called MA-XRF deconvolution. While there are several existing methods that have been proposed for MA-XRF deconvolution, they require a degree of manual intervention from the user that can affect the final results. The state-of-the-art AFRID approach can automatically deconvolute the datacube without user input, but it has a long processing time and does not exploit spatial dependency. In this paper, we propose two versions of a fast automatic deconvolution (FAD) method for MA-XRF datacubes collected from easel paintings with ADMM (alternating direction method of multipliers) and FISTA (fast iterative shrinkage-thresholding algorithm). The proposed FAD method not only automatically analyses the datacube and produces element distribution maps of high-quality with spatial dependency considered, but also significantly reduces the running time. The results generated on the MA-XRF datacubes collected from two easel paintings from the National Gallery, London, verify the performance of the proposed FAD method.
\end{abstract}


\IEEEpeerreviewmaketitle

\section{Introduction}
\label{sec:intro}

\IEEEPARstart{E}{asel} paintings, as one of the most iconic and influential forms of artistic heritage, are of great research value to heritage scientists. In order to study and analyse these valuable paintings, many non-invasive imaging and spectroscopic imaging techniques are used by heritage scientists, including X-radiography, infrared reflectrography, reflectance imaging spectroscopy and macro X-Ray fluorescence (MA-XRF) scanning. These methods generate a wealth of data and as a result, many researchers are working on developing different types of signal processing techniques to process the acquired datasets, see for instance \cite{huang2016computational,pizurica2015digital,huang2020multimodal,dai2017spatial,dai2019adaptive,sabetsarvestani2019artificial,8950399,pu2020connected,pu2021learning,pu2022mixed}.

In this paper, we focus on MA-XRF scanning, an increasingly important technique to analyse the presence and distribution of chemical elements present in easel paintings. For example, a knowledge of the different chemical elements present can help to identify the pigments used in the painting. Moreover, the distribution maps of different chemical elements can help to understand an artist's technique and working practice, characterise the paint layer stratigraphy and reveal hidden layers of the painting, which may contain concealed designs, modifications made during painting or even initial sketches that have been covered by the subsequent paint layers. 

\begin{figure}[t]
	\centering
	\includegraphics[width=0.7\linewidth]{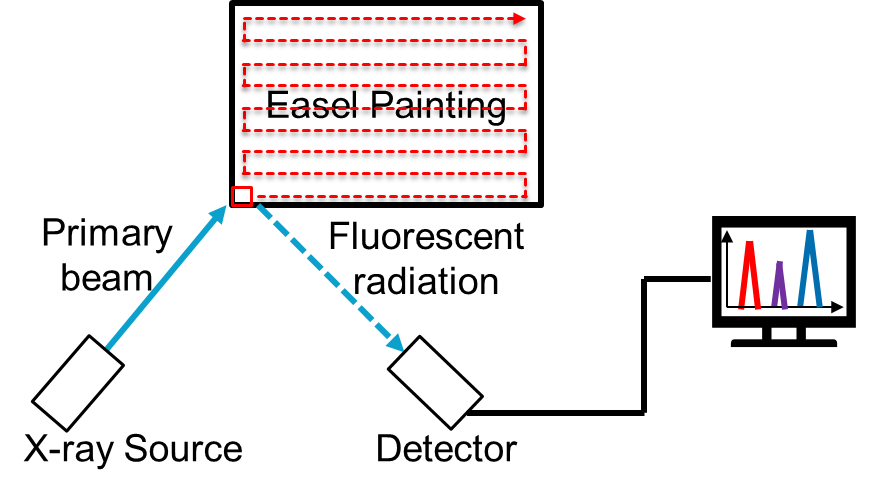}
	\vspace{-0.2cm}
	\caption{Schematic of the acquisition of an MA-XRF dataset from an easel painting.}
	\label{fig:XRF_device}
\end{figure}

An MA-XRF scanning instrument consists of an X-ray source and a detector, as shown in Fig. \ref{fig:XRF_device}. The X-ray source generates the primary X-ray beam and illuminates a sub-millimetre spot (also regarded as a pixel) of the painting. If the atoms of different chemical elements in the pixel receive enough energy from the X-ray beam, some of their inner shell electrons will be ejected and others from outer shells will fill the vacancies, at the same time releasing X-ray photons. The detector then collects the released photons and generates a spectrum for this pixel in terms of the energies of the photons. The X-ray source and detector are scanned across the painting, with the above process being repeated for all the pixels of the painting to produce an MA-XRF datacube. Since the atoms of each chemical element emit X-ray photons with a series of characteristic energies during the electronic transition, the resulting spectrum at each pixel, in theory, should be a series of lines at different energy levels, which are normally called element lines or characteristic X-ray lines. However, because of imperfect device performance and the statistical nature of the photon release process, each element line in the spectrum is broadened into a peak or `pulse' \cite{van2001handbook,beckhoff2007handbook,wilkinson1971breit}. This causes a problem in that the pulses due to different chemical elements may overlap in the spectrum if the energies of their element lines are sufficiently close. To identify the chemical elements that are present, we need to locate and separate the overlapping pulses from the XRF spectra and then relate them to the correct element lines. This critical step is called XRF deconvolution. 

Existing methods for deconvoluting the MA-XRF datacubes collected from easel paintings include AXIL (Analysis of X-ray spectra by Iterative least Squares) method \cite{van1977computer,vekemans1995comparison, vekemans1994analysis}, `PyMCA' (Python multichannel analyzer) software \cite{sole2007multiplatform}, `Datamuncher' software \cite{alfeld2015strategies}, Conover's method \cite{Conover2015}, and the proprietary commercial software supplied by Bruker Corporation with the M6 JETSTREAM MA-XRF scanning instrument \cite{alfeld2013mobile,Roald2017MAXRF}. However, all the methods mentioned above require users to manually input a list of chemical elements that are expected to be present in the painting, which can affect the resulting element maps. Different users' input can potentially lead to completely different distributions, making the deconvolution results less reliable. To address this problem, we have previously proposed \cite{yan2020revealing,yan2021prony} an automatic pixel-wise method (AFRID method) based on Finite Rate of Innovation (FRI) sampling theory \cite{vetterli2002sampling,uriguen2013fri}. After the necessary pre-processing steps, the AFRID method can automatically generate the distribution maps of the chemical elements that are present in the painting without the need for manual input from the user. The results showed that the AFRID method can detect weak elements, separate nearby element peaks and produce remarkable deconvolution results. Moreover, it also generates confidence maps for the chemical elements present, which provide additional information for interpretation of the data by the end user. However, due to the pixel-wise operation strategy, this approach has a relatively long processing time and does not exploit spatial dependency for each element map. Although this method only takes about 0.01 second to process the spectrum at each pixel, it takes several hours to deconvolute a datacube since an MA-XRF datacube normally consists of hundreds of thousands to millions of pixels. 

In this paper, we propose two versions of a fast automatic deconvolution (FAD) method for the MA-XRF datacubes collected from paintings, one with alternating direction method of multipliers (ADMM) and the other with fast iterative shrinkage-thresholding algorithm (FISTA). Not only do the proposed methods not require manual input by the user, making them fully automatic, but they also process the MA-XRF datacube in a global manner, which significantly reduces the processing time. We first model the XRF spectrum as a combination of multiple element characteristic X-ray lines convolved with a pulse shape. Then the locations (in energy channel) and amplitudes (in photon counts) of the element lines are estimated from the average and maximum spectra of an MA-XRF datacube (which are introduced in Eq. (\ref{eq:ave}) and (\ref{eq:max})). Finally we propose a strategy to leverage those detected element lines to estimate the amount of each chemical element present at each pixel in the painting. Having determined the locations and shapes of each of the estimated chemical element pulses, we formulate the spectrum collected at each pixel as a linear combination of several element pulses. Therefore, the deconvolution of an MA-XRF datacube is converted to retrieving the amplitudes of the individual element pulses (equivalent to the intensities of each element line), which is a matrix factorisation problem and can be solved with variations of ADMM and FISTA.

The remainder of this paper is organised as follows. We formulate the XRF deconvolution problem in Section \ref{sec:background} and introduce the proposed methods in Section \ref{sec:method}. The results are shown and compared with the ones from other methods in Section \ref{sec:numerical results}. Finally, the paper concludes in Section \ref{sec:conclusion}.

\section{Background}
\label{sec:background}
\subsection{MA-XRF Data}

The MA-XRF data included in this paper was acquired using the Bruker M6 JETSTREAM equipment \cite{alfeld2013mobile} at the National Gallery, London, with continuous scanning mode. The resulting MA-XRF data from an easel painting is a three-dimensional datacube, with two spatial and one energy dimensions. The range of the energy dimension is from 0 to 40 kilo-electron-volts (\si{\kilo\electronvolt}) and is divided into 4096 energy channels. As a result, the collected MA-XRF datacube can be represented by $\mathbf{Y}\in \mathbb{R}^{M \times H \times W}$, with $M=4096$ energy channels, $H$ pixels in height and $W$ pixels in width.

Each element line present in the collected XRF spectrum is due to the ionisation of the atoms of a chemical element and the energy of the photons released by that element's atoms equals the energy difference between the electronic shells (or sub-shells) between which electronic transitions occur. Since the atom of each element has unique atomic structure and atomic energy levels, the element lines of an element are produced in series (e.g. $\text{K}$-series, $\text{L}$-series, $\text{M}$-series, etc.) with unique characteristic energies. In this paper, we consider 7 dominant lines, specifically, $\text{K}_{\alpha}$, $\text{K}_{\beta}$, $\text{L}_l$, $\text{L}_{\alpha}$, $\text{L}_{\beta}$, $\text{L}_{\gamma}$ and $\text{M}_{\alpha}$, and a total of 34 elements potentially present in Old Master easel paintings and detectable using the Bruker M6 JETSTREAM equipment. The characteristic energies of these elements are summarised in a lookup table in Appendix \ref{apdx:A}.

\subsection{Problem Formulation}
\label{Sec:XRF_datacube}
As highlighted before, the element lines in XRF spectra are broadened into pulses due to imperfect device performance and the statistical nature of the photon emission process. The shape of the broadened pulse can be well modelled with a Gaussian function \cite{markowicz1993handbook,beckhoff2007handbook}, given by:
\begin{equation}
	\varphi[n] = \exp{\left( -\frac{n^2}{2{\sigma}^2} \right)},
\end{equation}
where the pulse width $\sigma$ increases with the pulse energy \cite{goldstein2017scanning}. Following the pre-processing and calibration steps in our previous work \cite{yan2021prony}, the pulse width can be calculated at any given location in the spectrum. Therefore, the collected XRF spectrum at a certain pixel can be modelled as a convolution of element lines at different locations in the spectrum and the Gaussian functions:
\begin{subequations}
    \begin{align}
        y[n]&=\sum_{k=1}^K a_k\delta(n-t_k)\otimes \varphi_k[n] \\
        &=\sum_{k=1}^K a_k \varphi_k[n-t_k] , \ n=0,1,...,M-1, \\
        &\text{with }  \varphi_k[n-t_k] = \exp{\left( -\frac{(n-t_k)^2}{2{\sigma_k}^2} \right)} ,
    \end{align}
\end{subequations}
where $K$ is the number of element lines present in the spectrum and $a_k$ and $t_k$ are the amplitudes and locations of the $k$-th element line. By constructing a matrix $\mathbf{S}\in \mathbb{R}^{M \times K}$ containing the shapes of the element pulses along each column:
\begin{align}
    \mathbf{S}=
    \begin{bmatrix}
        \vdots  & \vdots & \cdots & \vdots \\
        \varphi_1[n-t_1]  & \varphi_2[n-t_2] & \cdots & \varphi_k[n-t_k] \\
        \vdots & \vdots & \cdots & \vdots \\
    \end{bmatrix} ,
    \label{eq:S}
\end{align}
the spectrum at a certain pixel can then be formulated as a linear combination of these pulses:
\begin{equation}
    \mathbf{y}_{h,w}= \mathbf{S}\mathbf{a}_{h,w} , 
\end{equation}
where the column vector $\mathbf{y}_{h,w}\in \mathbb{R}^M$ is the XRF spectrum of pixel $(h,w)$ and $\mathbf{a}_{h,w}\in \mathbb{R}^K$ represents the amplitude of the pulses. If the matrix $\mathbf{S}$ includes all the element pulses that are present in the MA-XRF datacube, the whole datacube can also be represented as a linear combination of all element pulses:
\begin{equation}
    \mathbf{Y}= \mathbf{S}\mathbf{A},
\end{equation}
where $\mathbf{Y}\in \mathbb{R}^{M \times I}$ represent the MA-XRF datacube with all pixels rearranged in a column vector with $I=H\times W$ and with $\mathbf{A}\in \mathbb{R}^{K \times I}$ containing the amplitudes of the element pulses at all pixels. Under this model, the XRF deconvolution is formulated as the problem of estimating the pulse matrix $\mathbf{S}$ and the amplitude matrix $\mathbf{A}$ from MA-XRF datacube $\mathbf{Y}$, a bilinear inverse problem that is difficult to solve without considering the physics of the device and properties of the datacube. In this paper, we propose to estimate $\mathbf{S}$ leveraging our previous FRI-based Pulse Detection (FRIPD) algorithm in \cite{yan2021prony}. With the estimated $\mathbf{S}$, we then find $\mathbf{A}$ through a constrained optimisation. Finally, the desired distribution maps of element lines are obtained by rearranging each column of $\mathbf{A}$ to form an image of size $H\times W$.

\subsection{Challenges in XRF Deconvolution}

\begin{figure}[t]
	\centering
	\includegraphics[width=0.9\linewidth]{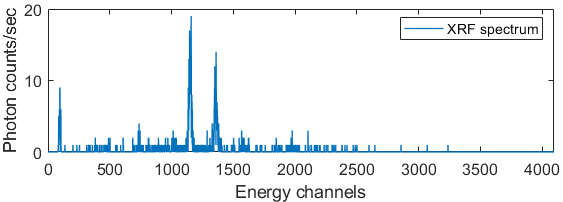}
	\caption{The spectrum at a sample pixel of the MA-XRF datacube collected from a painting with a Bruker M6 JETSTREAM instrument (10 \si{\milli\second} dwell time).}
	\label{fig:XRF_spectrum}
\end{figure}

XRF deconvolution, in particular for MA-XRF datacubes collected from paintings, presents four main challenges listed below:
\begin{enumerate}[leftmargin=*]
	\item As paintings are created using a combination of different pigments, many chemical elements exist at any given spot of the painting. For this reason, it is likely that element pulses overlap in the collected spectra, making it more difficult to deconvolute. 
	
	\item The individual pixel spectra acquired from a painting can be very noisy. This is because the time during which each pixel is illuminated by the X-ray beam, the so-called dwell time, affects the signal-to-noise ratio of the acquired spectrum at a certain pixel. While a longer dwell time results in a better SNR, it also makes it very time-consuming to scan the entire painting. In practice, therefore, a relatively short dwell time is normally employed. All the MA-XRF datacubes used in this paper were collected with a dwell time of 10 \si{\milli\second} and the individual pixel spectra are relatively noisy as shown in Fig. \ref{fig:XRF_spectrum}. 
	
	\item Due to the complex layer structure of a painting, differences in the amount of pigments used as well as the configuration and settings of the device and the sensitivity of the detector, the amplitudes of the element pulses in the XRF spectrum can vary from hundreds of photon counts to only a few. It is therefore common that a weak pulse in the spectrum may be (partially) covered by a much stronger pulse and successfully separating the weak pulse from the strong pulse can be difficult.
	
	\item Although the element pulses in the XRF spectrum can be well modelled with Gaussian functions, their widths are not fixed and increase as their energies increase \cite{Conover2015,goldstein2017scanning}, which brings additional difficulty to deconvoluting the XRF spectrum. From \cite{goldstein2017scanning}, the relationship between the full width at half maximum (FWHM) of the element pulse and its energy in electron-volts (\si{\electronvolt}) can be expressed by the following equation:
	\begin{equation}
		\text{FWHM}\propto \left(C^2E+N^2\right)^{1/2},
		\label{Eq:pulse_width}
	\end{equation}
	where $E$ is the energy of the pulse, $C$ is the uncertainty in the formation of charge carriers and $N$ is the electronic noise.
\end{enumerate}

\section{Proposed Method}
\label{sec:method}

\begin{figure*}[t]
	\centering
	\includegraphics[width=0.88\linewidth]{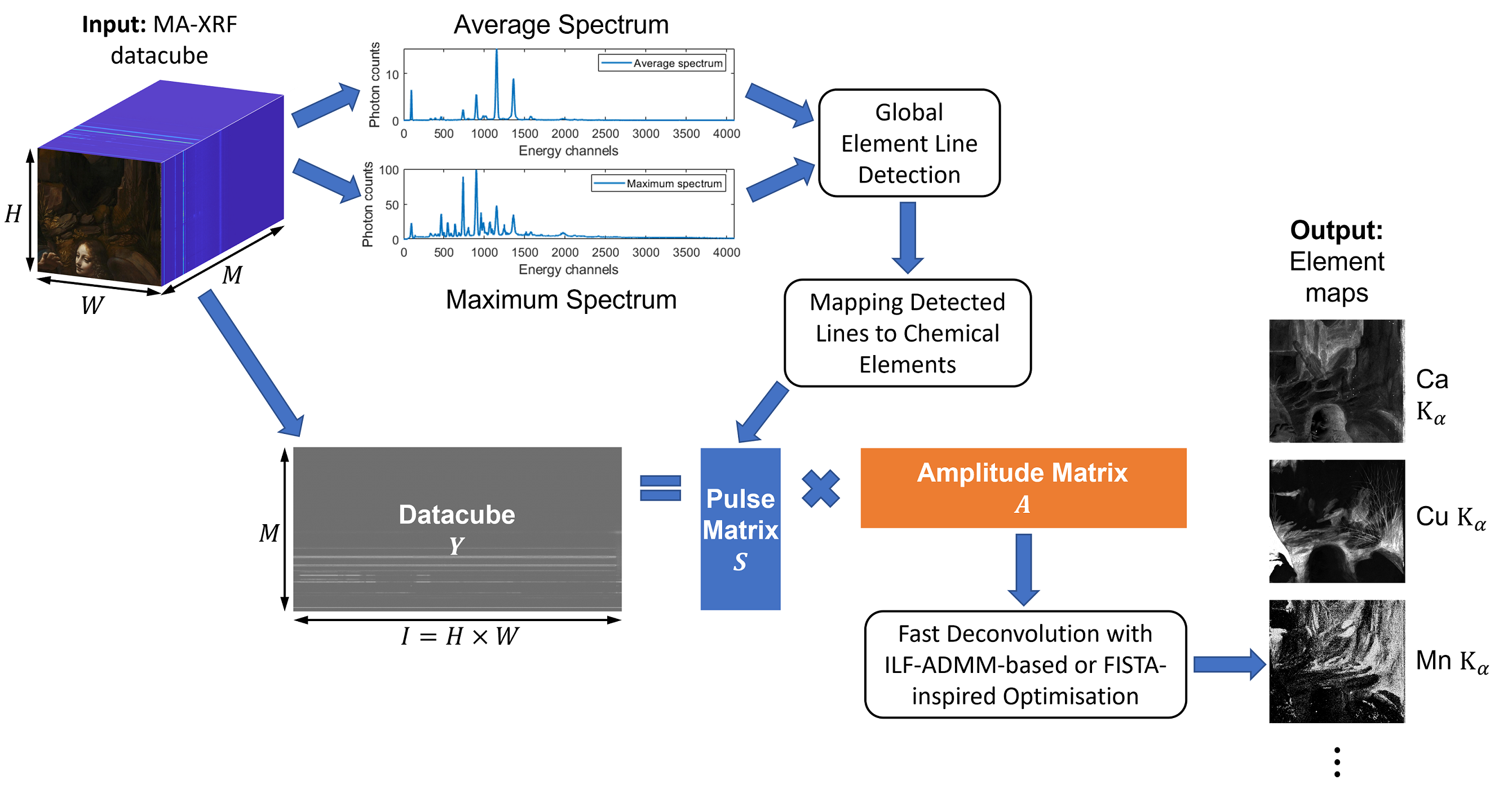}
	\vspace{-0.2cm}
	\caption{The framework of our proposed FAD method for deconvoluting the MA-XRF datacube of an easel painting, which takes the MA-XRF datacube as input and outputs the deconvoluted element maps. Firstly, the algorithm estimates the presence of chemical elements in the painting from the average and maximum spectra. Then, a matrix $\mathbf{S}$ containing the pulse shapes of the detected elements is constructed. The MA-XRF datacube is then modelled as a multiplication of the pulse matrix $\mathbf{S}$ and the amplitude matrix $\mathbf{A}$. Finally, the amplitude matrix is estimated with proposed two optimisation schemes based on ADMM and FISTA, from which the deconvoluted element maps can be extracted.}
	\label{fig:block_diagram}
	\vspace{-0.2cm}
\end{figure*}

In this section, we describe the proposed fast automatic deconvolution (FAD) method for MA-XRF datacubes collected from paintings. As the block diagram in Fig. \ref{fig:block_diagram} shows, the input is the MA-XRF datacube collected from an easel painting and the output is the set of deconvoluted element maps. The algorithm operates first on the average and maximum spectra of the datacube to produce an estimate of the elements present in the painting. Then, a pulse matrix is constructed, with which the MA-XRF deconvolution is modelled as a matrix factorisation inverse problem. Finally, we propose two optimisation schemes based on ADMM and FISTA, respectively, to solve the problem and extract the element maps from the datacube.

\subsection{Global Element Line Detection}
\label{sec:GELD}

We first describe how to detect the element lines of each of the chemical elements that are present in the whole MA-XRF datacube. The proposed method is based on FRI-based Pulse Detection (FRIPD) algorithm in \cite{yan2021prony}.

It has been discussed that the spectrum of a pixel in an MA-XRF datacube can be modelled as a convolution of the element lines and the Gaussian-shaped pulse function. If we divide the XRF spectrum into $J$ overlapping windows of a sufficiently small size, we can assume that the pulse shape within each window is the same. Under this assumption, the XRF spectrum can be modelled as a linear combination of several element lines convolved with the same pulse shape plus noise:
\begin{equation}
    \begin{aligned}
        y_{h,w,j}[n]=\sum_{k=1}^{K_{h,w,j}} a_{h,w,j,k} & \varphi_j[n-t_{h,w,j,k}]+\epsilon_{h,w,j}[n], \\ 
        n&=0,1,...,M-1,
    \end{aligned}
    \label{Eq:data_represetation}
\end{equation}
where $y_{h,w,j}[n]$ is the spectrum of the pixel $(h,w)$ in the $j$-th window region, $K_{h,w,j}$ is the number of element lines present in that window, $a_{h,w,j,k}$ and $t_{h,w,j,k}$ are the amplitudes and locations of the element lines, and $\epsilon_{h,w,j}[n]$ is the noise term. Under the model in (\ref{Eq:data_represetation}), the deconvolution problem reduces to that of retrieving the parameters $t_{h,w,j,k}$ and $a_{h,w,j,k}$ in the signal $y_{h,w,j}[n]$. 
In \cite{yan2021prony}, we solved this parameter estimation problem by connecting it to methods used in spectral estimation called Prony's method \cite{Prony1795} and the matrix pencil method \cite{hua1990matrix}. We showed that in this way we can retrieve the locations of the pulses accurately even in situation of having nearby or weak pulses. However, the downside of the AFRID method in \cite{yan2021prony} is that it operates pixel by pixel and this leads to a method that is relatively slow and does not exploit spatial dependency.

In contrast to our previous work, here we apply the FRIPD algorithm only on the average spectrum and maximum spectrum of the datacube to speed up the deconvolution process. The two spectra theoretically contain all the element lines that are present in the datacube and are routinely evaluated by MA-XRF end users during datacube acquisition and interrogation. The average spectrum $\bar{y}$ is generated by taking the average of the spectra at all pixels in the datacube:
\begin{equation}
    \bar{y}[n]=\frac{1}{W\times H}\sum_{w=1}^{W}\sum_{h=1}^{H} y_{h,w}[n], \ n=0,1,...,M-1, 
    \label{eq:ave}
\end{equation}
where $y_{h,w}[n]$ is the spectrum at pixel $(h,w)$ and $W\times H$ is the total number of pixels in the datacube. The maximum spectrum $\hat{y}$ is given by taking the maximum value of all the pixels in the datacube at each channel:
\begin{equation}
    \hat{y}[n]=\max\left( \left\{ y_{h,w}[n] \right\}_{(h,w)=(1,1)}^{(W,H)} \right), \ n=0,1,...,M-1.
    \label{eq:max}
\end{equation}
Moreover, both spectra within each window region can still be modelled in a form similar to (\ref{Eq:data_represetation}), given by:
\begin{equation}
    \bar{y}_j[n]=\sum_{k=1}^{K_{j}} \bar{a}_{j,k}\varphi[n-t_{j,k}]+\bar{\epsilon}_{j}[n],
    \label{Eq:data_represetation_2}
\end{equation}
and
\begin{equation}
    \hat{y}_j[n]=\sum_{k=1}^{K_{j}} \hat{a}_{j,k}\varphi[n-t_{j,k}]+\hat{\epsilon}_{j}[n], 
    \label{Eq:data_represetation_3}
\end{equation}
where $K_{j}$ is the total number of element lines that are present in any pixel of the datacube within window $j$. Therefore, $t_{j,k}$ represents the locations of those element lines, $\bar{a}_{j,k}$ and $\hat{a}_{j,k}$ represent the amplitudes of those element lines in the average and maximum spectrum, respectively, $\bar{\epsilon}_{j}[n]$ and $\hat{\epsilon}_{j}[n]$ are the noise terms. Given the representations in (\ref{Eq:data_represetation_2}) and (\ref{Eq:data_represetation_3}), the locations ($\{t_{k}\}_{k=1}^{K}$) and amplitudes ($\{\bar{a}_{k}\}_{k=1}^{K}$, $\{\hat{a}_{k}\}_{k=1}^{K}$) of all element lines can also be retrieved using the FRIPD algorithm, where $K=\sum_{j=1}^J K_{j}$ is the number of the total element lines present in the MA-XRF datacube.

Although the average and maximum spectra should both contain the element lines of all the chemical elements present in the datacube, the lines that FRIPD algorithm detects from the two spectra can be different. For example, while the average spectrum normally has a good signal-to-noise ratio, the averaging operation may reduce significantly the amplitude of the element lines of chemical elements that are only present in small areas in the painting or when present in small quantities. To solve this problem, we also use the maximum spectrum, from which the element lines of those trace elements are generally more evident. Applying the FRIPD algorithm on the two spectra leads to two sets of element lines at locations ($\{\bar{t}_{k}\}_{k=1}^{\bar{K}}$ and $\{\hat{t}_{k}\}_{k=1}^{\hat{K}}$) with amplitudes ($\{\bar{a}_{k}\}_{k=1}^{\bar{K}}$ and $\{\hat{a}_{k}\}_{k=1}^{\hat{K}}$) respectively, where $\bar{K}$ and $\hat{K}$ are the total numbers of detected element lines in the two spectra.

\begin{figure*}[t]
\centering
    \begin{subfigure}{0.32\linewidth}
        \centering
        \includegraphics[height=0.5\linewidth]{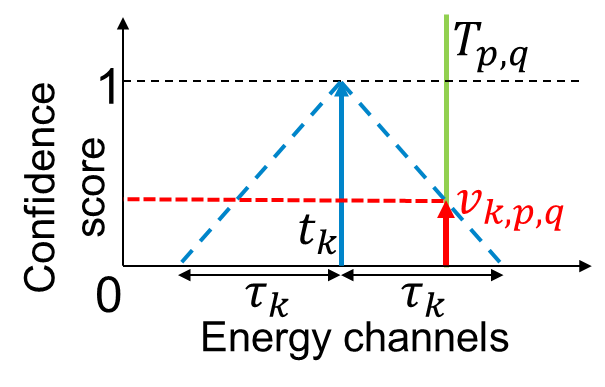}
        \vspace{-0.3cm}
        \caption{}
    \end{subfigure}
    \begin{subfigure}{0.32\linewidth}
        \centering
        \includegraphics[height=0.5\linewidth]{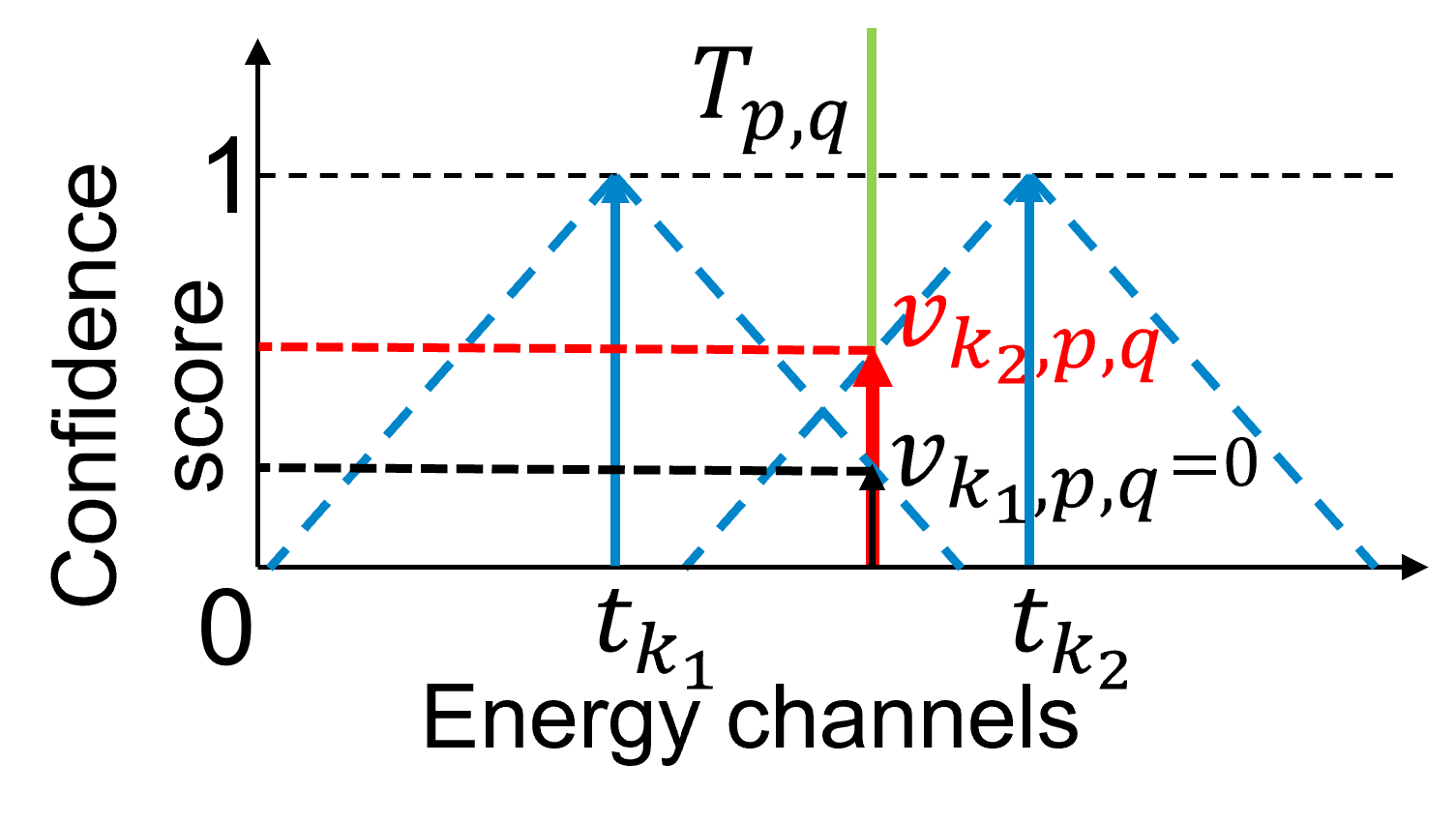}
        \vspace{-0.3cm}
        \caption{}
    \end{subfigure}
    \begin{subfigure}{0.32\linewidth}
        \centering
        \includegraphics[height=0.5\linewidth]{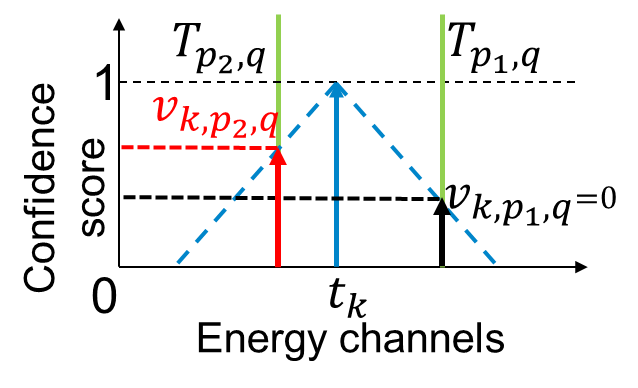}
        \vspace{-0.3cm}
        \caption{}
    \end{subfigure}
\caption{Computation of Line Confidence Score (LCS). Green lines represent the theoretical locations $T_{p,q}$ of the element lines. Blue arrows represent the locations $\{t_k\}$ of the detected pulses. Blue dashed lines represent the confidence score function defined in (\ref{eq:LCS}). Red arrows denote the computed LCS $v_{k,p,q}$ and black arrows in (b) and (c) are LCS set to zero with the two constraints. $v_{k,p,q}$ represents the LCS corresponding to the $k$-th detected line for the $q$-th element line of the $p$-th element. The value of $\tau_k$ depends on the amplitudes $a_k$ of the detected pulses and is defined in \cite{yan2021prony}.}
\label{fig:confidence}
\vspace{-0.1cm}
\end{figure*}

\subsection{Assignment of Detected Lines to Chemical Elements}
\label{sec:element_estimation}

Given the element lines detected from the average and maximum spectra, we need to assign them to the correct chemical element, according to the difference between the detected locations and the theoretical ones. A parameter, confidence score, is used here for quantifying the presence of chemical elements. It represents the confidence that we have correctly allocated a detected element line to a certain chemical element and is a value ranging from zero to one. 

\subsubsection{Line Confidence Score}
We first assign the lines (or pulses) detected from the spectra to the theoretical element lines by computing the difference between the detected locations and the theoretical ones, as follows:
\begin{equation}
    v_{k,p,q}=\max \left(  1 - \frac{|t_k-T_{p,q}|}{\tau_k} ,0 \right).
    \label{eq:LCS}
\end{equation}
Here $v_{k,p,q}$ is the line confidence score (LCS) computed for the $k$-th detected pulse and the $q$-th element line of the $p$-th element, $T_{p,q}$ is the theoretical location of the $q$-th element line of the $p$-th element. Remember that we consider only 34 elements and that each element has only 7 characteristic element lines ($\text{K}_{\alpha}$, $\text{K}_{\beta}$, $\text{L}_l$, $\text{L}_{\alpha}$, $\text{L}_{\beta}$, $\text{L}_{\gamma}$ and $\text{M}_{\alpha}$). Moreover, $\tau_k$ is an uncertainty factor indicating the maximum distance allowed between a detected pulse and its assigned element line, and it was derived depending on the amplitudes $a_k$ of the detected pulses based on Cram\'{e}r-Rao bounding techniques (see \cite{yan2021prony} for more details). A pulse with a larger amplitude will be given a smaller uncertainty $\tau_k$ because we are more confident to detect a pulse with larger amplitude. Although the statistics of the noise in the average spectrum and maximum spectrum are different from the noise in the spectrum at a single pixel, experiments show that the dependency of $\tau_k$ on amplitude $a_k$ still holds. When assigning the detected lines to the chemical element lines, we impose two constraints for better assignment: a) When an element line of a chemical element has non-zero LCS with more than one detected pulse, the highest LCS is kept and the others are set to zero. This is to ensure that only the closest detected pulse is assigned to each element line. b) When a detected pulse has non-zero LCS with respect to more than one element line of the same element, the highest LCS is kept and the others are set to zero. This is to ensure that one detected pulse can only be assigned to one element line of a certain element. The computation of the LCS with these two constraints is illustrated in Fig. \ref{fig:confidence}, after which an element line of a chemical element is assigned with at most one detected pulse. As a result,
\begin{equation}
    \tilde{v}_{p,q}=\sum_{k=1}^{K} v_{k,p,q},
    \label{Eq:3D_to_2D}
\end{equation}
can represent the LCS for the $q$-th element line of the $p$-th element.

\subsubsection{Element Confidence Score}
We then combine the line confidence scores (LCSs) of the various lines associated with a particular chemical element to estimate its presence in the XRF spectra. Since the $\alpha$-line is the most dominant in each line family of a chemical element, we compute the confidence score of a line family for an element, the so-called family confidence score (FCS), with more dependence on its $\alpha$-line, as the maximum between the LCS of the $\alpha$-line and the averaged LCS of all the element lines in this family. Due to the fact that the spectral range of an MA-XRF device is limited and the sensitivity to photons varies with the energy channel, the element lines of $\text{K}$, $\text{L}$ and $\text{M}$ families of a chemical element can be present in the spectrum independently. Hence, we define the confidence score of a chemical element, the so-called element confidence score (ECS), as the maximum among the FCS of the three families of that element. We also note that the element lines in $\text{K}$ and $\text{L}$ families of most elements are in the energy range where the XRF equipment has good sensitivity\cite{alfeld2013mobile}, making these lines more likely to be detected within the spectra. For this reason, if no element line in the $\text{K}$ or $\text{L}$ family of an element is detected, this element is considered not present in the datacube and its ECS is set to zero.

Following the strategy above and given the two sets of pulses detected from the average and maximum spectra, we can obtain two sets of chemical elements with non-zero ECS, denoted by $\bm{\bar{\Phi}}$ and $\bm{\hat{\Phi}}$, respectively. Finally, we consider all elements detected in either $\bm{\bar{\Phi}}$ or $\bm{\hat{\Phi}}$ as the chemical elements present in the datacube and use them to build the pulse matrix $\mathbf{S}$ in Eq. (\ref{eq:S}).

\subsection{Fast Deconvolution}
Given the detected chemical elements considered to be present in the datacube and given the theoretical locations of their element lines denoted with $\{t_{k}\}_{k=1}^{K}$, we can then build the matrix $\mathbf{S}$ as follows. Each element line is broadened into a Gaussian-shaped pulse in the spectrum, which can be formulated as:
\begin{equation}
    \varphi_k[n-t_k] = \exp{\left( -\frac{(n-t_k)^2}{2{\sigma_k}^2} \right)},
\end{equation}
whose width $\sigma_k$ can be calculated using Eq. (\ref{Eq:pulse_width}) as explained in \cite{yan2021prony}. Then we can construct the matrix $\mathbf{S}\in \mathbb{R}^{M \times K}$ containing all element pulses, given by:
\begin{align}
    \mathbf{S}=
    \begin{bmatrix}
        \vdots  & \vdots & \cdots & \vdots \\
        \varphi_1[n-t_1]  & \varphi_2[n-t_2] & \cdots & \varphi_k[n-t_k] \\
        \vdots & \vdots & \cdots & \vdots \\
    \end{bmatrix} .
\end{align}
As a result, the MA-XRF datacube $\mathbf{Y}\in \mathbb{R}^{M \times I}$ can be decomposed into a multiplication of the pulse matrix $\mathbf{S}$ and a matrix $\mathbf{A}\in \mathbb{R}^{K \times I}$ containing the amplitudes of these element pulses at all pixels in the datacube, expressed as:
\begin{equation}
    \mathbf{Y}= \mathbf{S}\mathbf{A}.
    \label{eq:Y=SA}
\end{equation}
Now, the problem of deconvoluting the MA-XRF datacube reduces to the problem of estimating the amplitude matrix $\mathbf{A}$ in (\ref{eq:Y=SA}): 
\begin{equation}
    \mathbf{\hat{A}}=\argmin_{\mathbf{A}}\frac{1}{2}\|\mathbf{Y}-\mathbf{S}\mathbf{A}\|^{2}_{F},
    \label{eq:inverse problem}
\end{equation}
where $\|\cdot\|_{F}$ denotes the Frobenius norm. Note that the pulse matrix $\mathbf{S}$ contains the element pulses of all the potential chemical elements present in the datacube, however, the chemical elements present at each pixel are a subset of those in the whole painting and may vary at different pixels, which makes the optimisation for (\ref{eq:inverse problem}) unreliable. We therefore introduce additional constraints to enhance the stability and robustness of the optimisation.

\subsection{Proposed Solutions}

In order to obtain reliable deconvoluted element maps, we impose a few constraints into the estimation problem. Since the amplitudes of the element pulses in XRF spectra are non-negative, we first impose a positivity constraint to the amplitude matrix $\mathbf{A}$. Moreover, the total-variation (TV) regularisation proposed in \cite{rudin1992nonlinear} can be used in image restoration and deconvolution problems to impose spatial smoothness and to reduce the noise in the reconstructed image \cite{babacan2008variational,figueiredo2010restoration}. We therefore impose TV to enhance the quality of the estimated element maps. More importantly, we also propose a physics-based constraint by considering that the intensities of some characteristic lines of an element cannot be larger than some of its other lines. As a result, the estimation problem in (\ref{eq:inverse problem}) becomes:
\begin{equation}
    \begin{aligned}
        \mathbf{\hat{A}}=&\argmin_{\mathbf{A}}\frac{1}{2}\|\mathbf{Y}-\mathbf{S}\mathbf{A}\|^{2}_{F}+  \mathcal{I}_{+}(\mathbf{A}) \\
        &+ \lambda\| \mathbf{D}_x \mathbf{A}\|_1 +\lambda\| \mathbf{D}_y \mathbf{A}\|_1 + \mathcal{P}(\mathbf{A}),
    \end{aligned}
    \label{eq:inverse problem 2}
\end{equation}
where $\mathcal{I}_{+}(\cdot)$ imposes the positivity constraint, $\mathbf{D}_x$ and $\mathbf{D}_y$ are the derivatives along $x$ and $y$ axes, which impose the TV regularisation on the spatial dimensions, $\lambda$ is the TV parameter that controls the spatial smoothness of the estimated element maps, $\|\cdot\|_{1}$ denotes the $\ell_1$ norm and $\mathcal{P}(\cdot)$ imposes the physical constraint.

\begin{algorithm}[t]
	\KwInput{MA-XRF datacube $\mathbf{Y}$ and pulse matrix $\mathbf{S}$}
	\KwOutput{Estimated pulse amplitude matrix $\mathbf{\hat{A}}$}
	Initialise matrices $\mathbf{A}^{(0)}=\mathbf{0}$, $\mathbf{Z}_1^{(0)}=\mathbf{0}$, $\mathbf{Z}_2^{(0)}=\mathbf{0}$, $\mathbf{Z}_3^{(0)}=\mathbf{0}$, $\mathbf{\tilde{Z}}_1^{(0)}=\mathbf{0}$, $\mathbf{\tilde{Z}}_2^{(0)}=\mathbf{0}$, $\mathbf{\tilde{Z}}_3^{(0)}=\mathbf{0}$ \\
	$m=1$\\
	\While{\text{not converged}}{
		$\mathbf{Z}_1^{(m)}=\text{prox}_{\|\cdot\|_{1},\lambda}(\mathbf{D}_x \mathbf{A}^{(m-1)}-\mathbf{\tilde{Z}}_1^{(m-1)})$ \label{Alg1:step4}\\
		$\mathbf{Z}_2^{(m)}=\text{prox}_{\|\cdot\|_{1},\lambda}(\mathbf{D}_y \mathbf{A}^{(m-1)}-\mathbf{\tilde{Z}}_2^{(m-1)})$\\
		$\mathbf{\hat{Z}}_3^{(m)}=\text{prox}_{\mathcal{I}_{+}}(\mathbf{A}^{(m-1)}-\mathbf{\tilde{Z}}_3^{(m-1)})$ \\
		$\mathbf{Z}_3^{(m)}=\text{prox}_{\mathcal{P}}(\mathbf{\hat{Z}}_3^{(m)})$ \label{Alg1:step7}\\
		$\mathbf{A}^{(m)}=
		\{\mathbf{S}^{\mathsf{T}} \mathbf{Y} + \rho_1 \mathbf{D}_x^{\mathsf{T}}(\mathbf{Z}_1^{(m)}+\mathbf{\tilde{Z}}_1^{(m-1)}) + \rho_2 \mathbf{D}_y^{\mathsf{T}}(\mathbf{Z}_2^{(m)}+\mathbf{\tilde{Z}}_2^{(m-1)}) + \rho_3 (\mathbf{Z}_3^{(m)}+\mathbf{\tilde{Z}}_3^{(m-1)}) + (\mathbf{L}-\mathbf{S}^{\mathsf{T}} \mathbf{S})\mathbf{A}^{(m-1)}
		\} \{\rho_1 \mathbf{D}_x^{\mathsf{T}} \mathbf{D}_x + \rho_2 \mathbf{D}_y^{\mathsf{T}} \mathbf{D}_y + \rho_3 \mathbf{I} + \mathbf{L}\}^{-1}$\label{eq:update_X}\\
		$\mathbf{\tilde{Z}}_1^{(m)}=\mathbf{\tilde{Z}}_1^{(m-1)}-\mathbf{D}_x \mathbf{A}^{(m)}+\mathbf{Z}_1^{(m)}$\\
		$\mathbf{\tilde{Z}}_2^{(m)}=\mathbf{\tilde{Z}}_2^{(m-1)}-\mathbf{D}_y \mathbf{A}^{(m)}+\mathbf{Z}_2^{(m)}$\\
		$\mathbf{\tilde{Z}}_3^{(m)}=\mathbf{\tilde{Z}}_3^{(m-1)}-\mathbf{A}^{(m)}+\mathbf{Z}_3^{(m)}$\\
		$m=m+1$ }
	$\mathbf{\hat{A}}=\mathbf{A}^{(m)}$
	\caption{Proposed optimisation Scheme for (\ref{eq:ADMM inverse problem}) with ILF-ADMM \cite{donati2019inner}}
	\label{Alg:ILF_ADMM}
\end{algorithm}
\begin{algorithm}[t]
\KwInput{MA-XRF datacube $\mathbf{Y}$ and pulse matrix $\mathbf{S}$}
\KwOutput{Estimated pulse amplitude matrix $\mathbf{\hat{A}}$}
Initialise coefficient $d_{(0)}=0$ \\
Initialise matrices $\mathbf{A}^{(-1)}=\mathbf{A}^{(0)}=\mathbf{0}$, $\mathbf{Z}_1^{(-1)}=\mathbf{Z}_1^{(0)}=\mathbf{0}$, $\mathbf{Z}_2^{(-1)}=\mathbf{Z}_2^{(0)}=\mathbf{0}$, so that $\mathbf{X}^{(-1)}=\left[\mathbf{A}^{(-1)},\mathbf{Z}_1^{(-1)},\mathbf{Z}_2^{(-1)}\right]^{\mathsf{T}}=\mathbf{0}$ and $\mathbf{X}^{(0)}=\left[\mathbf{A}^{(0)},\mathbf{Z}_1^{(0)},\mathbf{Z}_2^{(0)}\right]^{\mathsf{T}}=\mathbf{0}$ \\
$m=1$ \\
\While{\text{not converged}}{
$d_{(m)}=\frac{1+\sqrt{1+4{d_{(m-1)}}^2}}{2}$ \label{step:4}\\
$\mathbf{U}^{(m)}=\mathbf{X}^{(m-1)}+\frac{d_{(m-1)}-1}{d_{(m)}}(\mathbf{X}^{(m-1)}-\mathbf{X}^{(m-2)})$ \label{step:5}\\
$\mathbf{W}^{(m)}=\mathbf{U}^{(m)}-\frac{1}{\alpha}\mathbf{H}^{\mathsf{T}}(\mathbf{H}\mathbf{U}^{(m)}-\mathbf{\ddot{Y}})$ \label{step:6}\\
$\left[\mathbf{W}_1^{(m)},\mathbf{W}_2^{(m)},\mathbf{W}_3^{(m)}\right]^{\mathsf{T}}=\mathbf{W}^{(m)}$ \label{step:7}\\
$\mathbf{\tilde{A}}^{(m)}=\text{prox}_{\mathcal{I}_{+}}(\mathbf{W}_1^{(m)})$ \label{step:8}\\
$\mathbf{A}^{(m)}=\text{prox}_{\mathcal{P}}(\mathbf{\tilde{A}}^{(m)})$ \\
$\mathbf{Z}_1^{(m)}=\text{prox}_{\|\cdot\|_{1},\lambda}(\mathbf{W}_2^{(m)})$ \label{step:9}\\
$\mathbf{Z}_2^{(m)}=\text{prox}_{\|\cdot\|_{1},\lambda}(\mathbf{W}_3^{(m)})$ \label{step:10}\\
$\mathbf{X}^{(m)}=\left[{\mathbf{A}}^{(m)},\mathbf{Z}_1^{(m)},\mathbf{Z}_2^{(m)}\right]^{\mathsf{T}}$ \label{step:12}\\
$m=m+1$ \label{step:13}}
$\mathbf{\hat{A}}=\mathbf{A}^{(m)}$
\caption{Proposed FISTA-Inspired Optimisation Scheme for (\ref{eq:FISTA inverse problem})}
\label{Alg:FISTA}
\end{algorithm}

\subsubsection{ADMM-based Solution}
Alternating direction method of multipliers (ADMM) \cite{boyd2011distributed} is a popular method to solve the optimisation problem in (\ref{eq:inverse problem 2}). To increase the convergence speed, we adapt an inner-loop-free ADMM method (ILF-ADMM), which can avoid the inner gradient descent loops, proposed in \cite{donati2019inner} to solve (\ref{eq:inverse problem 2}). We need to introduce three auxiliary variables $\mathbf{Z}_1$, $\mathbf{Z}_2$ and $\mathbf{Z}_3$ so that (\ref{eq:inverse problem 2}) becomes:
\begin{equation}
    \begin{aligned}
	\mathbf{\hat{A}}=&\argmin_{\mathbf{A}}\frac{1}{2}\|\mathbf{Y}-\mathbf{S}\mathbf{A}\|^{2}_{F}+  \mathcal{I}_{+}(\mathbf{Z}_3) \\
	&+ \mathcal{P}(\mathbf{Z}_3) + \lambda\| \mathbf{Z}_1\|_1 +\lambda\| \mathbf{Z}_2\|_1 , \\
	\text{s.t.} \ \ \ \ & \ \mathbf{Z}_1 =\mathbf{D}_x \mathbf{A}, \
	\mathbf{Z}_2=\mathbf{D}_y \mathbf{A}, \
	\mathbf{Z}_3=\mathbf{A}.
    \end{aligned}
    \label{eq:ADMM inverse problem}
\end{equation}
Following the ILF-ADMM scheme \cite{donati2019inner}, then we need to iteratively update all the variables and the optimisation scheme is summarised in Algorithm \ref{Alg:ILF_ADMM}. Here $\mathbf{\tilde{Z}}_1$, $\mathbf{\tilde{Z}}_2$ and $\mathbf{\tilde{Z}}_3$ are dual variables used in the ADMM loop, while $\rho_1$, $\rho_2$ and $\rho_3$ are importance coefficients that control different constraints. In steps \ref{Alg1:step4}-\ref{Alg1:step7}, $\text{prox}_{\mathcal{I}_{+}}$, $\text{prox}_{\mathcal{P}}$ and $\text{prox}_{\|\cdot\|_{1},\lambda}$ are proximal operators that ensure the estimated $\mathbf{Z}_1^{(m)}$, $\mathbf{Z}_2^{(m)}$ and $\mathbf{Z}_3^{(m)}$ meet the constraints for $\mathbf{Z}_1$, $\mathbf{Z}_2$ and $\mathbf{Z}_3$ in (\ref{eq:ADMM inverse problem}), respectively. The proximal operator $\text{prox}_{\|\cdot\|_{1},\lambda}$ with $\ell_1$ constraints is equivalent to the element-wise soft thresholding operators, expressed as:
\begin{equation}
    \text{prox}_{\|\cdot\|_{1},\lambda}(z)=
    \begin{cases}
        z-\lambda, \ \text{if} \  z>\lambda, \\
        0, \ \text{if} \ |z|\leq\lambda, \\
        z+\lambda, \ \text{if} \  z<-\lambda.
    \end{cases} 
\end{equation}
Then the proximal operator $\text{prox}_{\mathcal{I}_{+}}$ with non-negative constraint simply sets the negative entries to zero:
\begin{equation}
    \text{prox}_{\mathcal{I}_{+}}(z)=
    \begin{cases}
        z, \ \text{if} \  z>0, \\
        0, \ \text{if} \ z\leq 0.
    \end{cases} 
\end{equation}
Note that the proximal operator $\text{prox}_{\mathcal{P}}$ for the physical constraint is proposed based on the fact that the amplitude of the $\alpha$ line in the $\text{L}$ family of an element is normally larger than the other lines and the amplitude of the $\alpha$ line in the $\text{K}$ family is normally at least twice the amplitude of the other lines. To simplify the computation, we approximate the proximal operator for the physical constraint as follows:
\begin{equation}
    \text{prox}_{\mathcal{P}}(z_{p,q}) =
    \begin{cases}
        \min(z_{p,q},z_{p,\text{L}\alpha}), \ \text{if} \ q \in \text{L family}, \\
        \min(z_{p,q},\frac{1}{2}z_{p,\text{K}\alpha}), \ \text{if} \ q \in \text{K family},
    \end{cases}
    \label{eq:adjustment}
\end{equation}
where $z_{p,q}$ represents the estimated amplitudes of the $q$-th characteristic line of the $p$-th chemical element at a certain pixel. In step \ref{eq:update_X}, $\mathbf{L}$ is a scalar matrix added to avoid the inner gradient descent loops in traditional ADMM, given by $\mathbf{L}=\|\mathbf{S}\|^2 \mathbf{I}$. By adding $\mathbf{L}$, the matrix $(\rho_1 \mathbf{D}_x^{\mathsf{T}} \mathbf{D}_x + \rho_2 \mathbf{D}_y^{\mathsf{T}} \mathbf{D}_y + \rho_3 \mathbf{I} + \mathbf{L})$ in step \ref{eq:update_X} is circulant so that it is equivalent to a circular convolution, which can be transferred into element-wise multiplication in the Fourier domain.  As a result, the equation in step \ref{eq:update_X} for updating $\mathbf{A}$ can be solved using fast Fourier transform instead of gradient descent, given by:
\begin{equation}
	\begin{aligned}
		\mathbf{A}^{(m)}= & \ \mathcal{F}^{-1} \{ \mathcal{F}(\mathbf{G})/[\rho_1{\mathcal{F}(\mathbf{d}_x)}^* \cdot \mathcal{F}(\mathbf{d}_x) \\ 
		&+ \rho_2{\mathcal{F}(\mathbf{d}_y)}^* \cdot \mathcal{F}(\mathbf{d}_y) + \rho_3 + \|\mathbf{S}\|^2] \} ,
	\end{aligned}
\end{equation}
with
\begin{equation}
	\begin{aligned}
		\mathbf{G}= & \ \mathbf{S}^{\mathsf{T}} \mathbf{Y} + \rho_1 \mathbf{D}_x^{\mathsf{T}}(\mathbf{Z}_1^{(m)}+\mathbf{\tilde{Z}}_1^{(m-1)}) \\ 
		& +\rho_2 \mathbf{D}_y^{\mathsf{T}}(\mathbf{Z}_2^{(m)}+ \mathbf{\tilde{Z}}_2^{(m-1)}) + \rho_3 (\mathbf{Z}_3^{(m)}+\mathbf{\tilde{Z}}_3^{(m-1)}) \\
		&+ (\mathbf{L}-\mathbf{S}^{\mathsf{T}} \mathbf{S})\mathbf{A}^{(m-1)},
	\end{aligned}
\end{equation}
where $\cdot$ and $/$ represent the element-wise multiplication and division, $\mathcal{F}$ and $\mathcal{F}^{-1}$ represent the fast Fourier transform and inverse fast Fourier transform operations, $(\cdot)^*$ is the complex conjugate, $\mathbf{d}_x$ and $\mathbf{d}_y$ are the convolution kernels of the derivative operations $\mathbf{D}_x$ and $\mathbf{D}_y$. Since ILF-ADMM can avoid using inner gradient descent loops, it is much faster than the traditional ADMM.

\subsubsection{FISTA-based Solution}
The ILF-ADMM method still requires computation of the auxiliary and dual variables in the iterative loop, which still makes the whole process complex and computationally expensive. We therefore also propose a solution with fewer auxiliary variables based on the idea of the fast iterative shrinkage-thresholding algorithm (FISTA) \cite{beck2009fast}, which has lower computational complexity. If we use two auxiliary variables $\mathbf{Z}_1$ and $\mathbf{Z}_2$ to represent the derivatives of $\mathbf{A}$ along $x$ and $y$ axes:
\begin{align}
    \mathbf{Z}_1&=\mathbf{D}_x \mathbf{A}, \\
    \mathbf{Z}_2&=\mathbf{D}_y \mathbf{A},
\end{align}
then (\ref{eq:inverse problem 2}) can be simplified as follows:
\begin{equation}
\begin{aligned}
    \mathbf{\hat{A}}=\argmin_{\mathbf{A},\mathbf{Z}_1,\mathbf{Z}_2}
    \frac{1}{2}\left\|
    \begin{bmatrix}
        \rho\mathbf{Y} \\
        \mathbf{0} \\
        \mathbf{0}
    \end{bmatrix}
    -
    \begin{bmatrix}
        \rho\mathbf{S} & \mathbf{0} & \mathbf{0} \\
        \mathbf{D}_x & -\mathbf{I} & \mathbf{0} \\
        \mathbf{D}_y & \mathbf{0} & -\mathbf{I}
    \end{bmatrix} 
    \begin{bmatrix}
        \mathbf{A} \\
        \mathbf{Z}_1 \\
        \mathbf{Z}_2
    \end{bmatrix}
    \right\|^{2}_{F} \\
    + \mathcal{I}_{+}(\mathbf{A}) + \mathcal{P}(\mathbf{A}) +\lambda\| \mathbf{Z}_1\|_1 +\lambda\| \mathbf{Z}_2\|_1,
\end{aligned}
\label{eq:inverse problem 3}
\end{equation}
where $\mathbf{0}$ is the zero matrix and $\mathbf{I}$ is the identity matrix. Parameter $\rho$ controls the tradeoff between finding the optimal $\mathbf{A}$ with all non-negative entries from the datacube $\mathbf{Y}$ and imposing the spatial smoothness to the estimated maps. In our experiments, we set $\rho=1$. To simplify notations, we also re-write (\ref{eq:inverse problem 3}) as follows:
\begin{align}
    \mathbf{\hat{A}}=&\argmin_{\mathbf{A},\mathbf{Z}_1,\mathbf{Z}_2}
    \frac{1}{2}\left\|
    \mathbf{\ddot{Y}}-\mathbf{H}\mathbf{X}
    \right\|^{2}_{F} 
    + \mathcal{I}_{+}(\mathbf{X}) \nonumber \\ 
    &+ \mathcal{P}(\mathbf{A}) +\lambda\| \mathbf{Z}_1\|_1 +\lambda  \| \mathbf{Z}_2\|_1 , 
    \label{eq:FISTA inverse problem}\\
    \text{with} \ \mathbf{\ddot{Y}}=
    \begin{bmatrix}
        \mathbf{Y} \\
        \mathbf{0} \\
        \mathbf{0}
    \end{bmatrix}&, \
    \mathbf{H}=
    \begin{bmatrix}
        \mathbf{S} & \mathbf{0} & \mathbf{0} \\
        \mathbf{D}_x & -\mathbf{I} & \mathbf{0} \\
        \mathbf{D}_y & \mathbf{0} & -\mathbf{I}
    \end{bmatrix} \text{and} \ 
    \mathbf{X}=
    \begin{bmatrix}
        \mathbf{A} \\
        \mathbf{Z}_1 \\
        \mathbf{Z}_2
    \end{bmatrix} . \nonumber
\end{align}
Now the optimisation problem in (\ref{eq:FISTA inverse problem}) can be solved using a variation of FISTA \cite{beck2009fast} with details explained in Algorithm \ref{Alg:FISTA}.
In particular, in steps \ref{step:4}-\ref{step:6} the iterative optimisation is based on a specific linear combination of the previous two points ($\mathbf{X}^{(m-1)}$ and $\mathbf{X}^{(m-2)}$) to have a faster convergence. Gradient descent is applied in step \ref{step:6} to gradually find the local minimum of the cost function.

\section{Numerical Results}
\label{sec:numerical results}

\begin{figure}[t]
	\centering
	\includegraphics[width=0.6\linewidth]{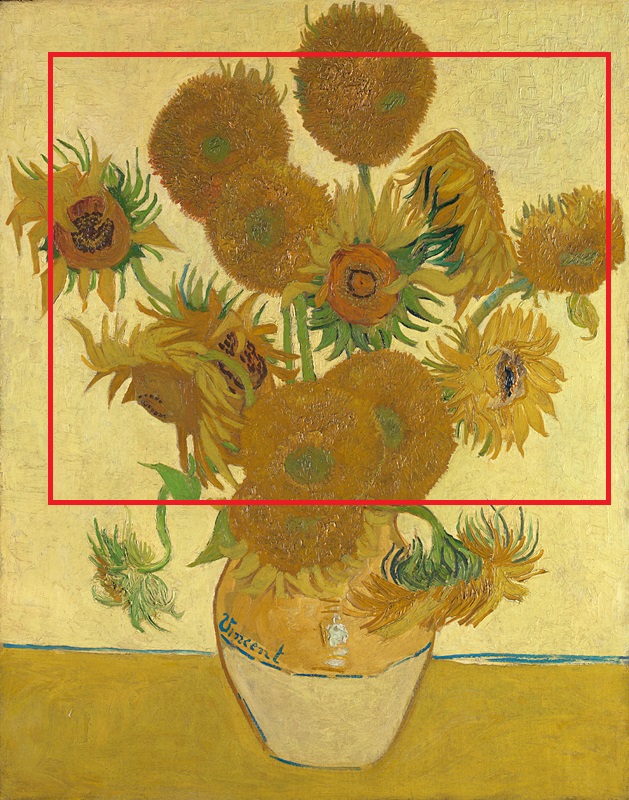}
	\caption{Vincent van Gogh, \emph{Sunflowers} (NG3863), 1888. Oil on canvas \cite{NG_Technical_37, hendriks2019van}. \copyright  The National Gallery, London. Highlighted is the region scanned with a Bruker M6 JETSTREAM instrument (580 \si{\micro\meter} spot size, 580 \si{\micro\meter} step size, 10 \si{\milli\second} dwell time operated with X-ray tube reduced to 450\si{\micro\ampere}).}
	\label{fig:sunflowers}
	\vspace{-0.2cm}
\end{figure}
\begin{figure}[t]
	\centering
	\includegraphics[width=0.8\linewidth]{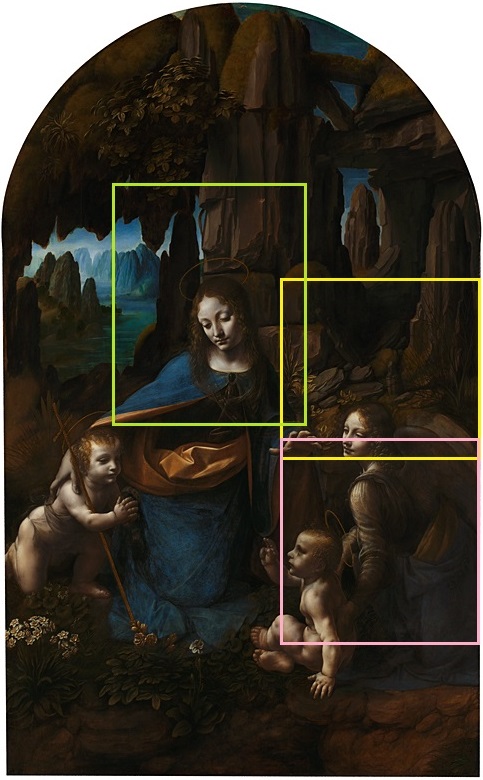}
	\caption{Leonardo da Vinci, \emph{The Virgin with the Infant Saint John the Baptist adoring the Christ Child accompanied by an Angel (`The Virgin of the Rocks')} (NG1093), about 1491/2-9 and 1506-8. Oil on poplar, thinned and cradled \cite{NG_Technical_32}. \copyright  The National Gallery, London. Highlighted are three regions (d10 in yellow, d11 in pink and d12 in green) scanned with a Bruker M6 JETSTREAM instrument (350 \si{\micro\meter} spot size, 350 \si{\micro\meter} step size and 10 \si{\milli\second} dwell time).}
	\vspace{-0.2cm}
	\label{fig:leonardo_oil_painting}
\end{figure}
\begin{figure*}[t]
	\centering
	\begin{subfigure}{0.161\linewidth}
		\centering
		\includegraphics[width=1\linewidth]{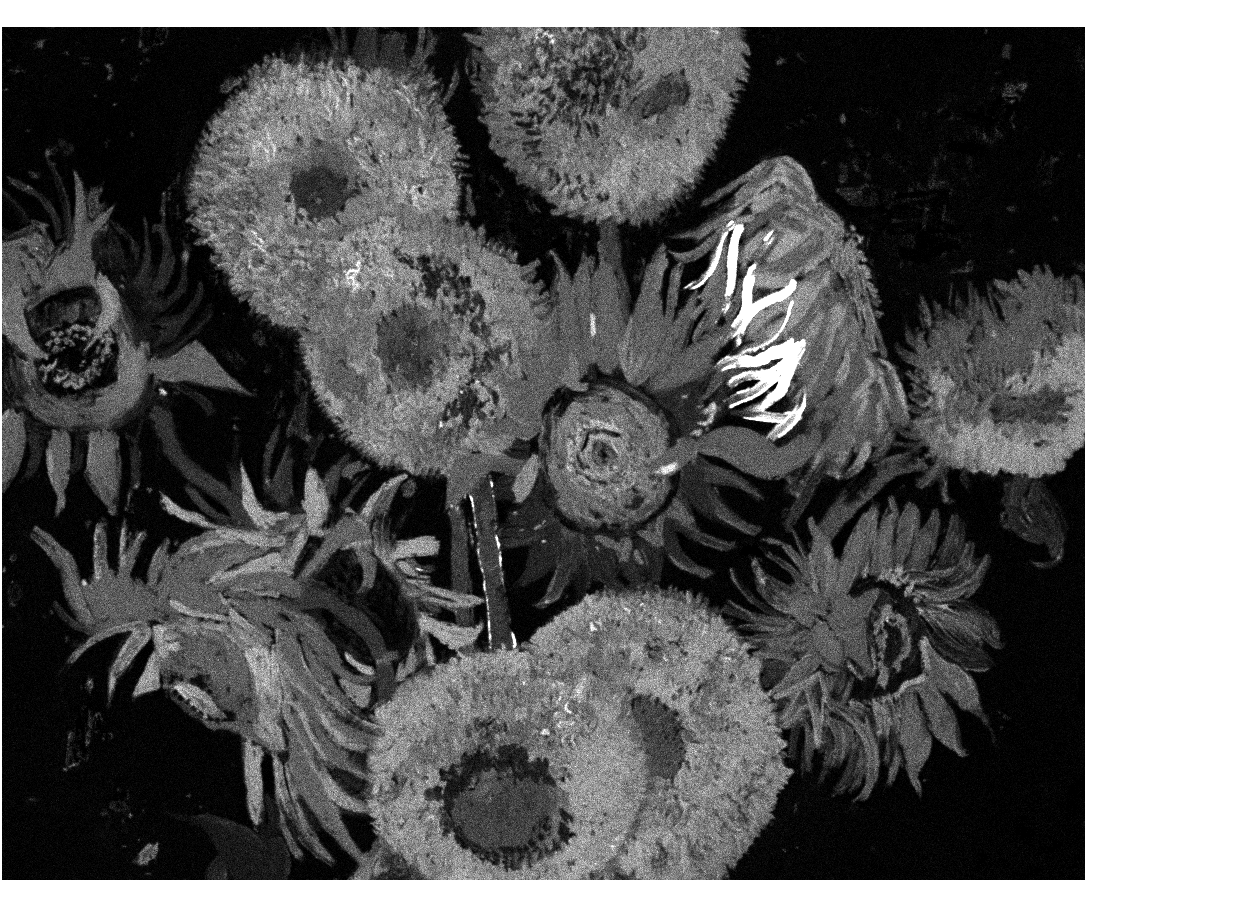}
	\end{subfigure}
	\begin{subfigure}{0.161\linewidth}
		\centering
		\includegraphics[width=1\linewidth]{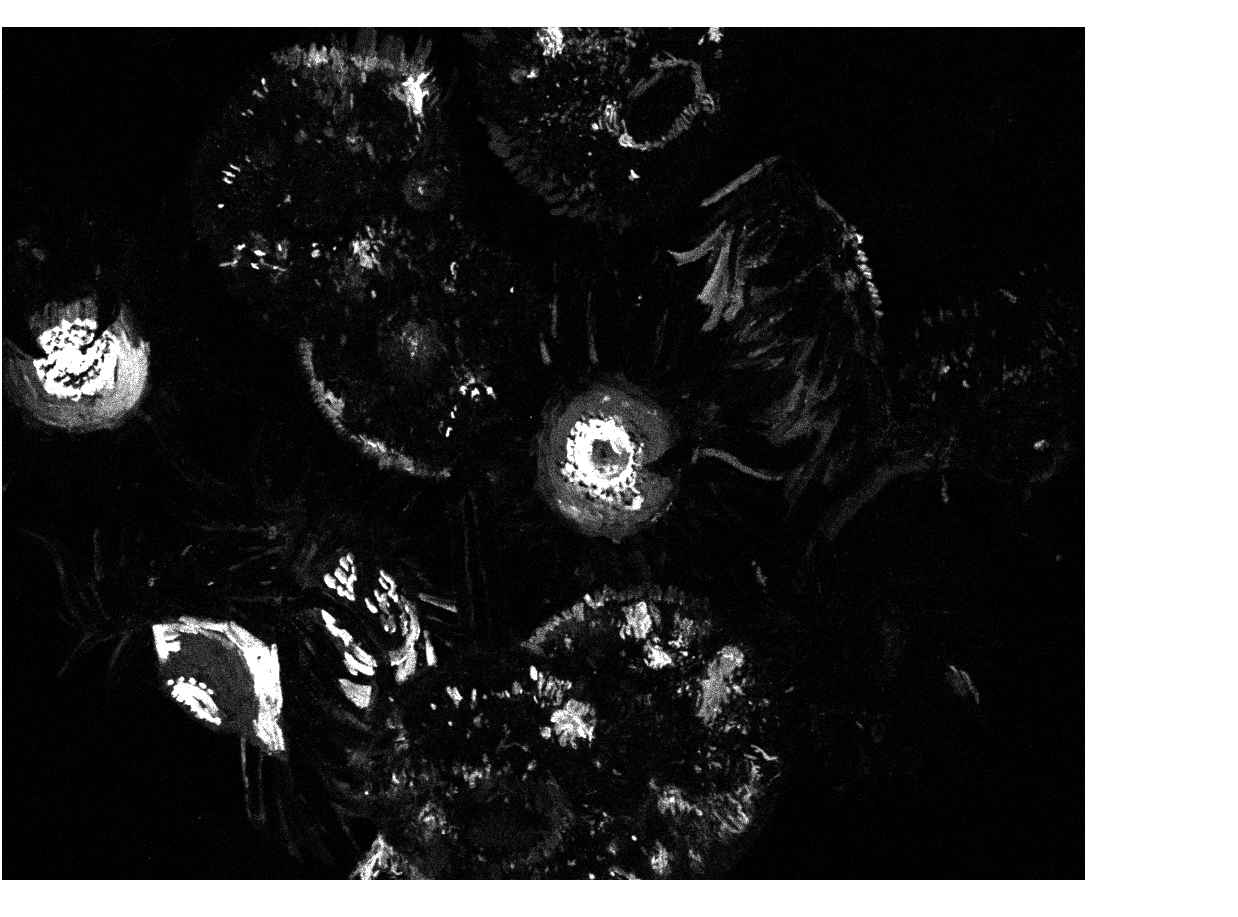}
	\end{subfigure}
	\begin{subfigure}{0.161\linewidth}
		\centering
		\includegraphics[width=1\linewidth]{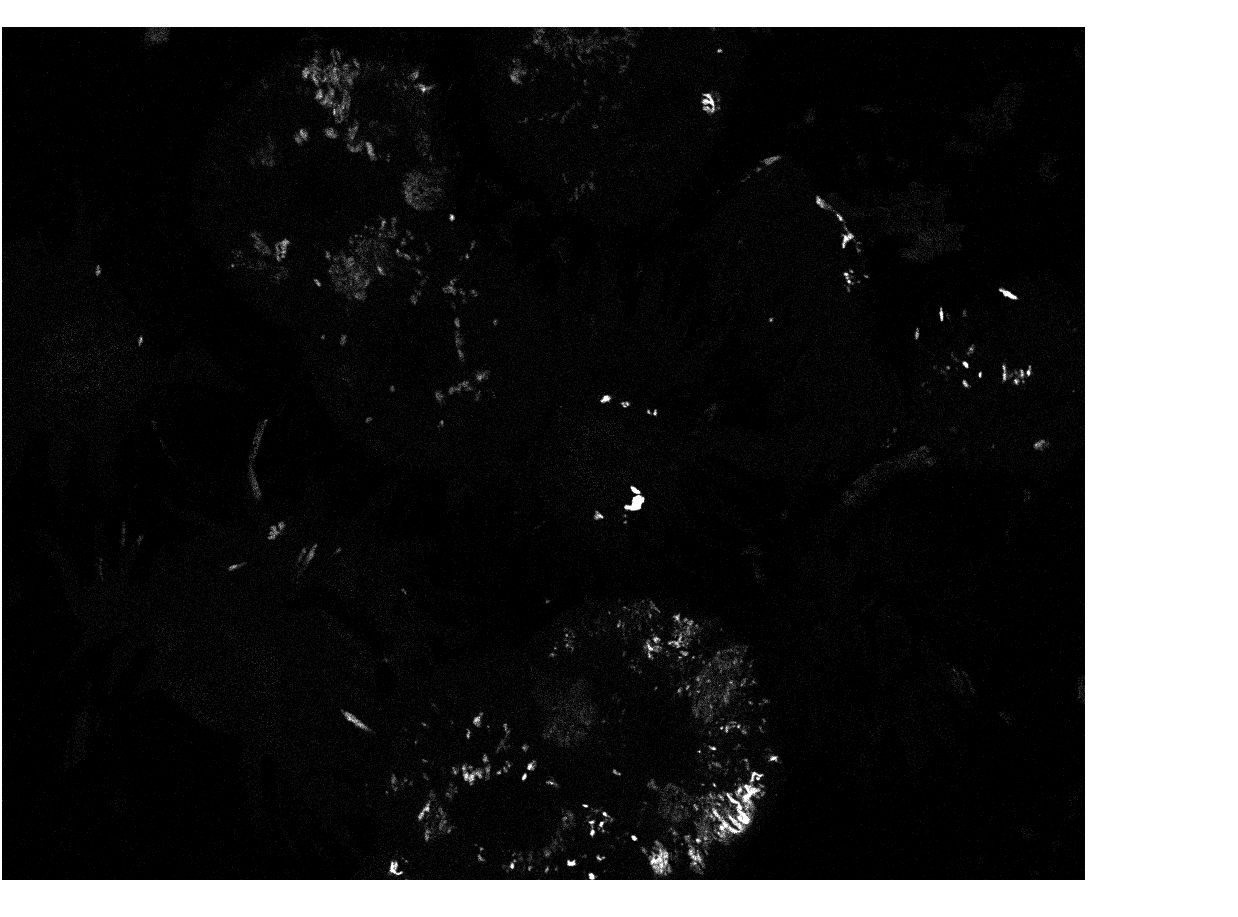}
	\end{subfigure}
	\begin{subfigure}{0.161\linewidth}
		\centering
		\includegraphics[width=1\linewidth]{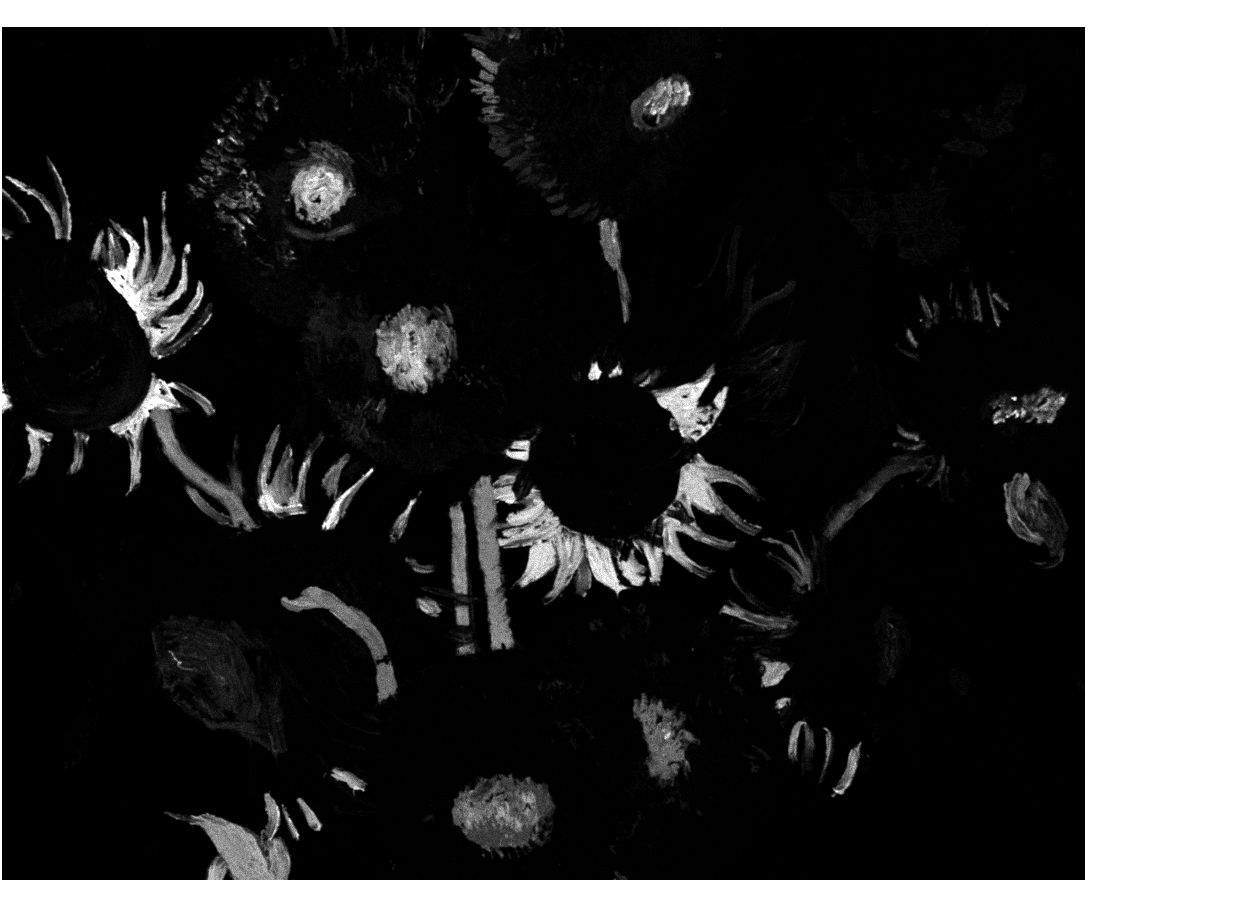}
	\end{subfigure}
	\begin{subfigure}{0.161\linewidth}
		\centering
		\includegraphics[width=1\linewidth]{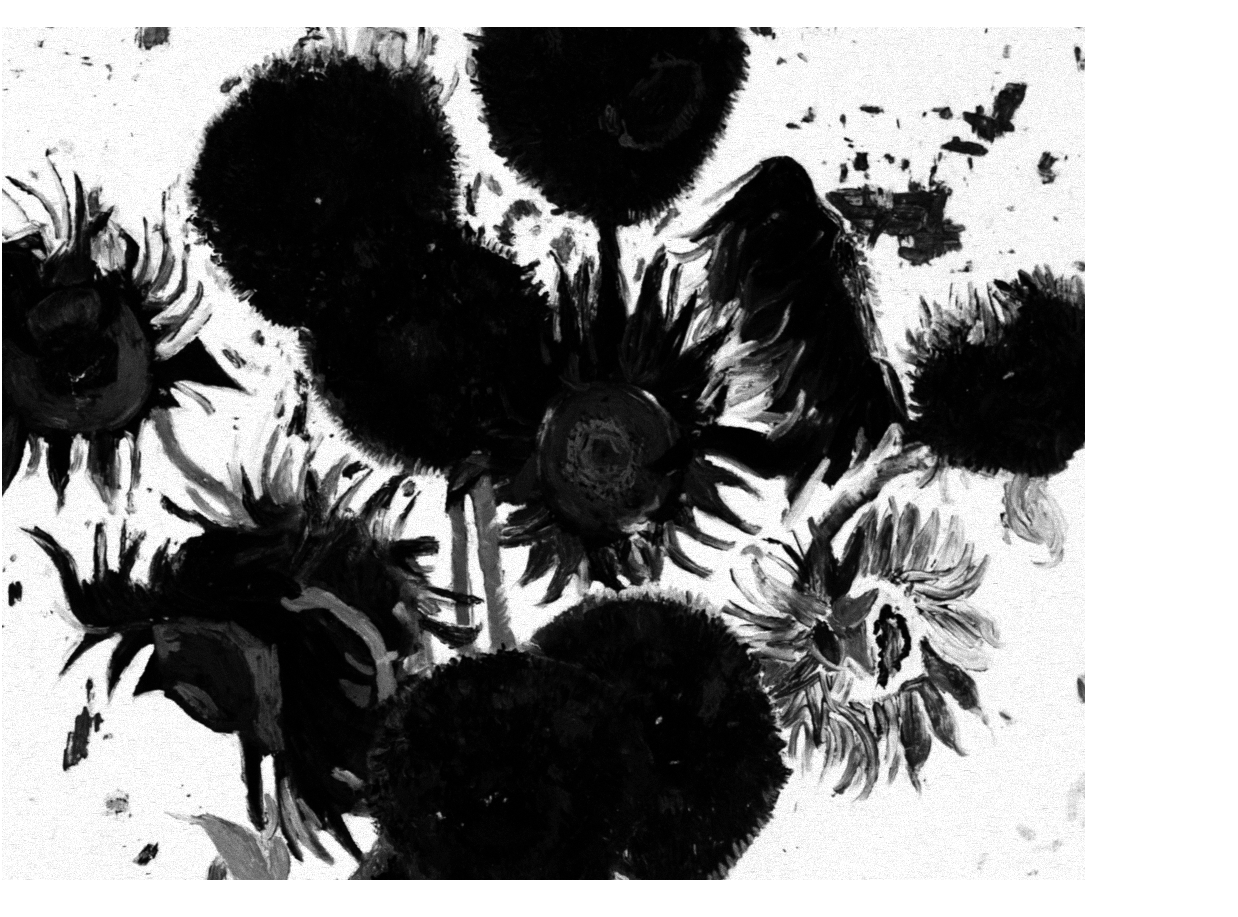}
	\end{subfigure}
	\begin{subfigure}{0.161\linewidth}
		\centering
		\includegraphics[width=1\linewidth]{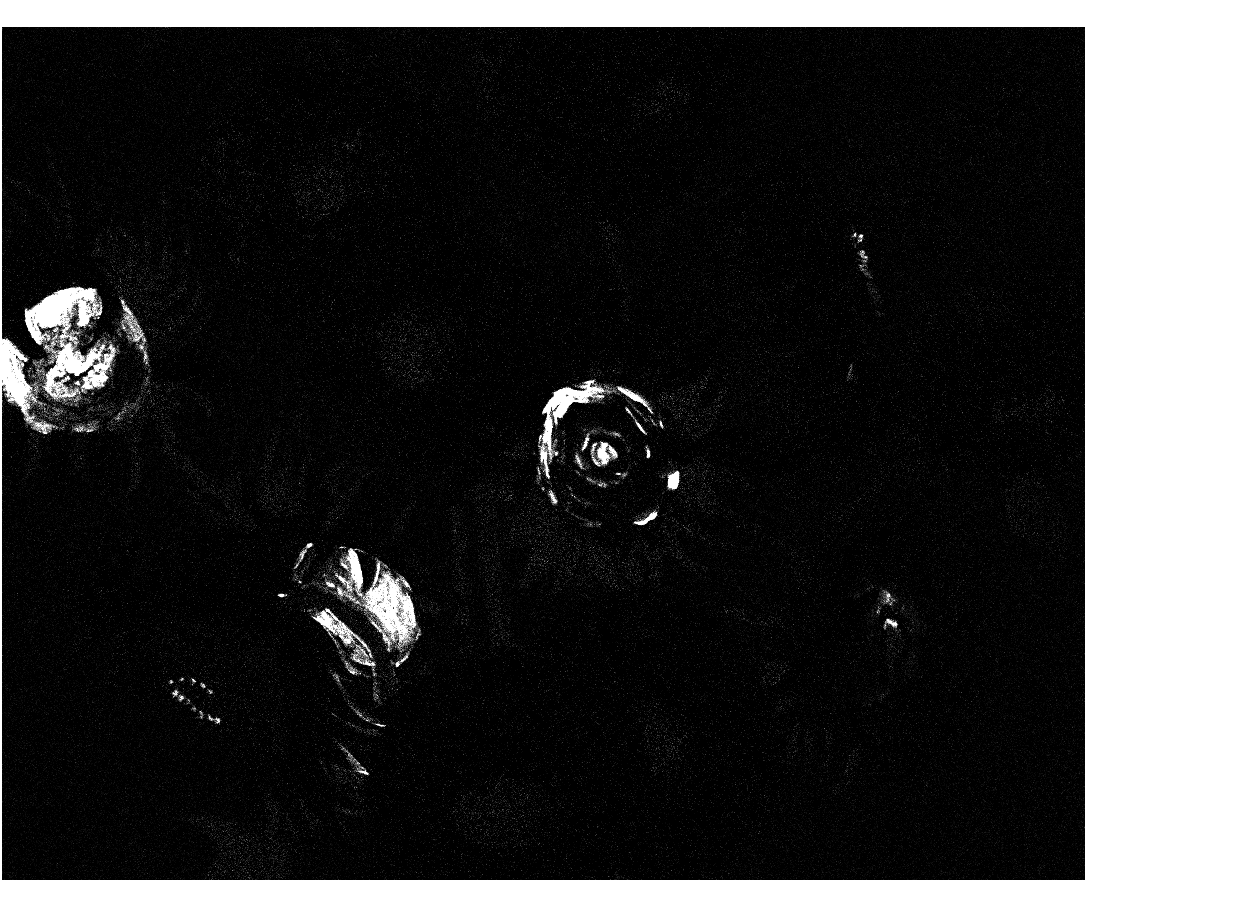}
	\end{subfigure}
	\\
	\begin{subfigure}{0.161\linewidth}
		\centering
		\includegraphics[width=1\linewidth]{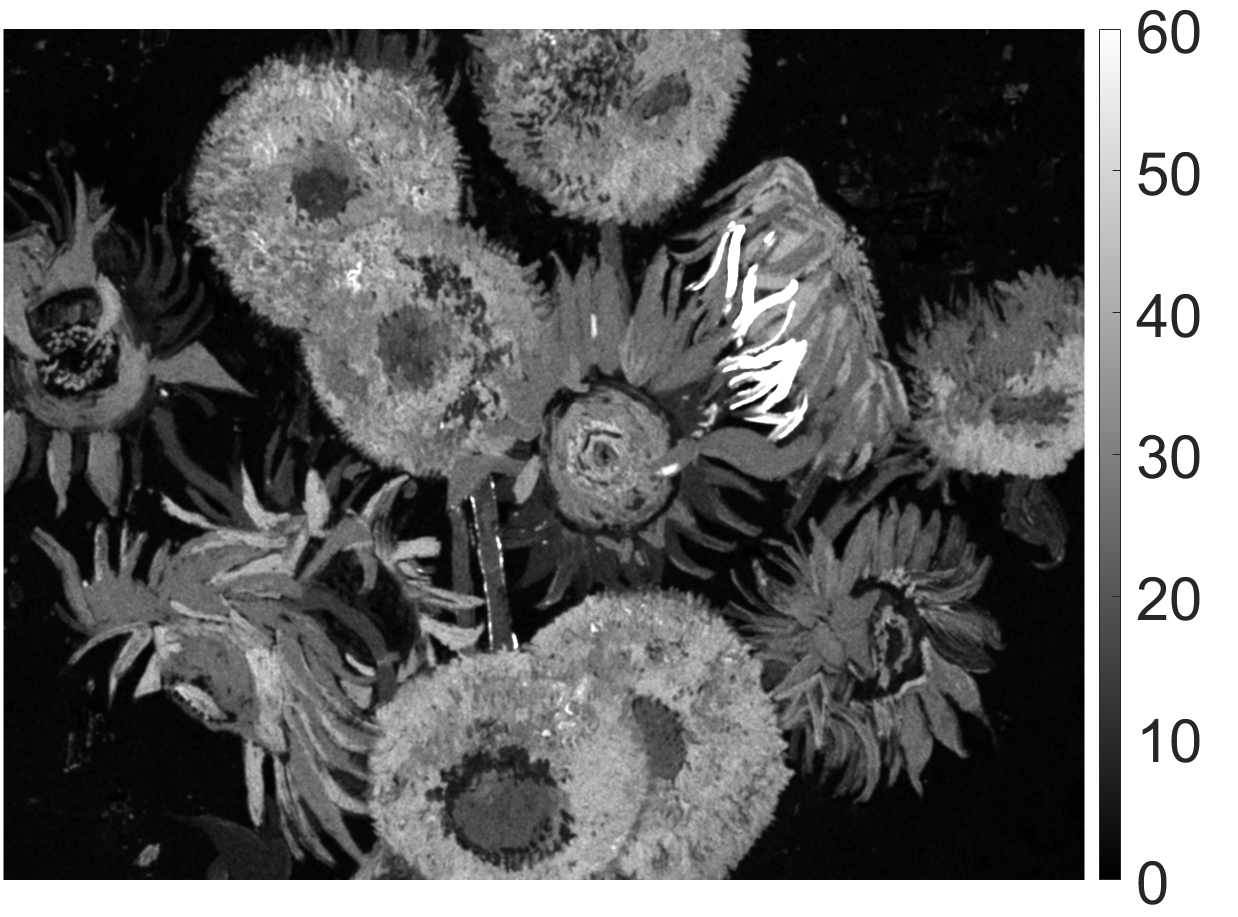}
		\vspace{-0.6cm}
		\caption{Cr $\text{K}_{\alpha}$}
		\vspace{-0.1cm}
		\label{fig:NG3863_d1_Cr_K_alpha_FISTA}
	\end{subfigure}
	\begin{subfigure}{0.161\linewidth}
		\centering
		\includegraphics[width=1\linewidth]{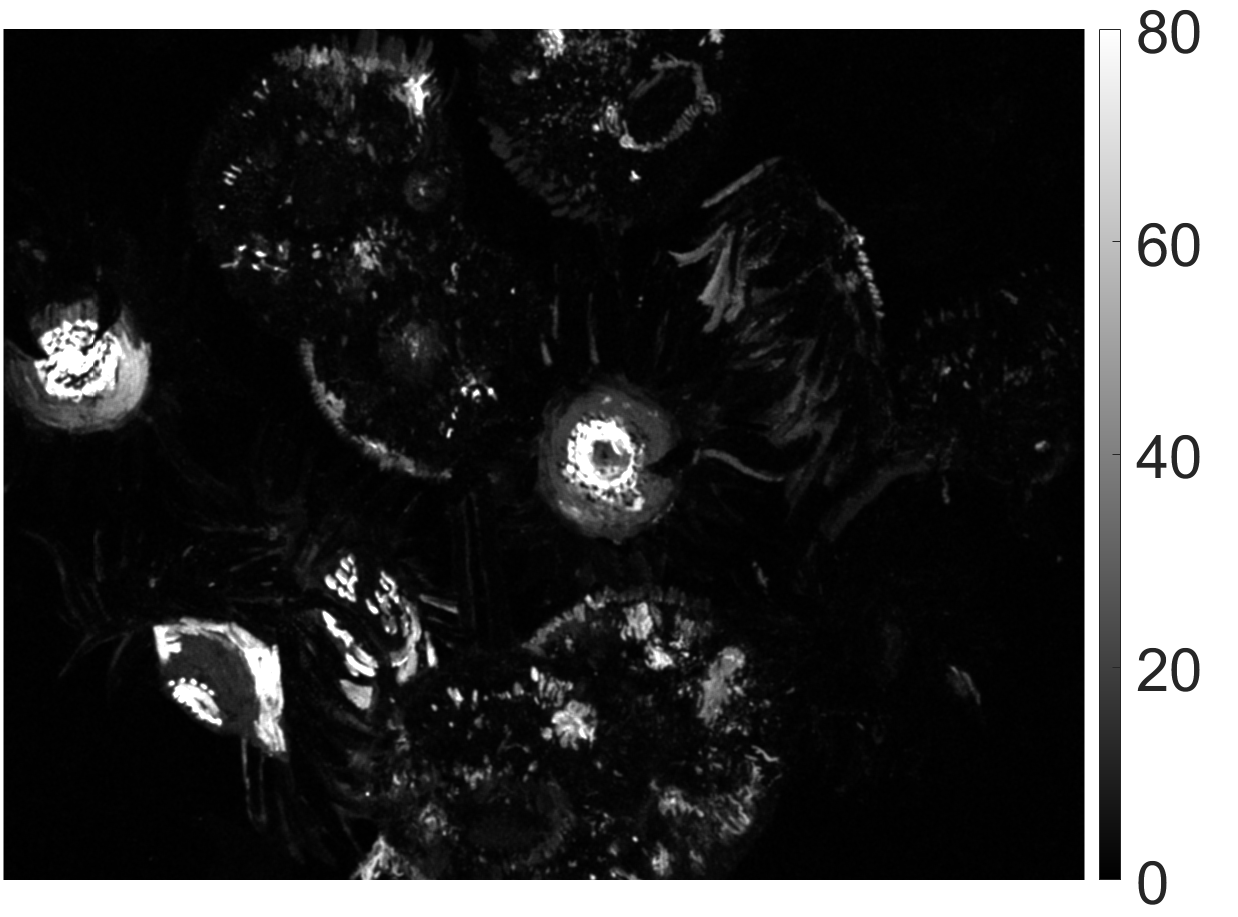}
		\vspace{-0.6cm}
		\caption{Fe $\text{K}_{\alpha}$}
		\vspace{-0.1cm}
		\label{fig:NG3863_d1_Fe_K_alpha_FISTA}
	\end{subfigure}
	\begin{subfigure}{0.161\linewidth}
		\centering
		\includegraphics[width=1\linewidth]{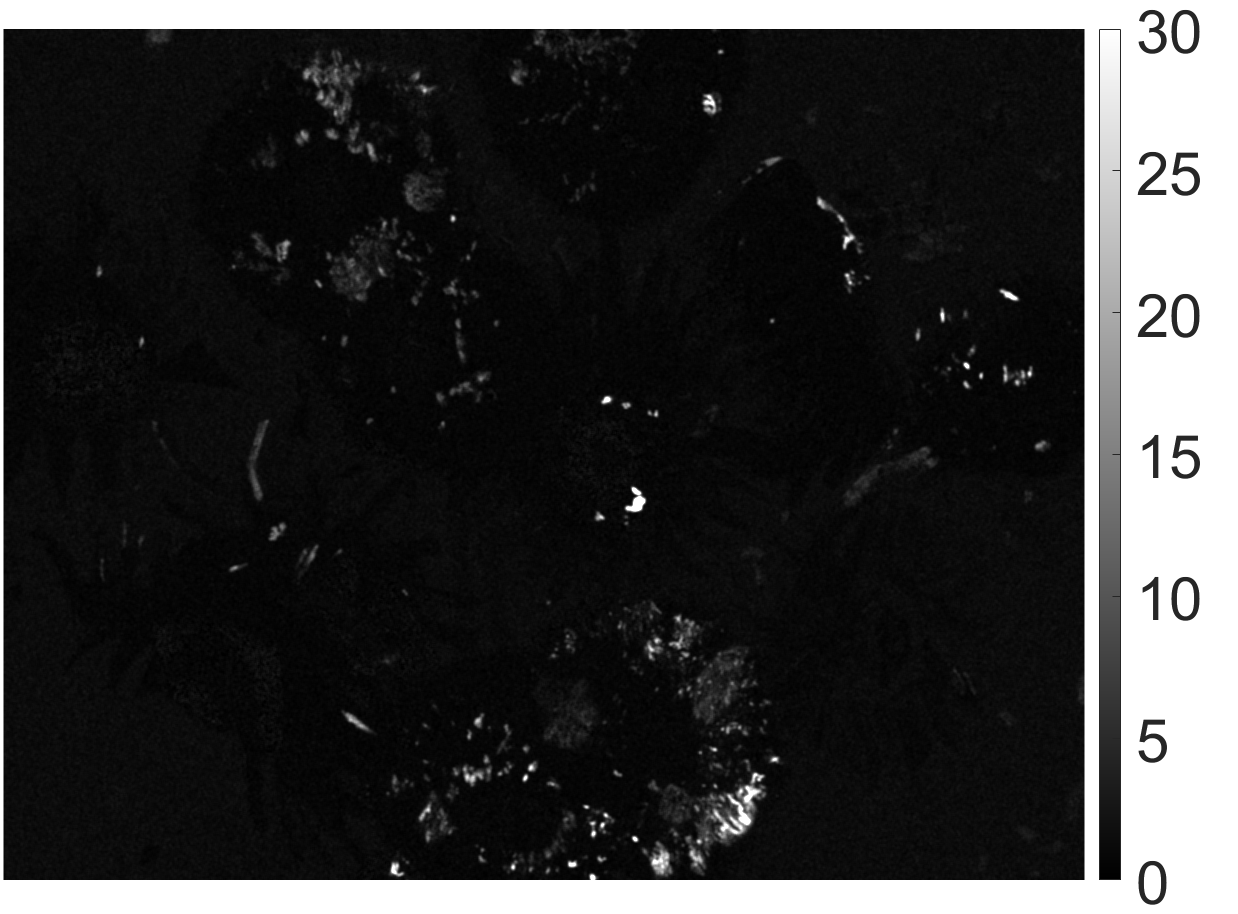}
		\vspace{-0.6cm}
		\caption{Co $\text{K}_{\alpha}$}
		\vspace{-0.1cm}
		\label{fig:NG3863_d1_Pb_L_beta_FISTA}
	\end{subfigure}
	\begin{subfigure}{0.161\linewidth}
		\centering
		\includegraphics[width=1\linewidth]{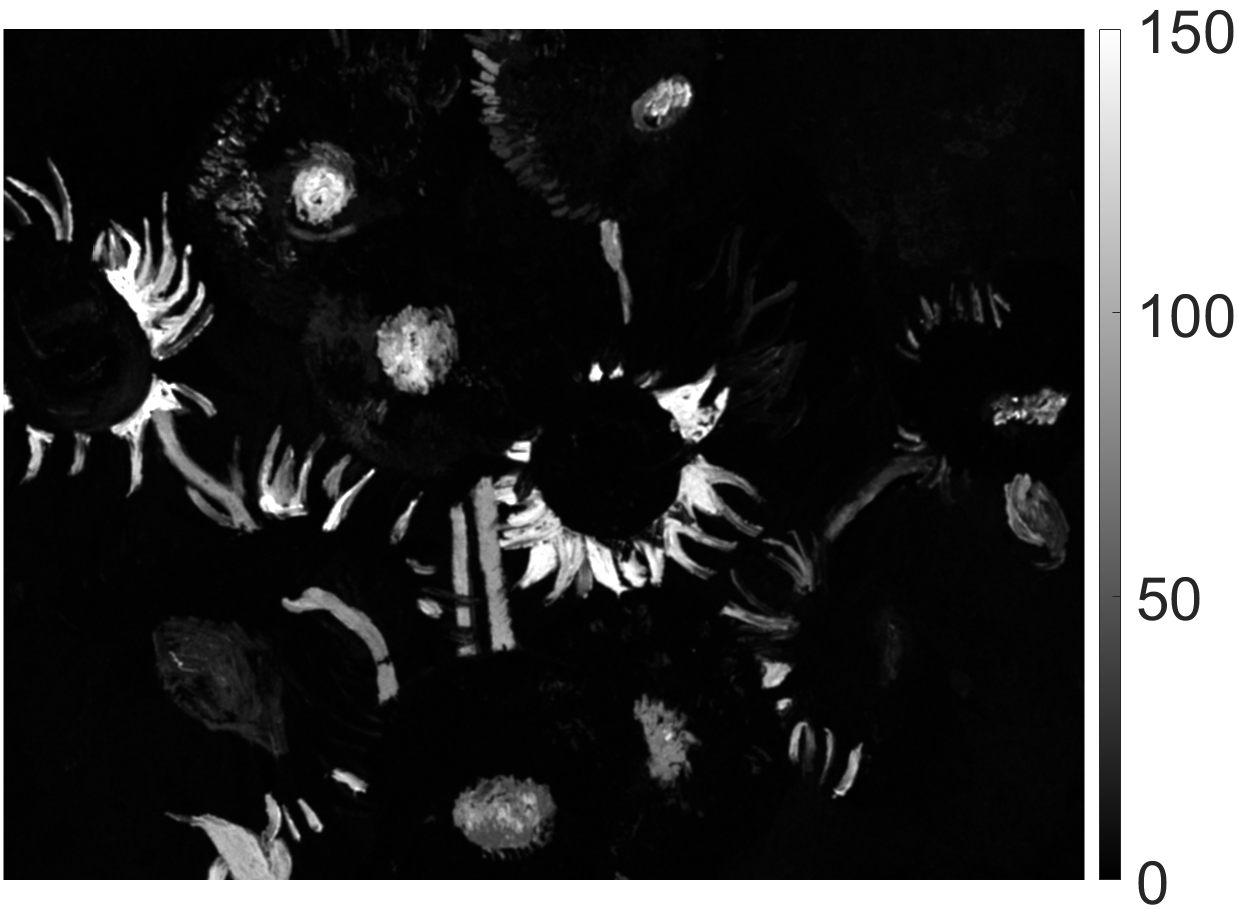}
		\vspace{-0.6cm}
		\caption{Cu $\text{K}_{\alpha}$}
		\vspace{-0.1cm}
		\label{fig:NG3863_d1_Cu_K_alpha_FISTA}
	\end{subfigure}
	\begin{subfigure}{0.161\linewidth}
		\centering
		\includegraphics[width=1\linewidth]{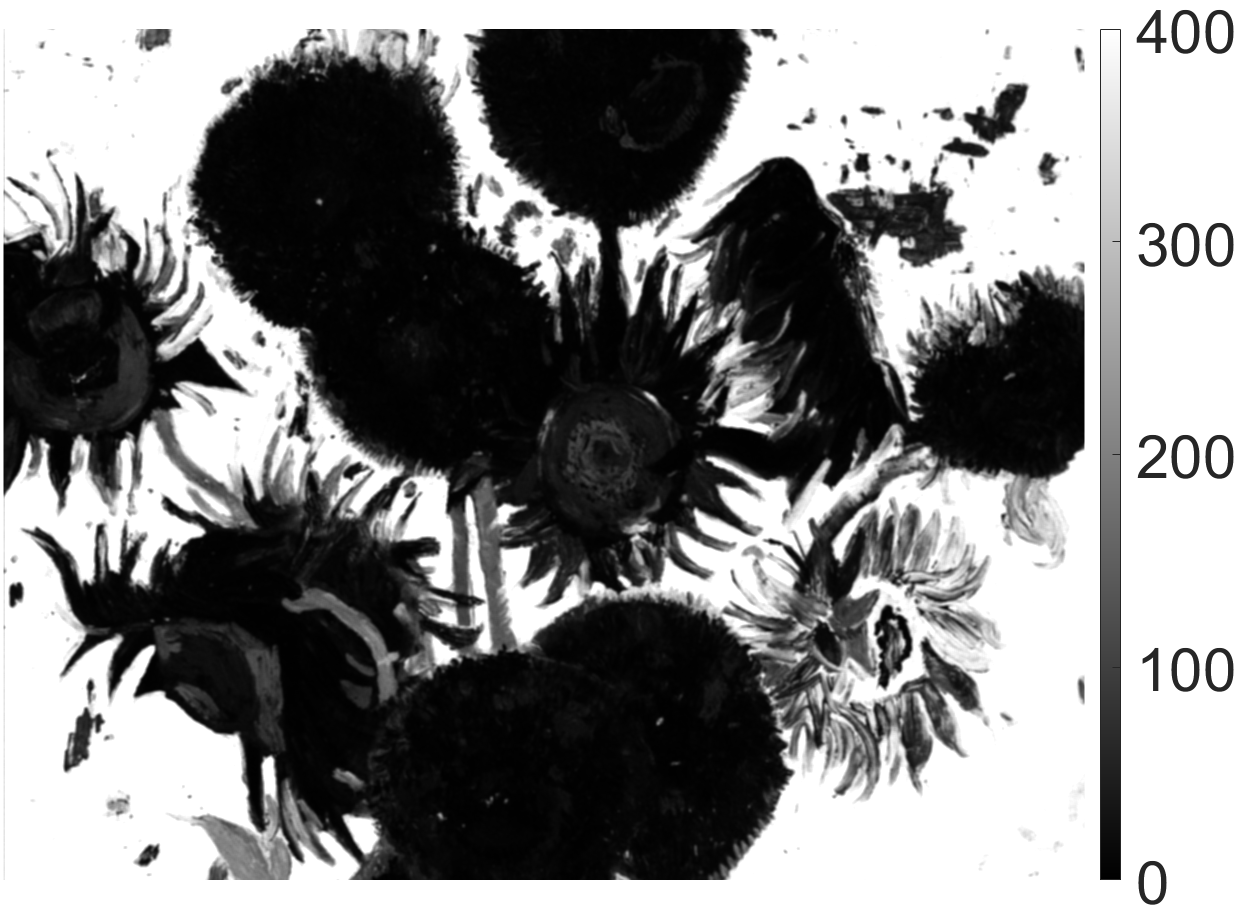}
		\vspace{-0.6cm}
		\caption{Zn $\text{K}_{\alpha}$}
		\vspace{-0.1cm}
		\label{fig:NG3863_d1_Zn_K_alpha_FISTA}
	\end{subfigure}
	\begin{subfigure}{0.161\linewidth}
		\centering
		\includegraphics[width=1\linewidth]{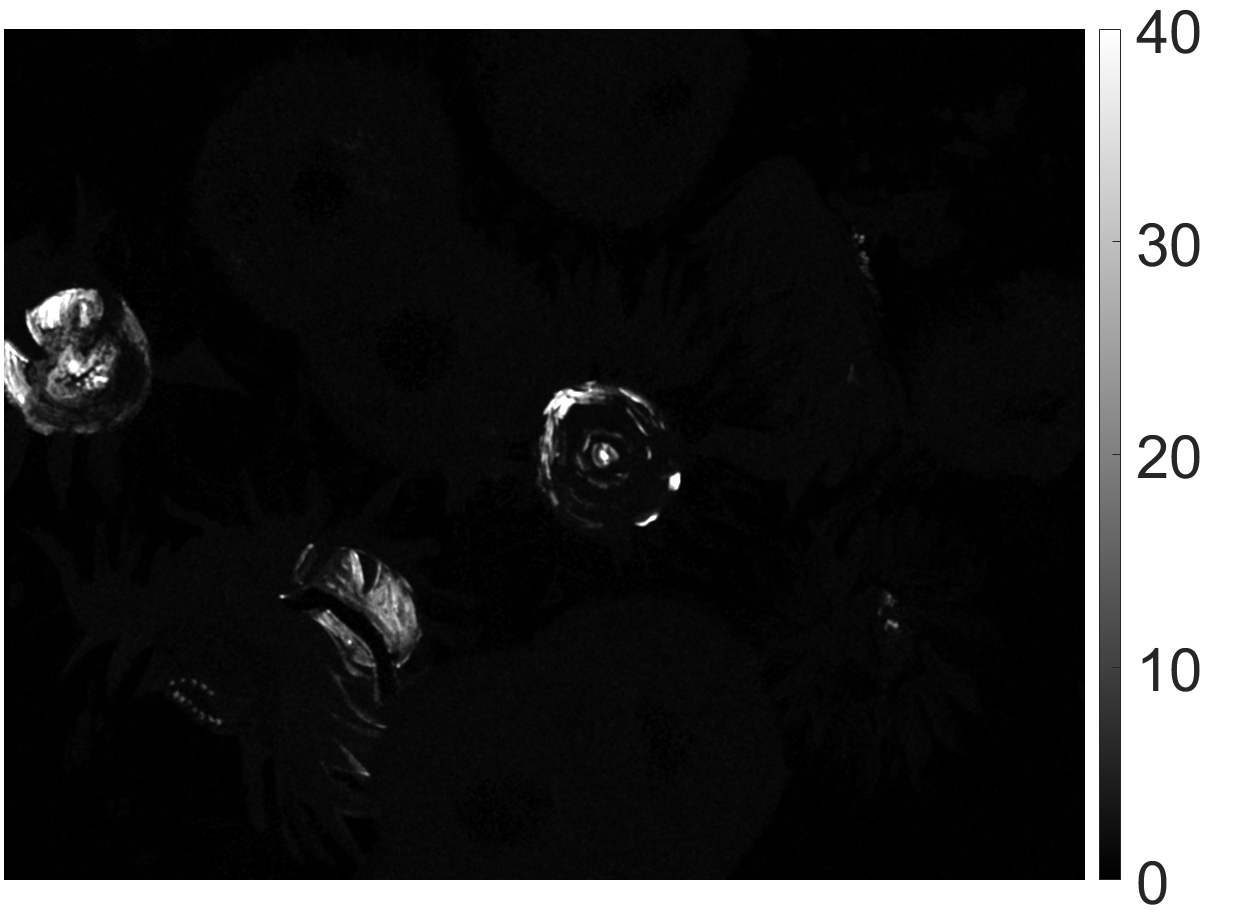}
		\vspace{-0.6cm}
		\caption{Br $\text{K}_{\alpha}$}
		\vspace{-0.1cm}
		\label{fig:NG3863_d1_Br_K_alpha_FISTA}
	\end{subfigure}
	\caption{First row: selection of deconvoluted element distribution maps generated using the Bruker M6 software with user input for the scanned region of \emph{Sunflowers}. Second row: equivalent selection of element distribution maps generated automatically with no user input using the proposed FAD method with the FISTA-inspired optimisation scheme (Algorithm \ref{Alg:FISTA}) for the same region of \emph{Sunflowers}. (a) Chromium (Cr) $\text{K}_{\alpha}$, (b) Iron (Fe) $\text{K}_{\alpha}$, (c) Cobalt (Co) $\text{K}_{\alpha}$, (d) Copper (Cu) $\text{K}_{\alpha}$, (e) Zinc (Zn) $\text{K}_{\alpha}$, (f) Bromine (Br) $\text{K}_{\alpha}$.}
	\label{fig:NG3863 elemental maps}
\end{figure*}

\begin{figure}[t]
\vspace{0.1cm}
\captionsetup[subfigure]{aboveskip=4.6pt,belowskip=6pt}
    \begin{multicols}{2}
    \centering
    \hfill
    \begin{subfigure}[t]{0.88\linewidth}
        \includegraphics[width=1\linewidth]{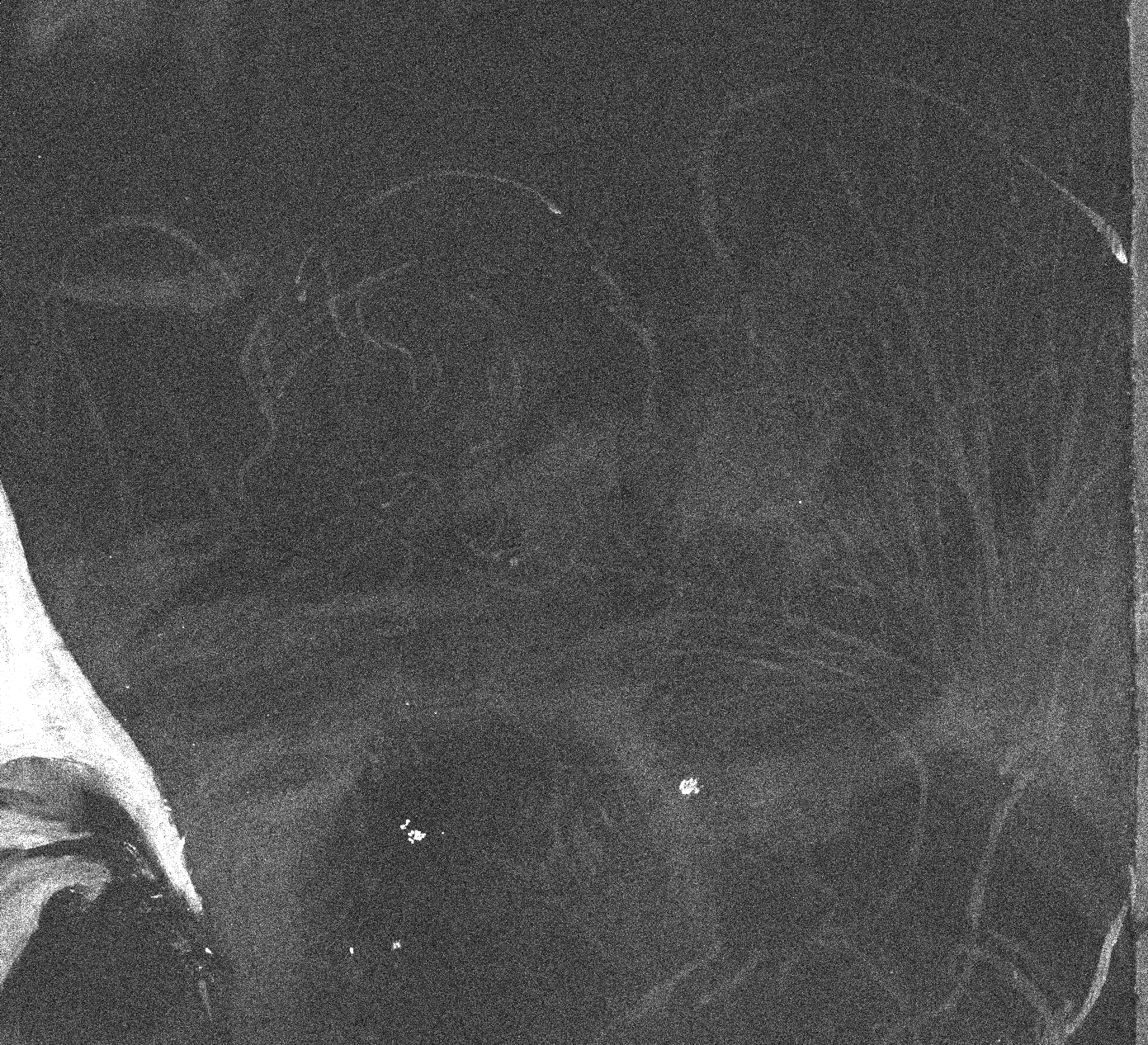}
        \vspace{-0.5cm}
        \caption{Zn $\text{K}_{\alpha}$ ROI}
        \vspace{-0.1cm}
        \label{fig:NG1093 Zn bruker}
    \end{subfigure}
    \par
    \hfill
    \begin{subfigure}[t]{0.88\linewidth}
        \includegraphics[width=1\linewidth]{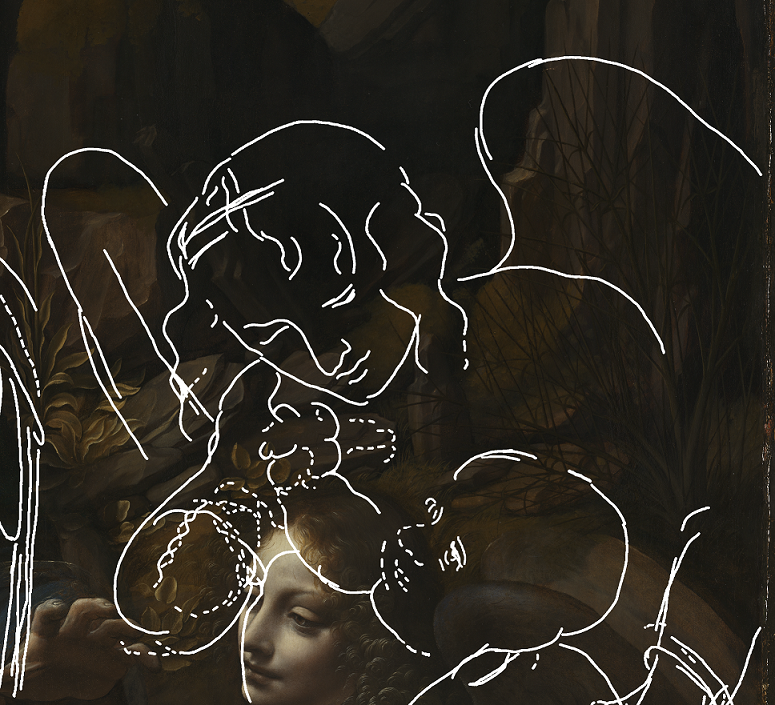}
        \vspace{-0.5cm}
        \caption{Tracing of underdrawings}
        \vspace{-0.1cm}
        \label{fig:NG1093 underdrawing}
    \end{subfigure}
    \par
    \hfill
    \begin{subfigure}[t]{0.88\linewidth}
        \includegraphics[width=1\linewidth]{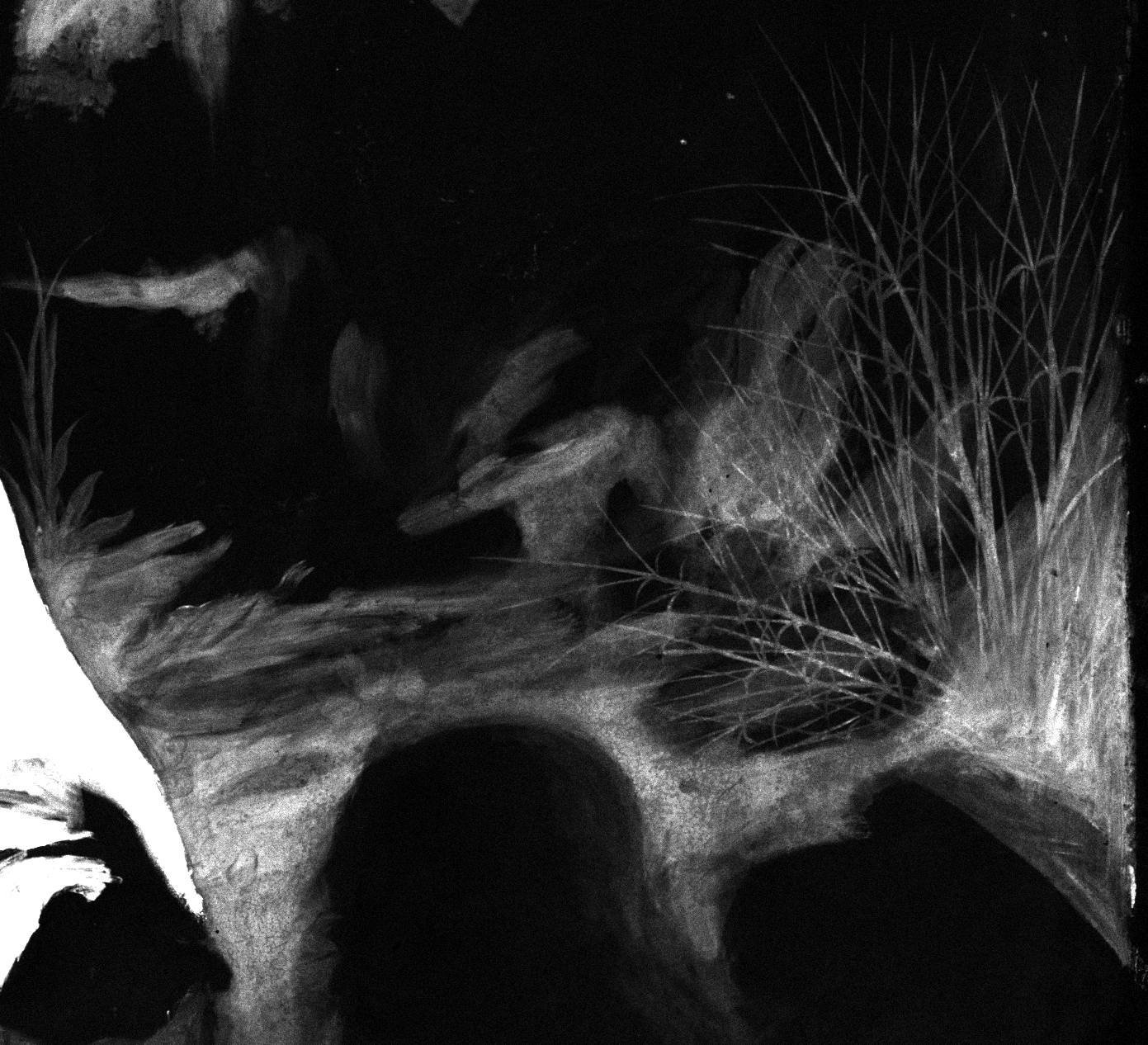}
        \vspace{-0.5cm}
        \caption{Cu $\text{K}_{\alpha}$}
        \vspace{-0.2cm}
        \label{fig:NG1093 Cu bruker}
    \end{subfigure}
    \par
    \begin{subfigure}[t]{0.88\linewidth}
        \includegraphics[width=1\linewidth]{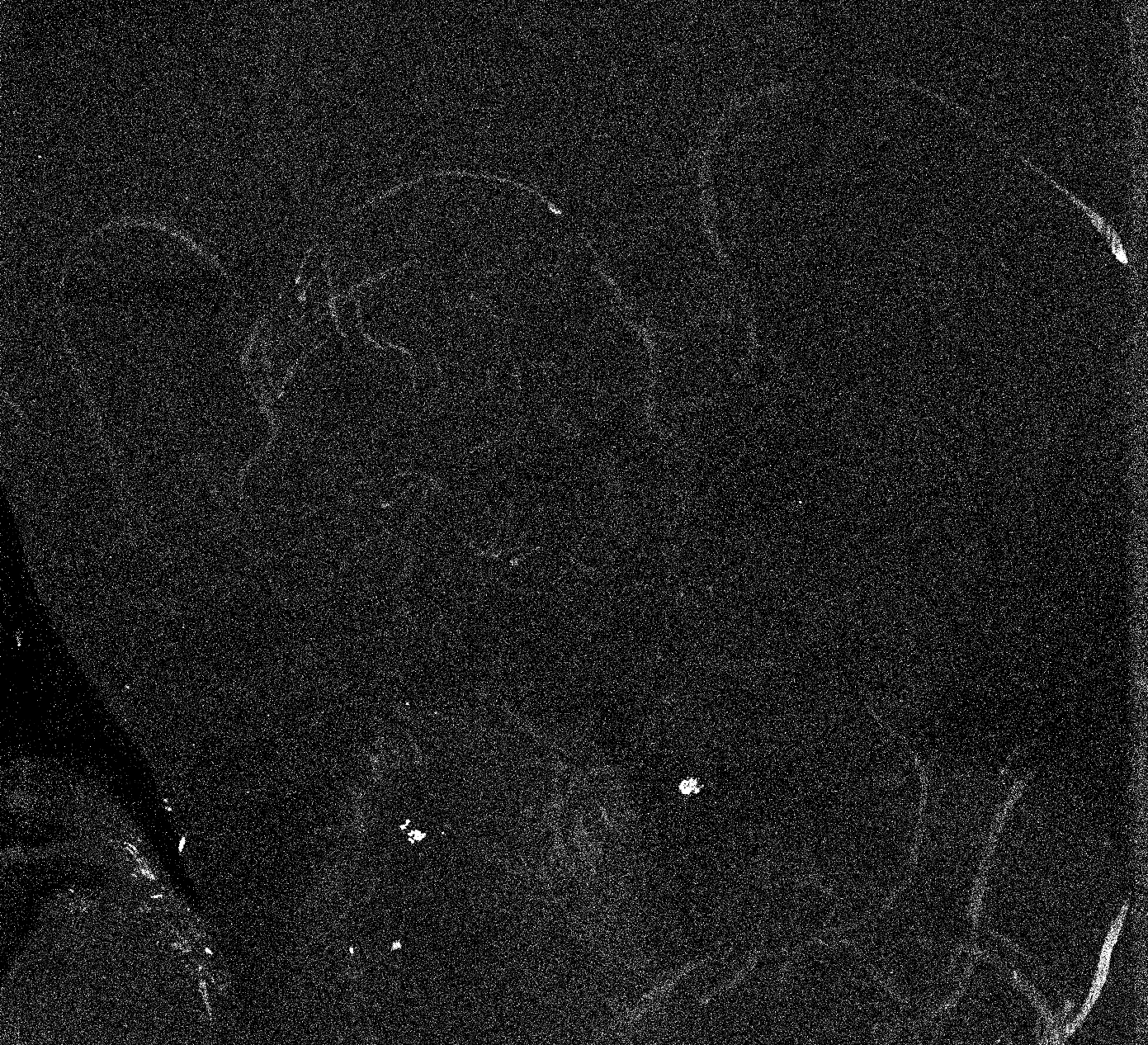}
        \vspace{-0.5cm}
        \caption{Zn $\text{K}_{\alpha}$}
        \vspace{-0.1cm}
        \label{fig:NG1093 Zn-Cu bruker}
    \end{subfigure}
    \hfill
    \par
    \begin{subfigure}[t]{0.88\linewidth}
        \includegraphics[width=1\linewidth]{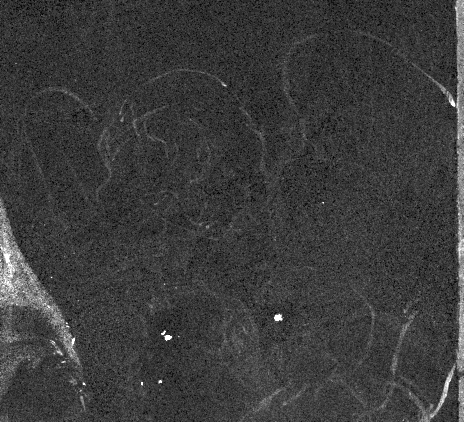}
        \vspace{-0.5cm}
        \caption{Zn $\text{K}_{\alpha}$ ROI minus Cu $\text{K}_{\alpha}$}
        \vspace{-0.1cm}
        \label{fig:NG1093 Zn-Cu manually bruker}
    \end{subfigure}
    \hfill
    \par
    \begin{subfigure}[t]{0.88\linewidth}
        \includegraphics[width=1\linewidth]{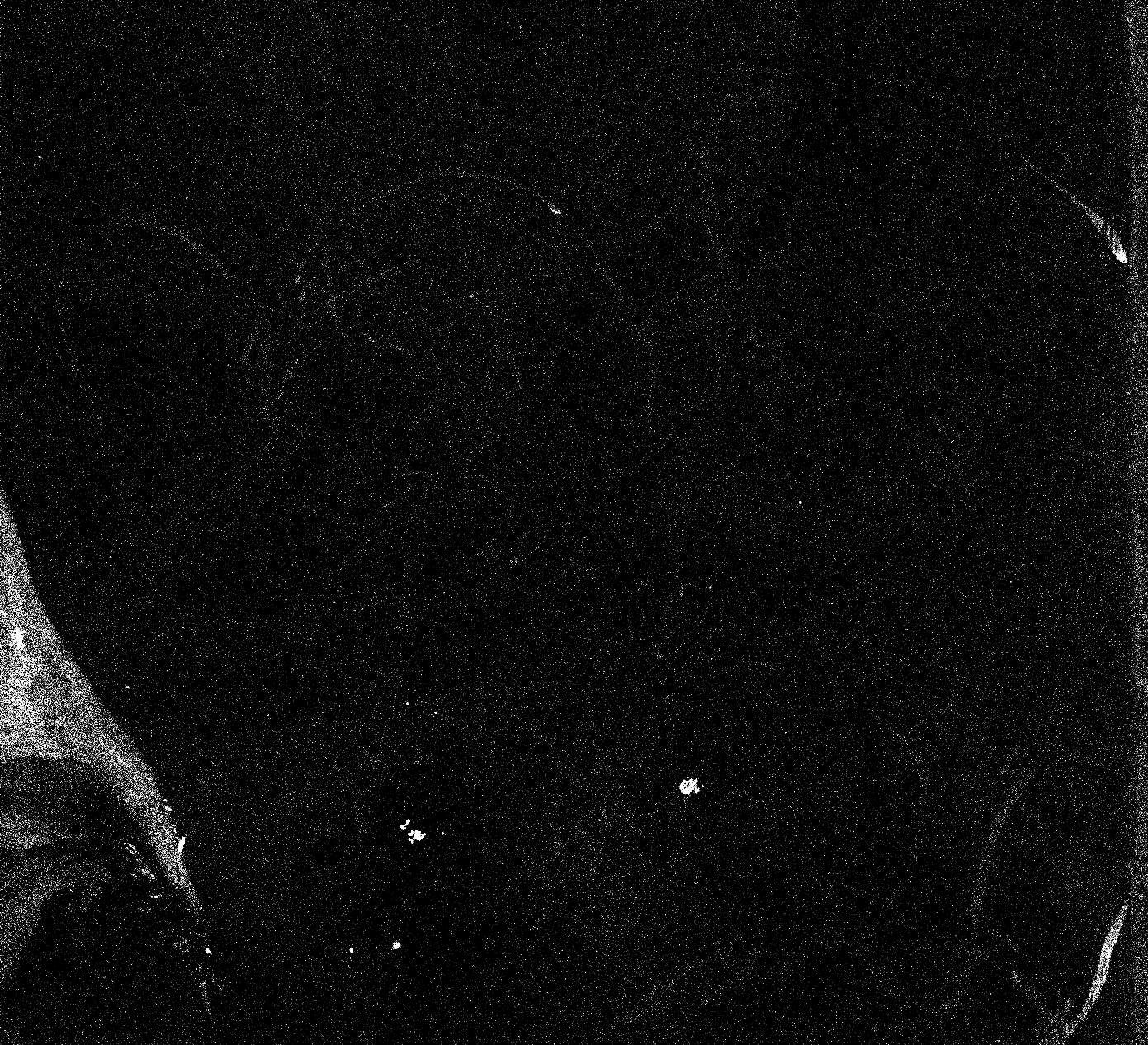}
        \vspace{-0.5cm}
        \caption{Zn $\text{K}_{\alpha}$ with Ni added}
        \vspace{-0.2cm}
        \label{fig:NG1093 Zn_Ni bruker}
    \end{subfigure}
  	\hfill
    \end{multicols}
\vspace{-0.3cm}
\caption{Element distribution maps produced by Bruker M6 software for the scanned region d10 (yellow outline) of \emph{`The Virgin of the Rocks'}. (a) Zn $\text{K}_{\alpha}$ region of interest (ROI), (b)
manual tracing of detected underdrawing lines, white, overlaid onto the visible image of the painting, (c) Cu $\text{K}_{\alpha}$ after deconvolution, (d) Zn $\text{K}_{\alpha}$ after deconvolution without Ni added in the element list, (e) Zn $\text{K}_{\alpha}$ ROI with Cu $\text{K}_{\alpha}$ manually subtracted and 3-by-3 pixel binning, (f) Zn $\text{K}_{\alpha}$ after deconvolution with Ni added in the element list.}
\label{fig:NG1093 Bruker elemental maps}
\vspace{-0.2cm}
\end{figure}

\begin{figure}[t]
\captionsetup[subfigure]{aboveskip=2pt,belowskip=3pt}
    \begin{multicols}{2}
    \centering
      \begin{subfigure}[t]{1\linewidth}
        \includegraphics[width=1\linewidth]{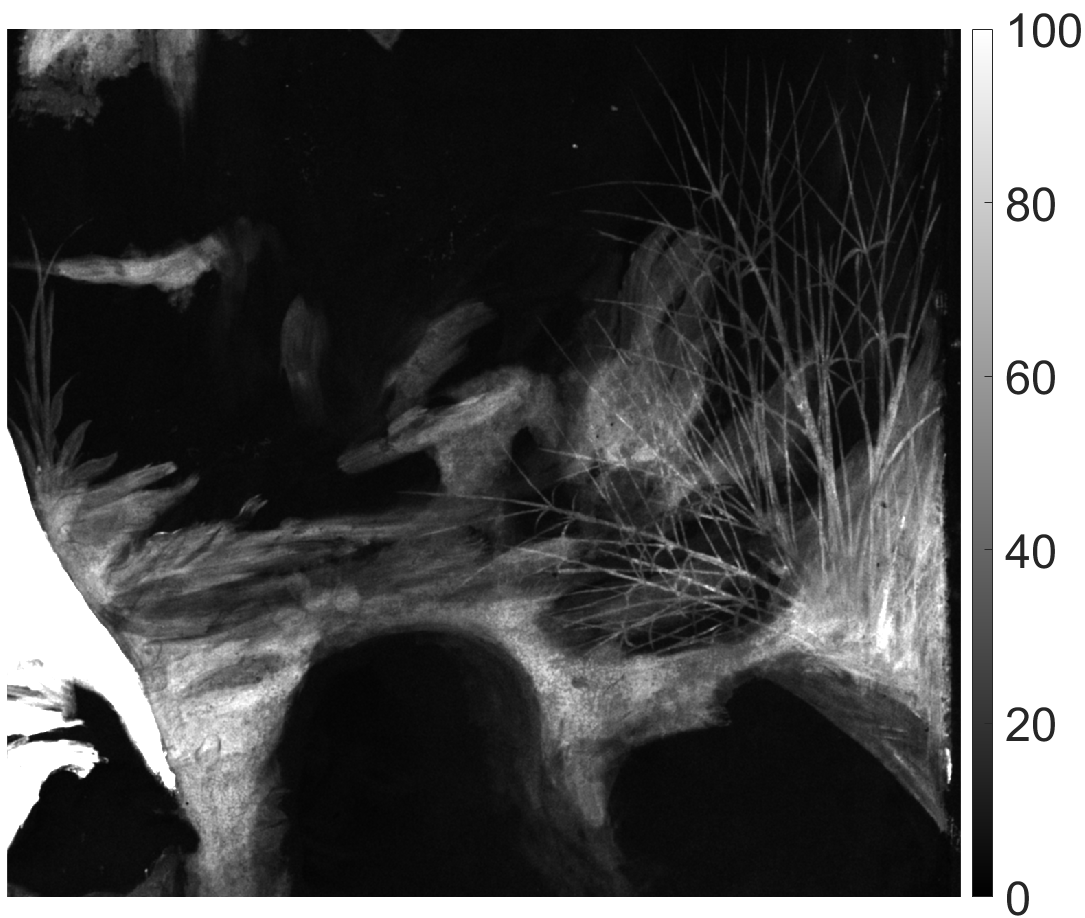}
        \vspace{-0.5cm}
        \caption{Cu $\text{K}_{\alpha}$}
        \vspace{-0.1cm}
        \label{fig:NG1093_d10_Cu_K_alpha_FISTA}
      \end{subfigure}
      \par
      \begin{subfigure}[t]{1\linewidth}
        \includegraphics[width=1\linewidth]{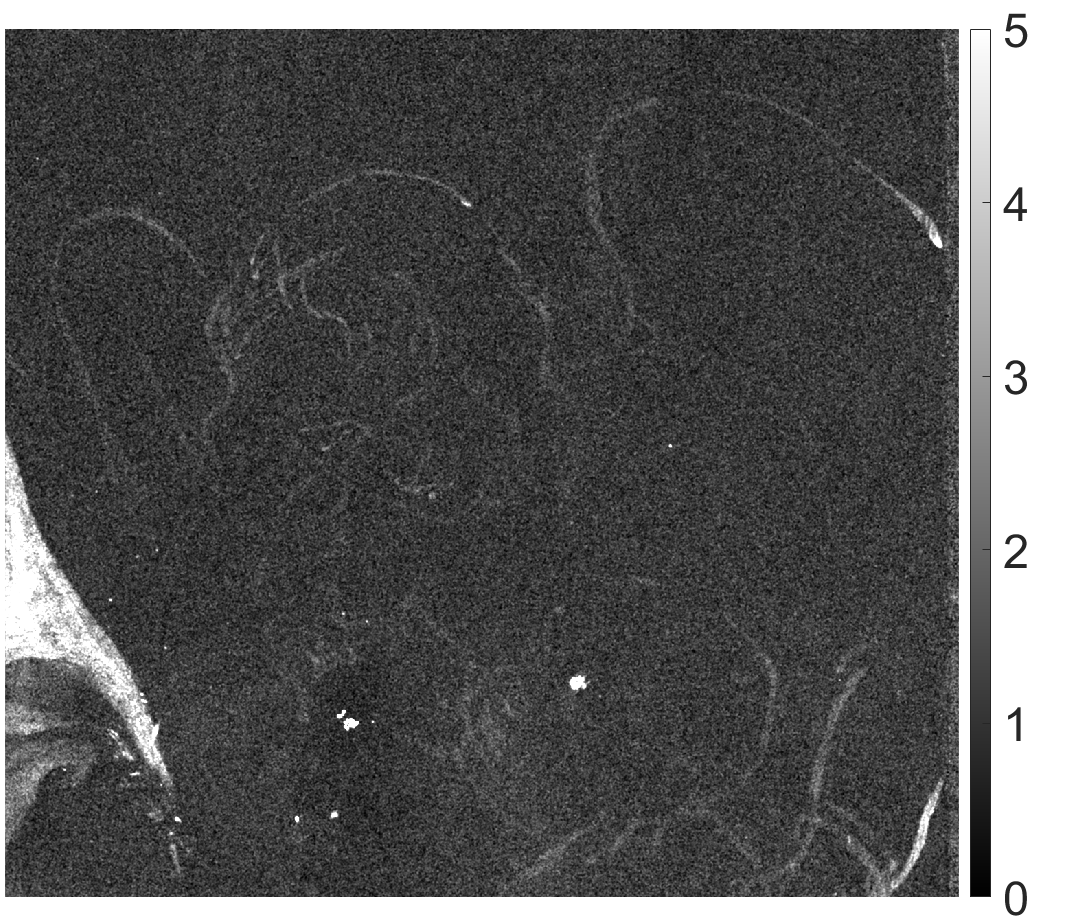}
        \vspace{-0.5cm}
        \caption{Zn $\text{K}_{\alpha}$}
        \vspace{-0.1cm}
        \label{fig:NG1093_d10_Zn_K_alpha_FISTA}
      \end{subfigure}
      \par
      \begin{subfigure}[t]{1\linewidth}
        \includegraphics[width=1\linewidth]{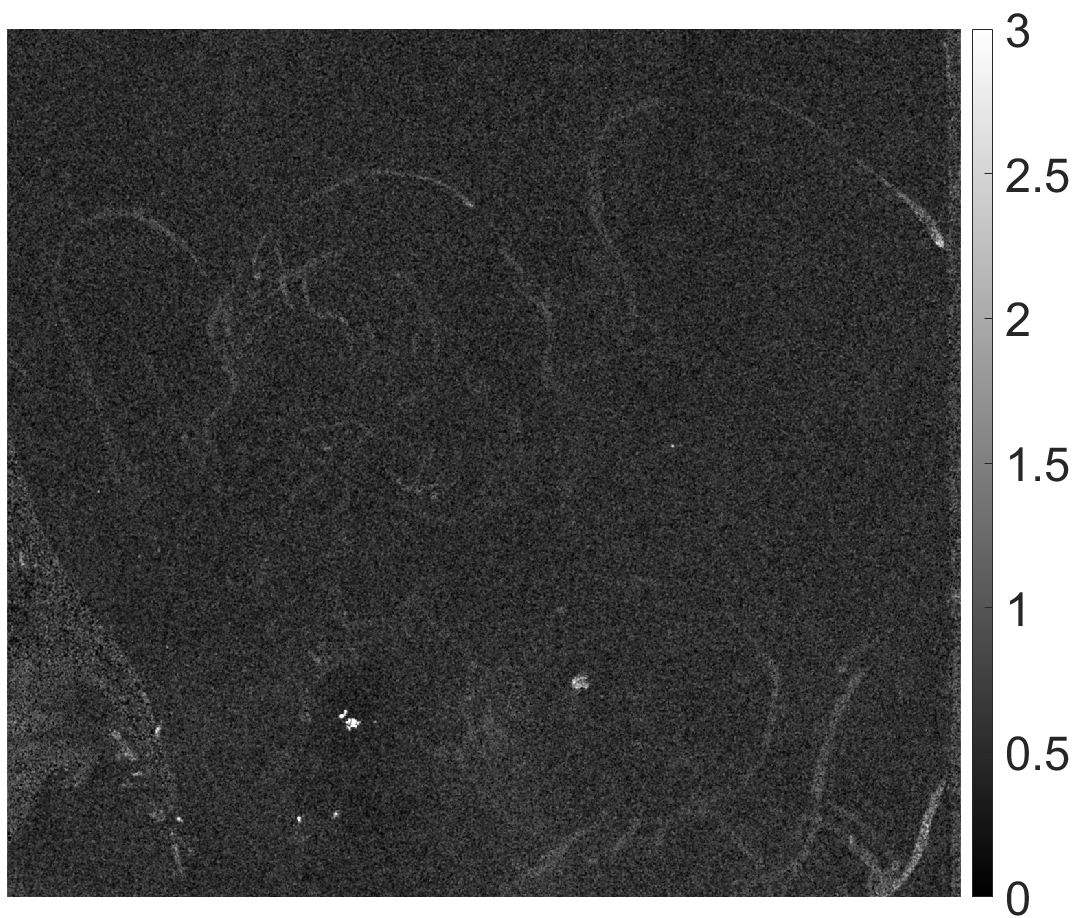}
        \vspace{-0.5cm}
        \caption{Zn $\text{K}_{\beta}$}
        \vspace{-0.2cm}
        \label{fig:NG1093_d10_Zn_K_beta_FISTA}
      \end{subfigure}
      \par
      \begin{subfigure}[t]{1\linewidth}
        \includegraphics[width=1\linewidth]{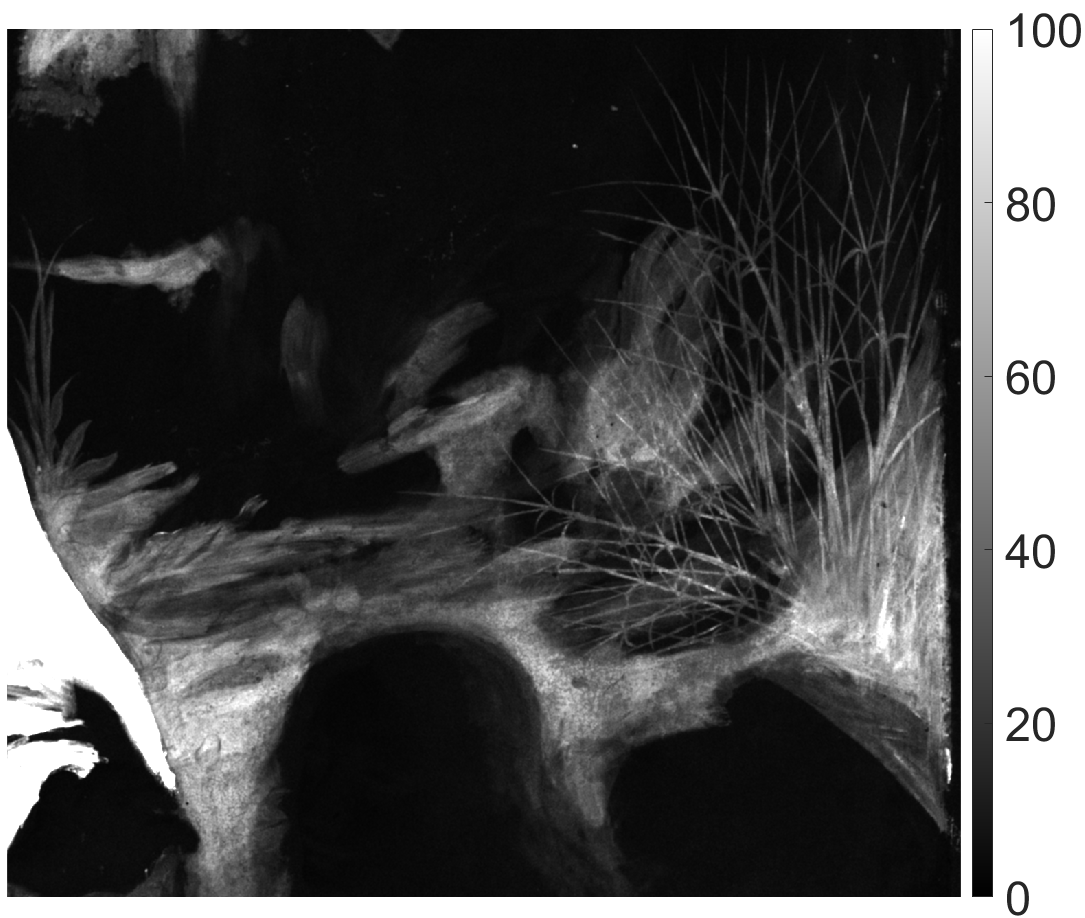}
        \vspace{-0.5cm}
        \caption{Cu $\text{K}_{\alpha}$ w/o constraint (\ref{eq:adjustment})}
        \vspace{-0.1cm}
        \label{fig:NG1093_d10_Cu_K_alpha_FISTA_no_check}
      \end{subfigure}
      \par
      \begin{subfigure}[t]{1\linewidth}
        \includegraphics[width=1\linewidth]{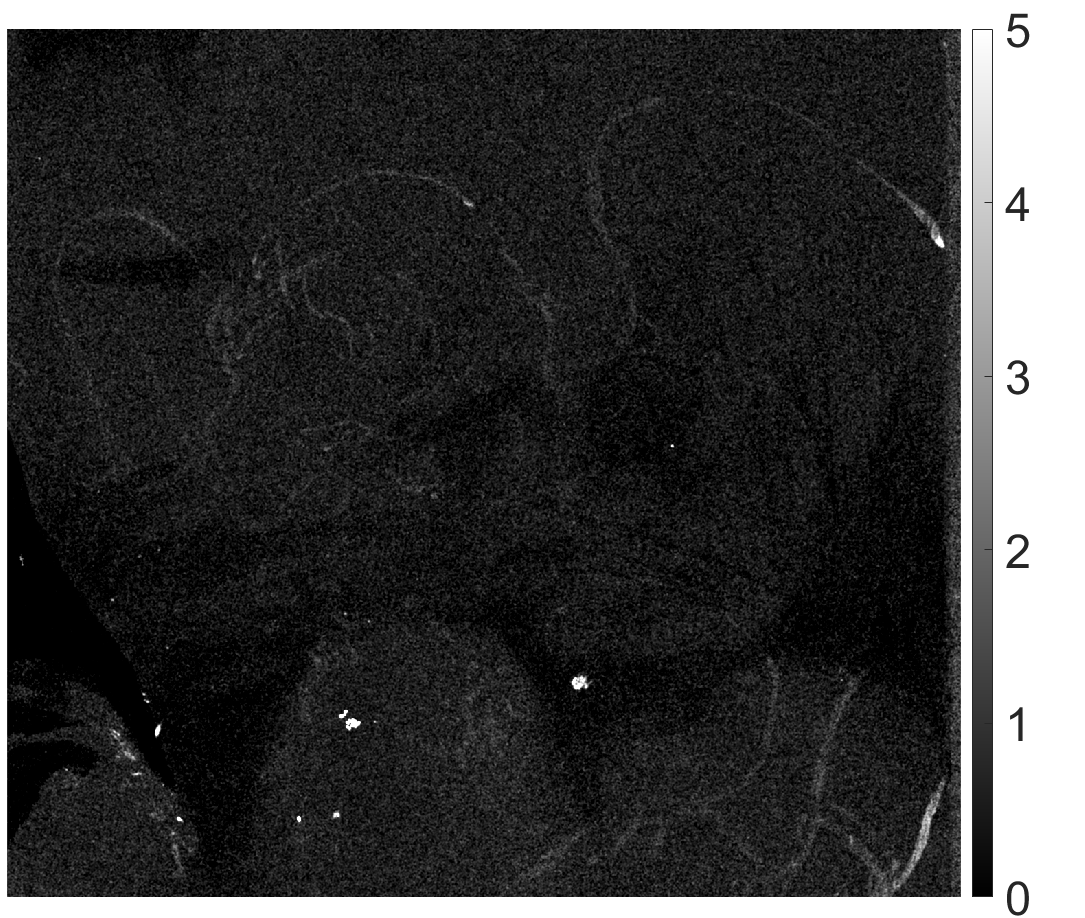}
        \vspace{-0.5cm}
        \caption{Zn $\text{K}_{\alpha}$ w/o constraint (\ref{eq:adjustment})}
        \vspace{-0.1cm}
        \label{fig:NG1093_d10_Zn_K_alpha_FISTA_no_check}
      \end{subfigure}
      \par
      \begin{subfigure}[t]{1\linewidth}
        \includegraphics[width=1\linewidth]{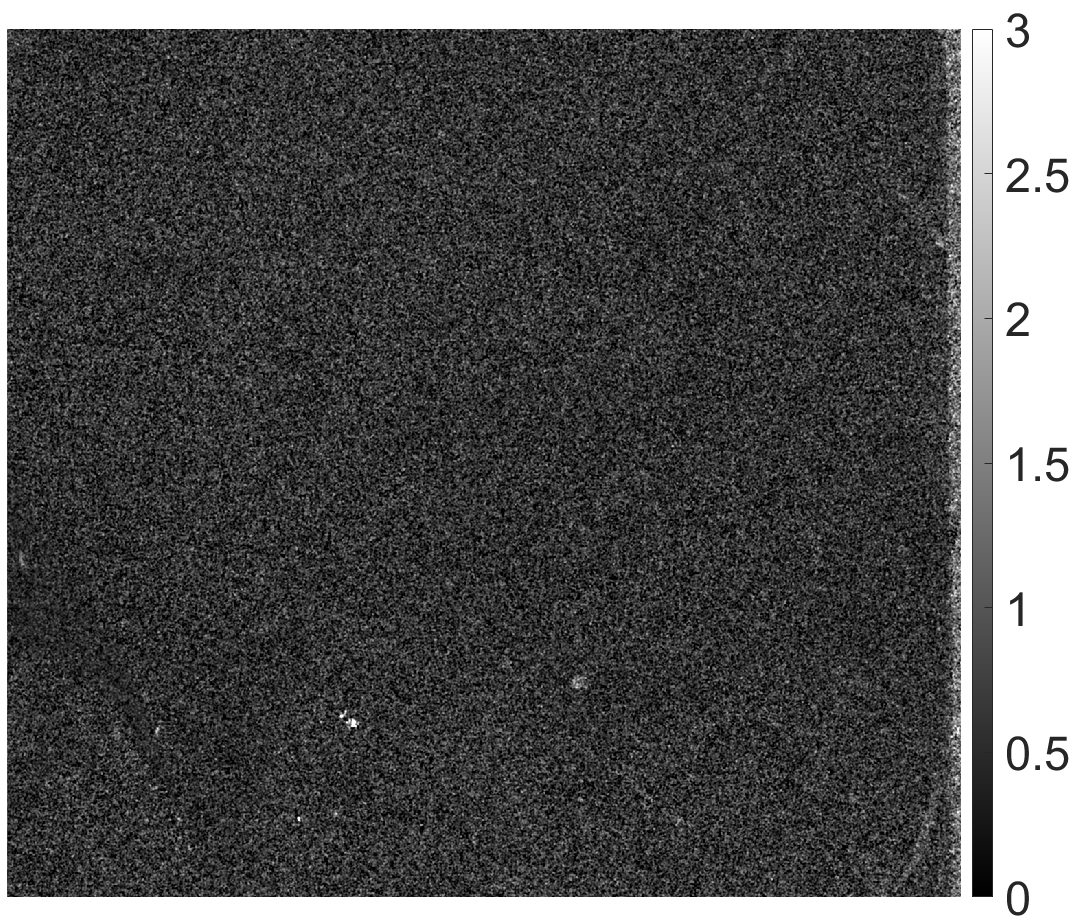}
        \vspace{-0.5cm}
        \caption{Zn $\text{K}_{\beta}$ w/o constraint (\ref{eq:adjustment})}
        \vspace{-0.2cm}
        \label{fig:NG1093_d10_Zn_K_beta_FISTA_no_check}
      \end{subfigure}
    \end{multicols}
\vspace{-0.2cm}
\caption{Element distribution maps produced by the proposed FAD method with FISTA-inspired optimisation scheme (Algorithm \ref{Alg:FISTA}) for the scanned region d10 (yellow outline) of \emph{`The Virgin of the Rocks'}. (a) Cu $\text{K}_{\alpha}$, (b) Zn $\text{K}_{\alpha}$, (c) Zn $\text{K}_{\beta}$, (d) Cu $\text{K}_{\alpha}$ without the physical constraint (\ref{eq:adjustment}), (e) Zn $\text{K}_{\alpha}$ without the physical constraint (\ref{eq:adjustment}), (f) Zn $\text{K}_{\beta}$ without the physical constraint (\ref{eq:adjustment}).}
\label{fig:NG1093 d10 FISTA}
\end{figure}
\begin{figure}[t]
	\captionsetup[subfigure]{aboveskip=2pt,belowskip=3pt}
	\begin{multicols}{2}
		\centering
		\begin{subfigure}[t]{1\linewidth}
			\includegraphics[width=1\linewidth]{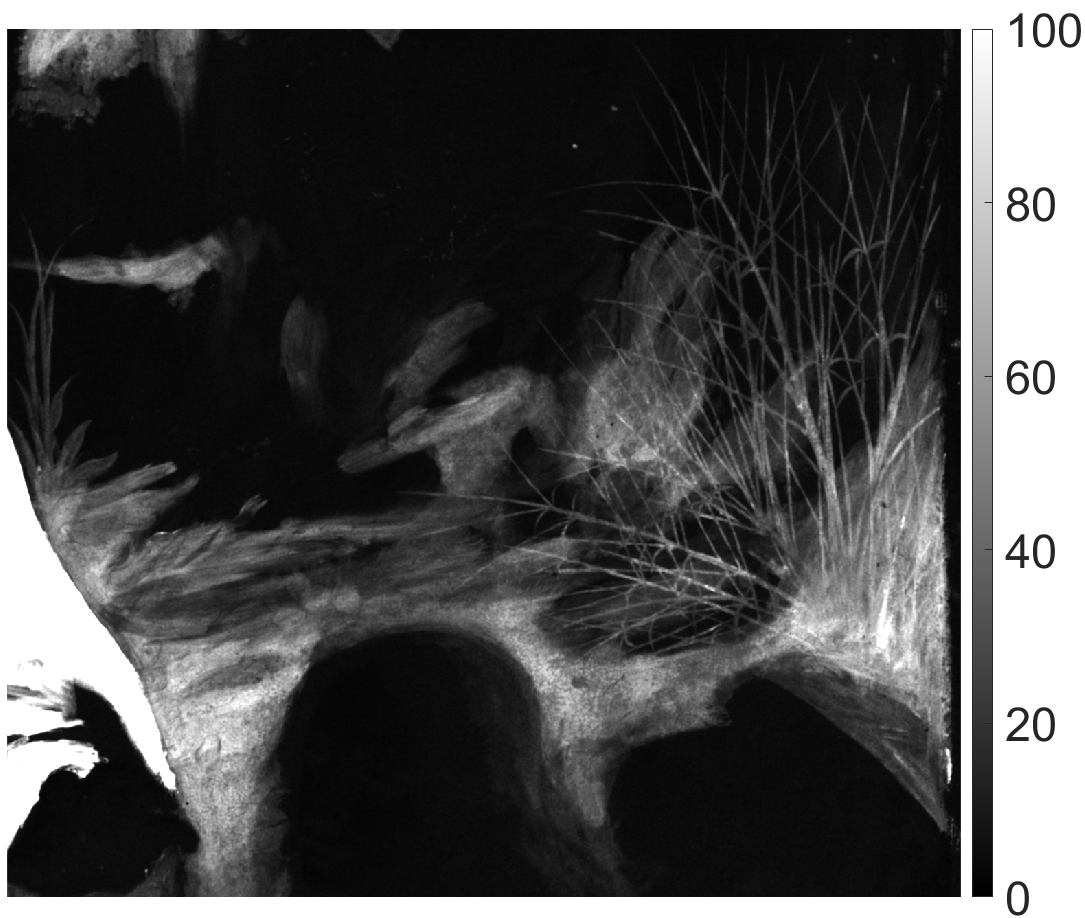}
			\vspace{-0.5cm}
            \caption{Cu $\text{K}_{\alpha}$}
            \vspace{-0.1cm}
			\label{fig:NG1093_d10_Cu_K_alpha_ADMM}
		\end{subfigure}
		\par
		\begin{subfigure}[t]{1\linewidth}
			\includegraphics[width=1\linewidth]{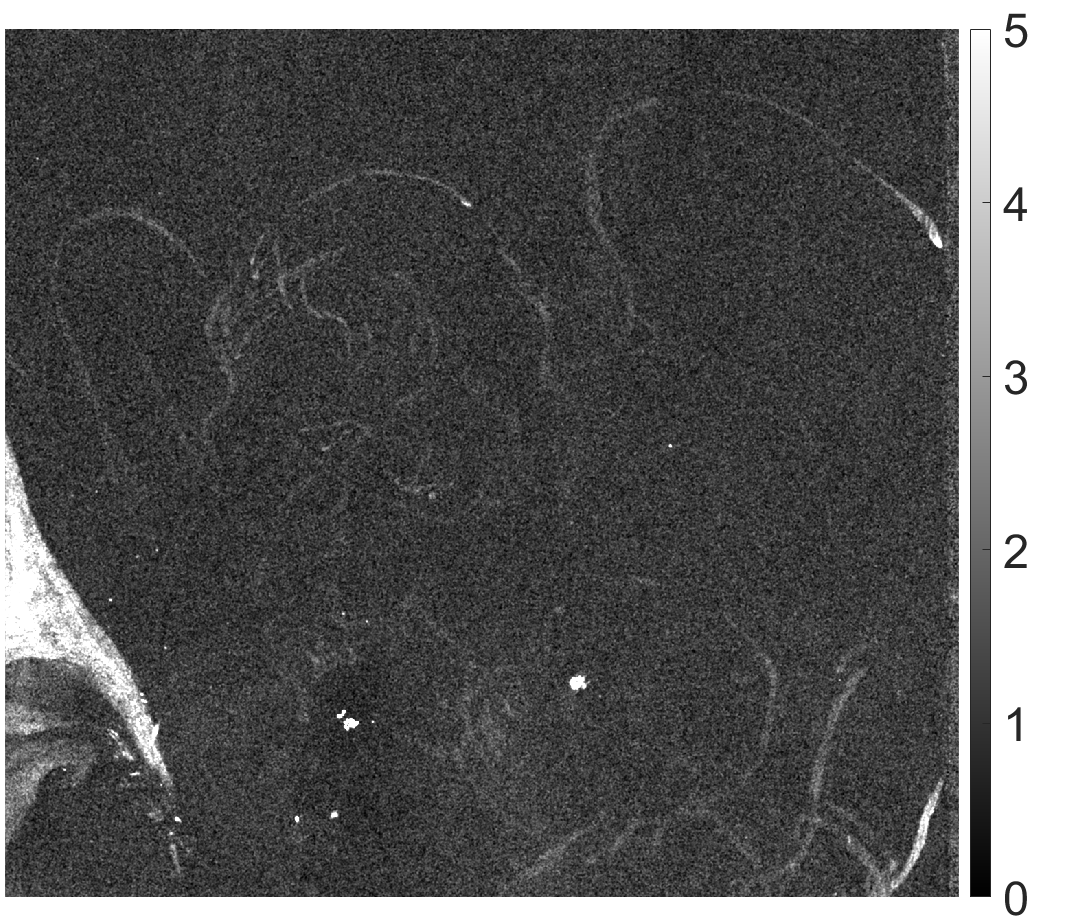}
			\vspace{-0.5cm}
            \caption{Zn $\text{K}_{\alpha}$}
            \vspace{-0.1cm}
			\label{fig:NG1093_d10_Zn_K_alpha_ADMM}
		\end{subfigure}
		\par
		\begin{subfigure}[t]{1\linewidth}
			\includegraphics[width=1\linewidth]{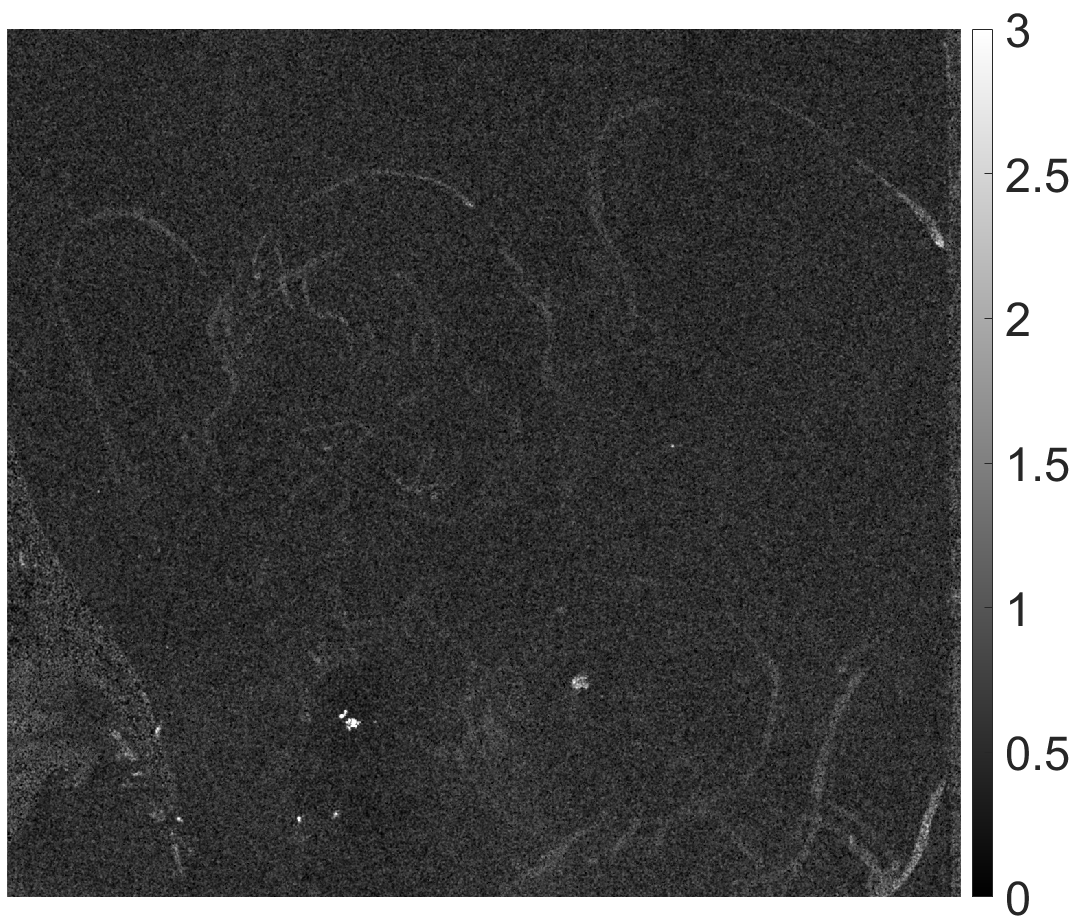}
			\vspace{-0.5cm}
            \caption{Zn $\text{K}_{\beta}$}
            \vspace{-0.1cm}
			\label{fig:NG1093_d10_Zn_K_beta_ADMM}
		\end{subfigure}
		\par
		\begin{subfigure}[t]{1\linewidth}
			\includegraphics[width=1\linewidth]{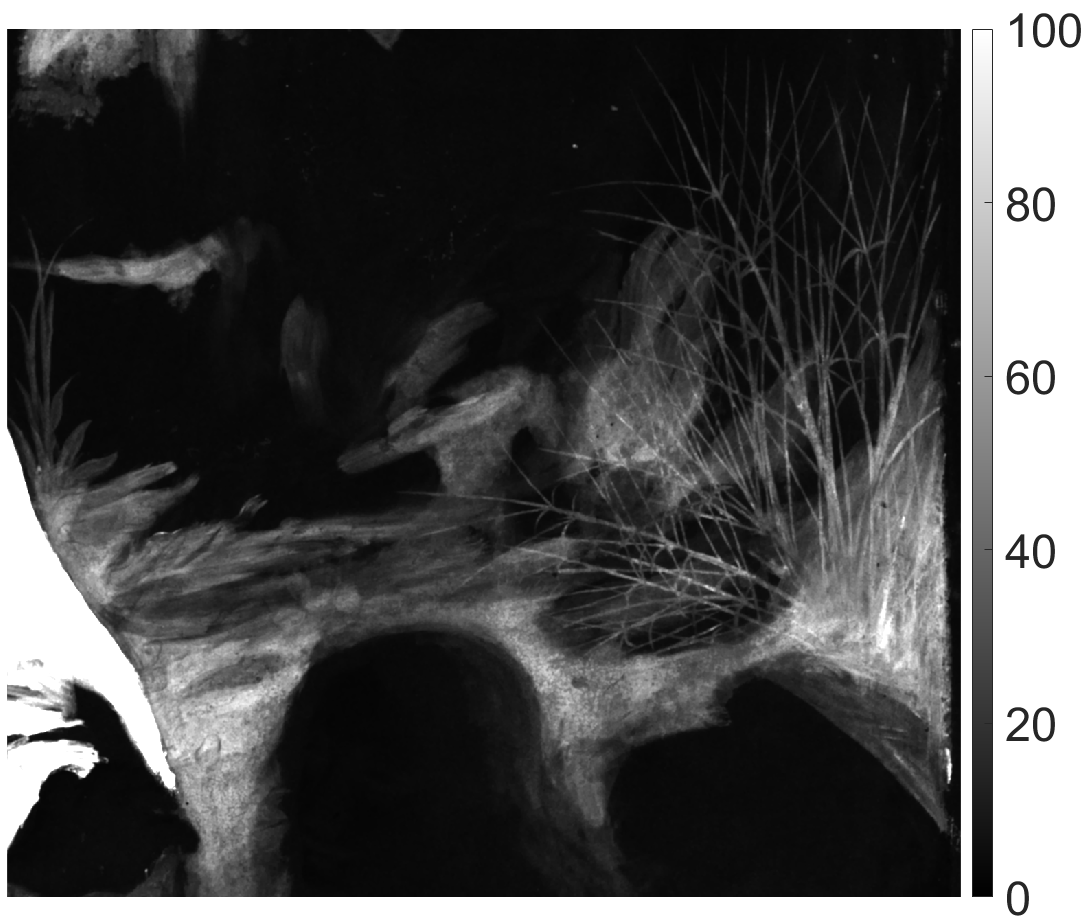}
			\vspace{-0.5cm}
            \caption{Cu $\text{K}_{\alpha}$ w/o constraint (\ref{eq:adjustment})}
            \vspace{-0.1cm}
			\label{fig:NG1093_d10_Cu_K_alpha_ADMM_no_check}
		\end{subfigure}
		\par
		\begin{subfigure}[t]{1\linewidth}
			\includegraphics[width=1\linewidth]{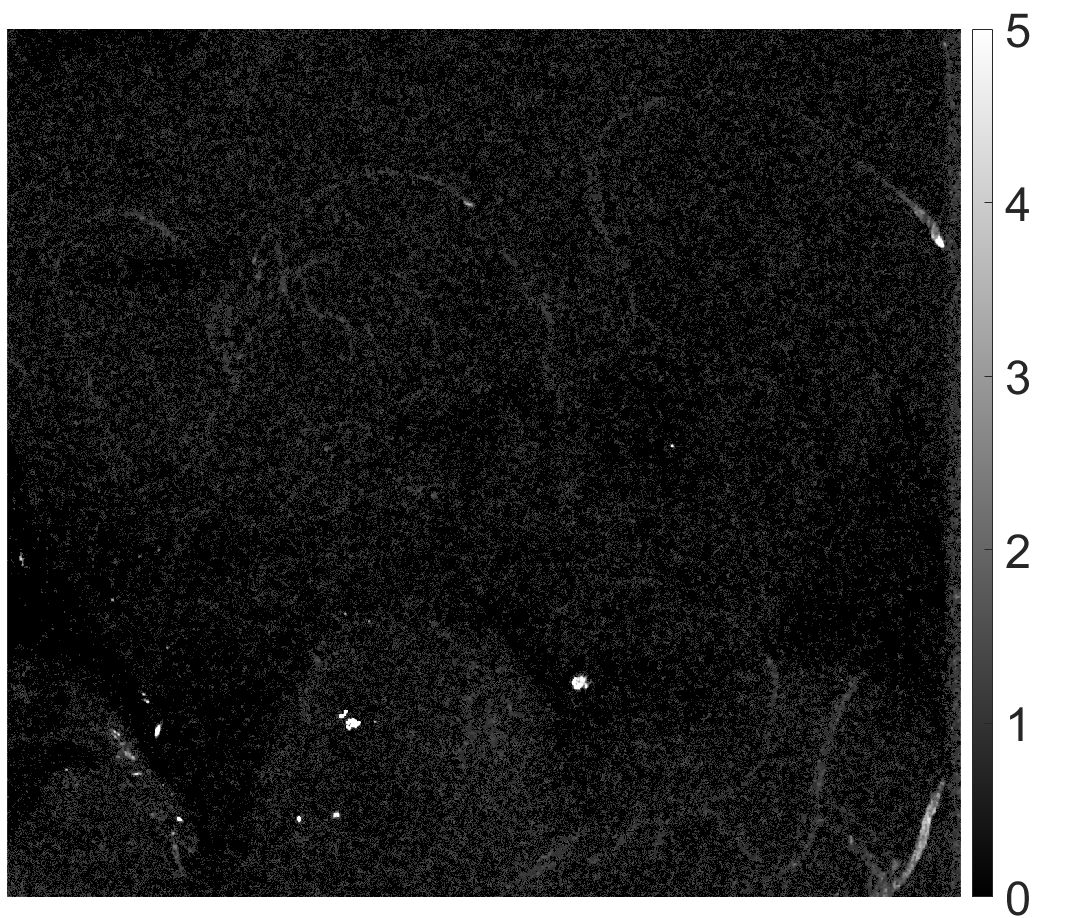}
			\vspace{-0.5cm}
            \caption{Zn $\text{K}_{\alpha}$ w/o constraint (\ref{eq:adjustment})}
            \vspace{-0.1cm}
			\label{fig:NG1093_d10_Zn_K_alpha_ADMM_no_check}
		\end{subfigure}
		\par
		\begin{subfigure}[t]{1\linewidth}
			\includegraphics[width=1\linewidth]{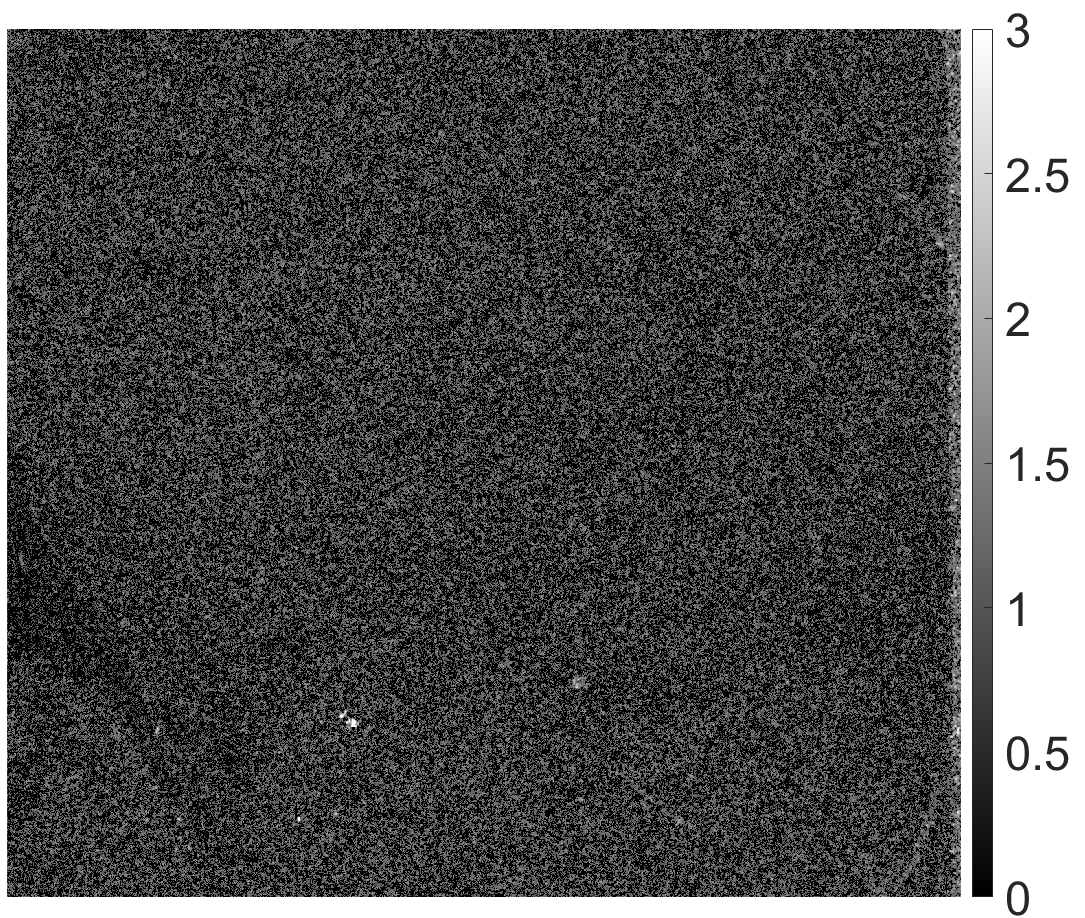}
			\vspace{-0.5cm}
            \caption{Zn $\text{K}_{\beta}$ w/o constraint (\ref{eq:adjustment})}
            \vspace{-0.1cm}
			\label{fig:NG1093_d10_Zn_K_beta_ADMM_no_check}
		\end{subfigure}
	\end{multicols}
	\vspace{-0.2cm}
	\caption{Element distribution maps produced by the proposed FAD method with ILF-ADMM optimisation scheme (Algorithm \ref{Alg:ILF_ADMM}) for the scanned region d10 (yellow outline) of \emph{`The Virgin of the Rocks'}. (a) Cu $\text{K}_{\alpha}$, (b) Zn $\text{K}_{\alpha}$, (c) Zn $\text{K}_{\beta}$, (d) Cu $\text{K}_{\alpha}$ without the physical constraint (\ref{eq:adjustment}), (e) Zn $\text{K}_{\alpha}$ without the physical constraint (\ref{eq:adjustment}), (f) Zn $\text{K}_{\beta}$ without the physical constraint (\ref{eq:adjustment}).}
	\label{fig:NG1093 d10 ADMM}
	\vspace{-0.2cm}
\end{figure}
\begin{figure}[t]
	\captionsetup[subfigure]{aboveskip=2pt,belowskip=3pt}
	\begin{multicols}{2}
		\centering
		\begin{subfigure}[t]{1\linewidth}
			\includegraphics[width=1\linewidth]{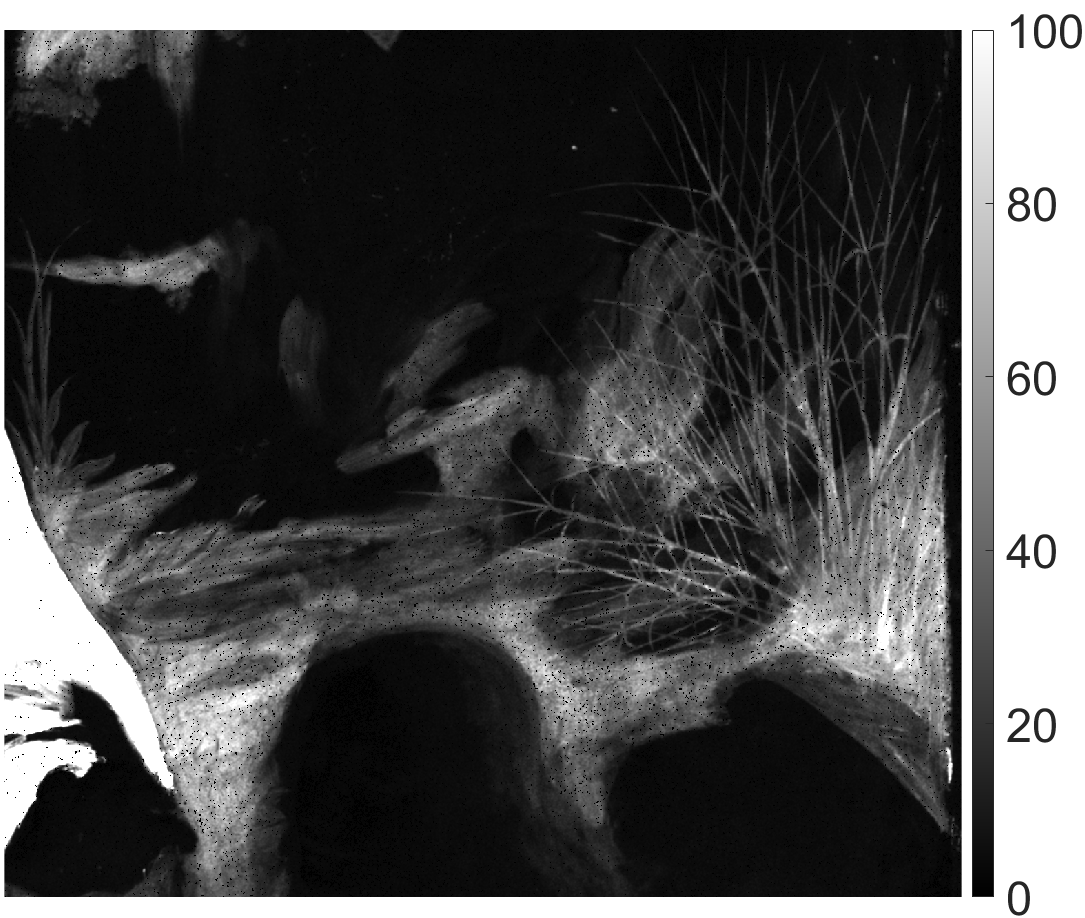}
			\vspace{-0.5cm}
            \caption{Cu $\text{K}_{\alpha}$ quantity map}
            \vspace{-0.1cm}
			\label{fig:NG1093_d10_Cu_K_alpha_FRI_quantity}
		\end{subfigure}
		\par
		\begin{subfigure}[t]{1\linewidth}
			\includegraphics[width=1\linewidth]{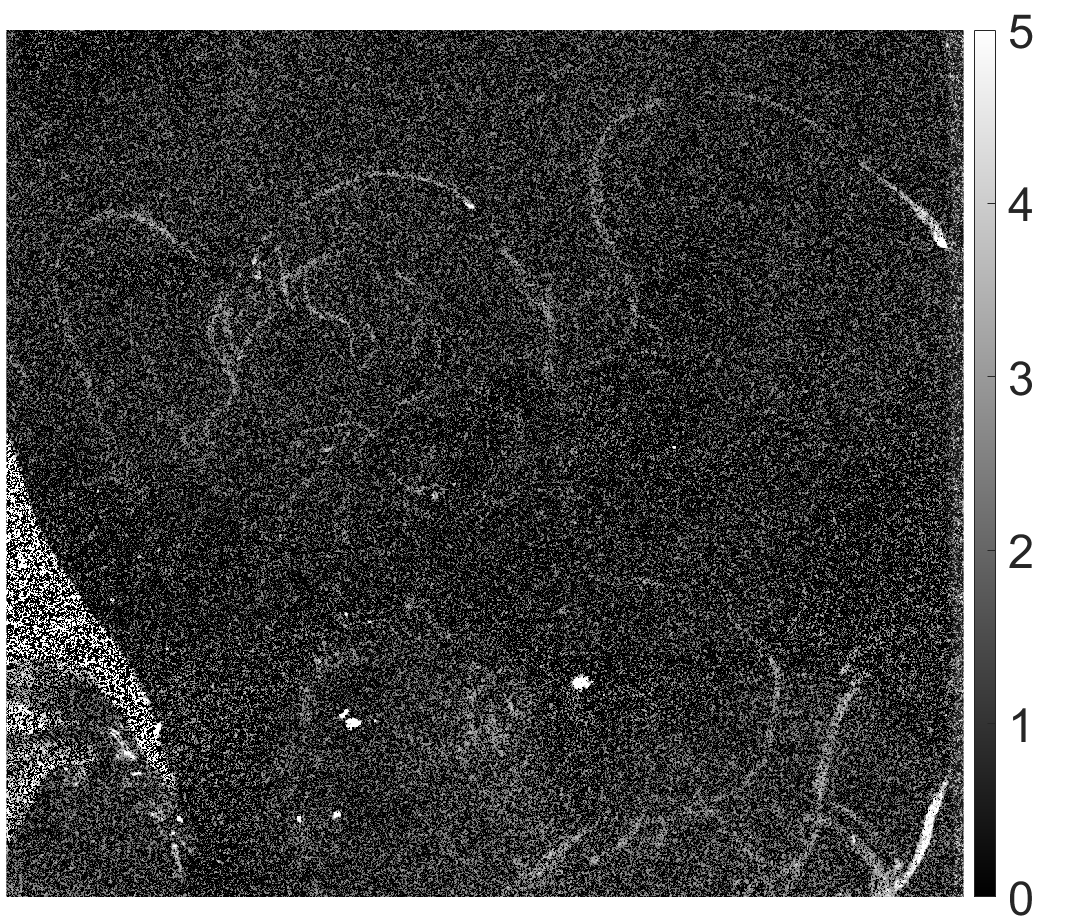}
			\vspace{-0.5cm}
            \caption{Zn $\text{K}_{\alpha}$ quantity map}
            \vspace{-0.1cm}
			\label{fig:NG1093_d10_Zn_K_alpha_FRI_quantity}
		\end{subfigure}
		\par
		\begin{subfigure}[t]{1\linewidth}
			\includegraphics[width=1\linewidth]{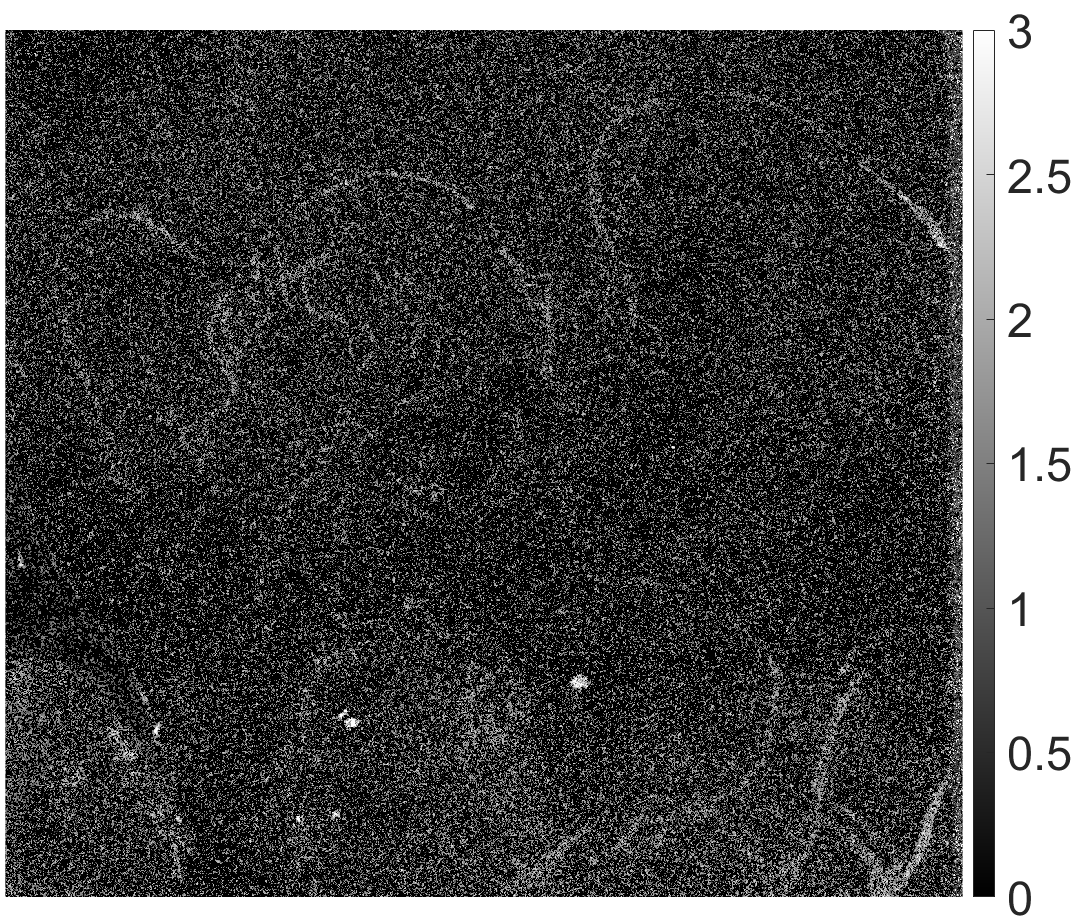}
			\vspace{-0.5cm}
            \caption{Zn $\text{K}_{\beta}$ quantity map}
            \vspace{-0.1cm}
			\label{fig:NG1093_d10_Zn_K_beta_FRI_quantity}
		\end{subfigure}
		\par
		\begin{subfigure}[t]{1\linewidth}
			\includegraphics[width=1\linewidth]{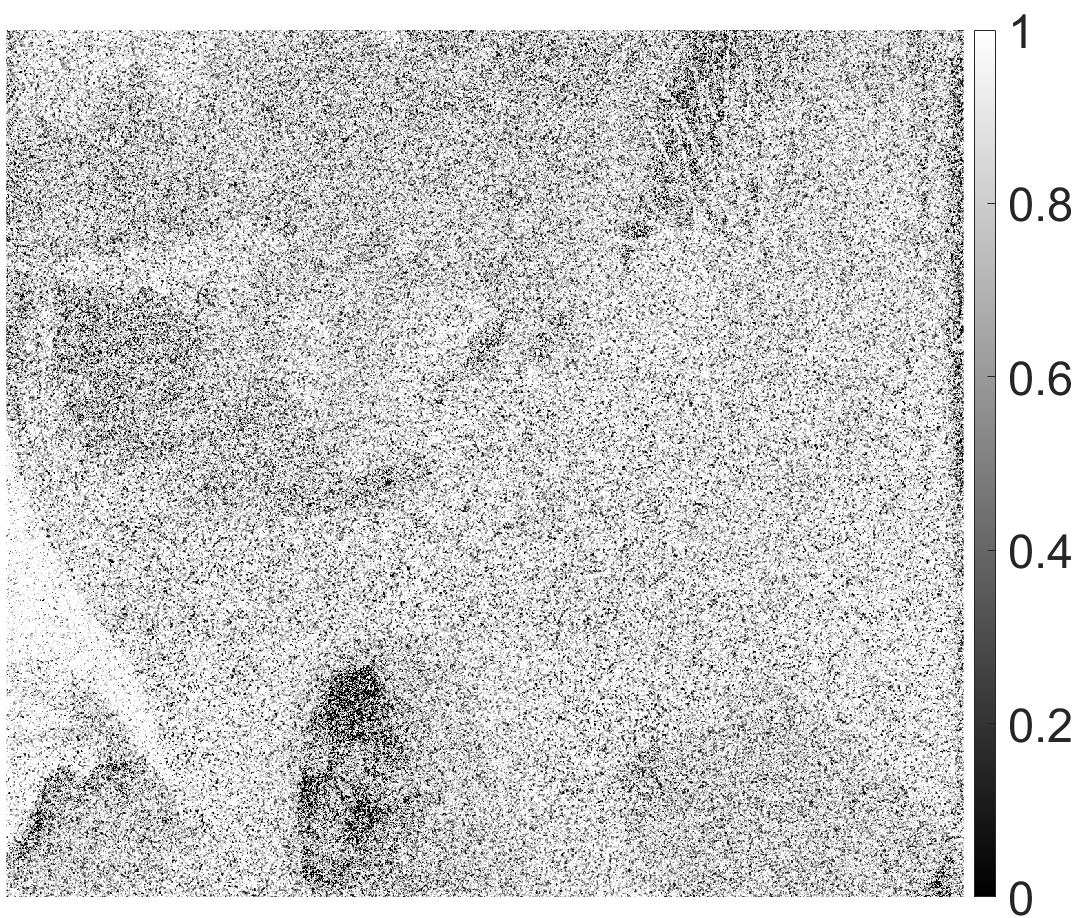}
			\vspace{-0.5cm}
            \caption{Cu $\text{K}_{\alpha}$ confidence map}
            \vspace{-0.1cm}
			\label{fig:NG1093_d10_Cu_K_alpha_FRI_confidence}
		\end{subfigure}
		\par
		\begin{subfigure}[t]{1\linewidth}
			\includegraphics[width=1\linewidth]{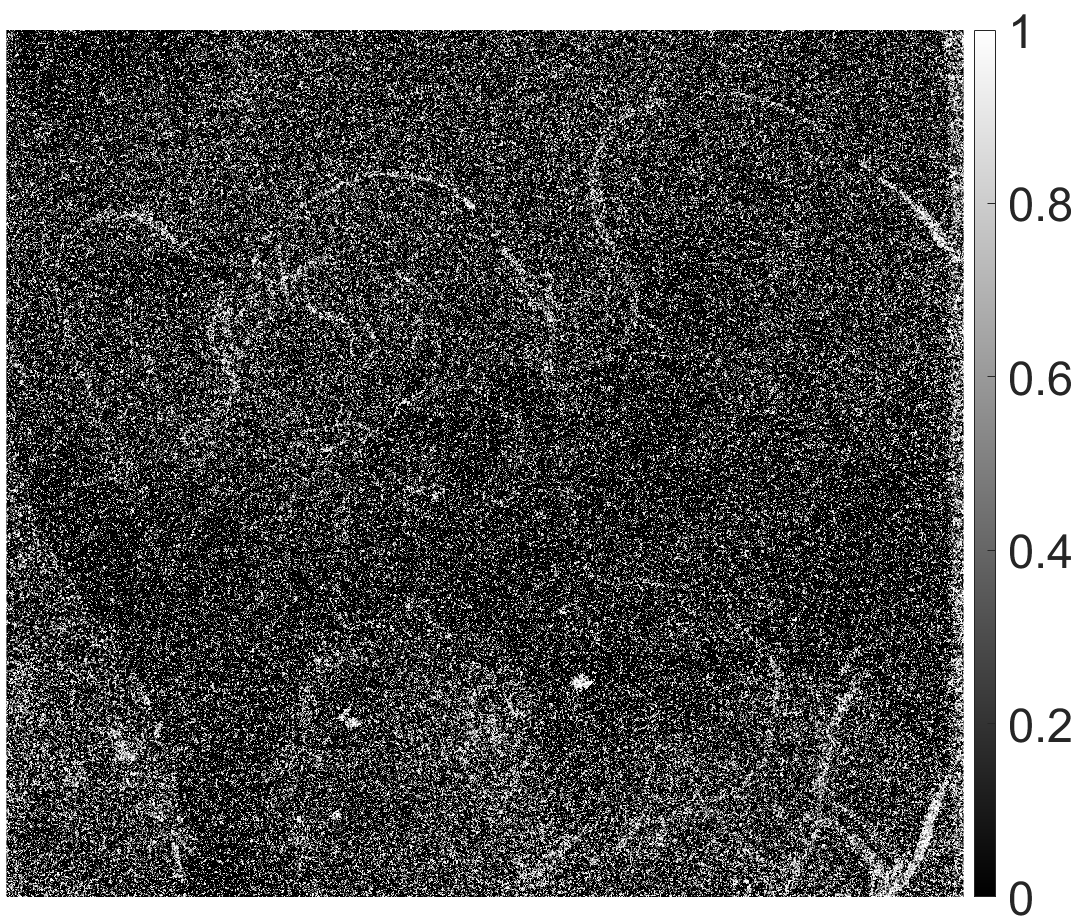}
			\vspace{-0.5cm}
            \caption{Zn $\text{K}_{\alpha}$ confidence map}
            \vspace{-0.1cm}
			\label{fig:NG1093_d10_Zn_K_alpha_FRI_confidence}
		\end{subfigure}
		\par
		\begin{subfigure}[t]{1\linewidth}
			\includegraphics[width=1\linewidth]{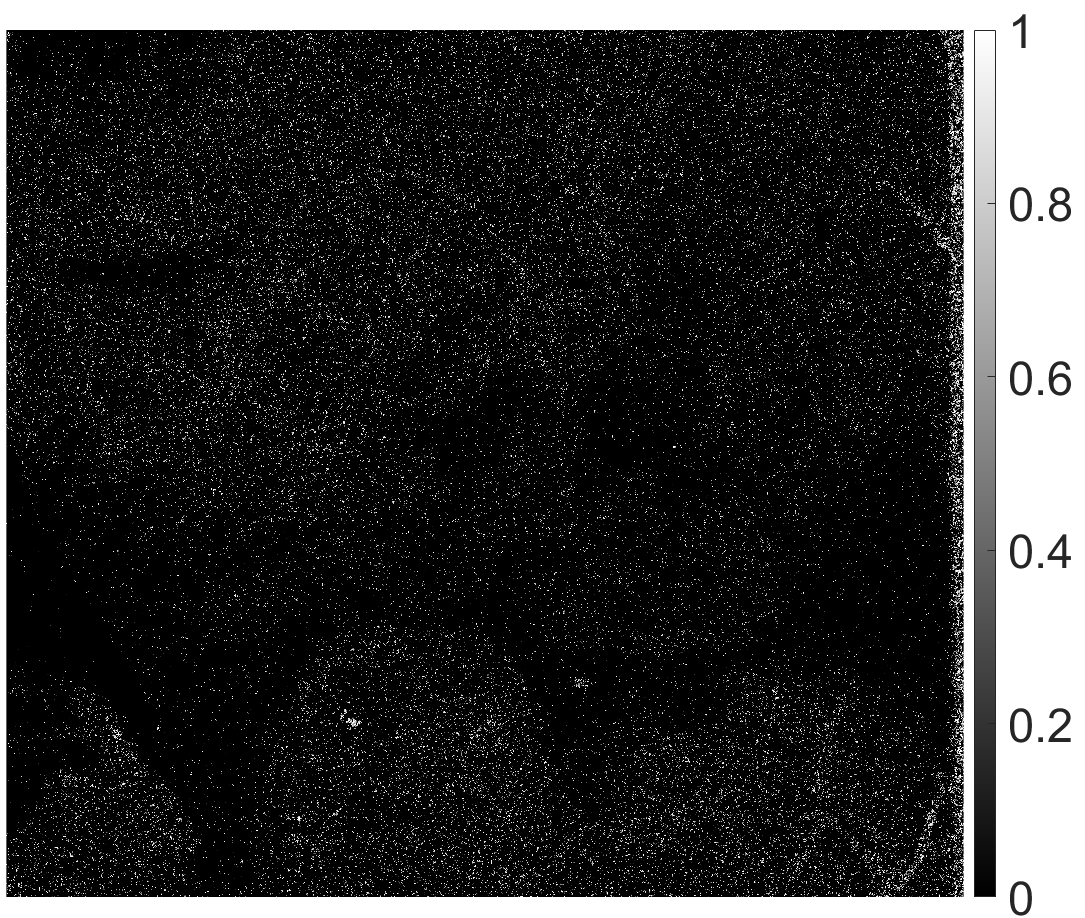}
			\vspace{-0.5cm}
            \caption{Zn $\text{K}_{\beta}$ confidence map}
            \vspace{-0.1cm}
			\label{fig:NG1093_d10_Zn_K_beta_FRI_confidence}
		\end{subfigure}
	\end{multicols}
	\vspace{-0.2cm}
	\caption{Element distribution maps produced by the AFRID method \cite{yan2021prony} for the scanned region d10 (yellow outline) of \emph{`The Virgin of the Rocks'}. (a) Cu $\text{K}_{\alpha}$ quantity map, (b) Zn $\text{K}_{\alpha}$ quantity map, (c) Zn $\text{K}_{\beta}$ quantity map, (d) Cu $\text{K}_{\alpha}$ confidence map, (e) Zn $\text{K}_{\alpha}$ confidence map, (f) Zn $\text{K}_{\beta}$ confidence map.}
	\label{fig:NG1093 d10 FRI}
	\vspace{-0.2cm}
\end{figure}
\begin{figure*}[t]
\centering
    \hspace*{\fill}
    \begin{subfigure}{0.32\linewidth}
        \centering
        \includegraphics[width=1\linewidth]{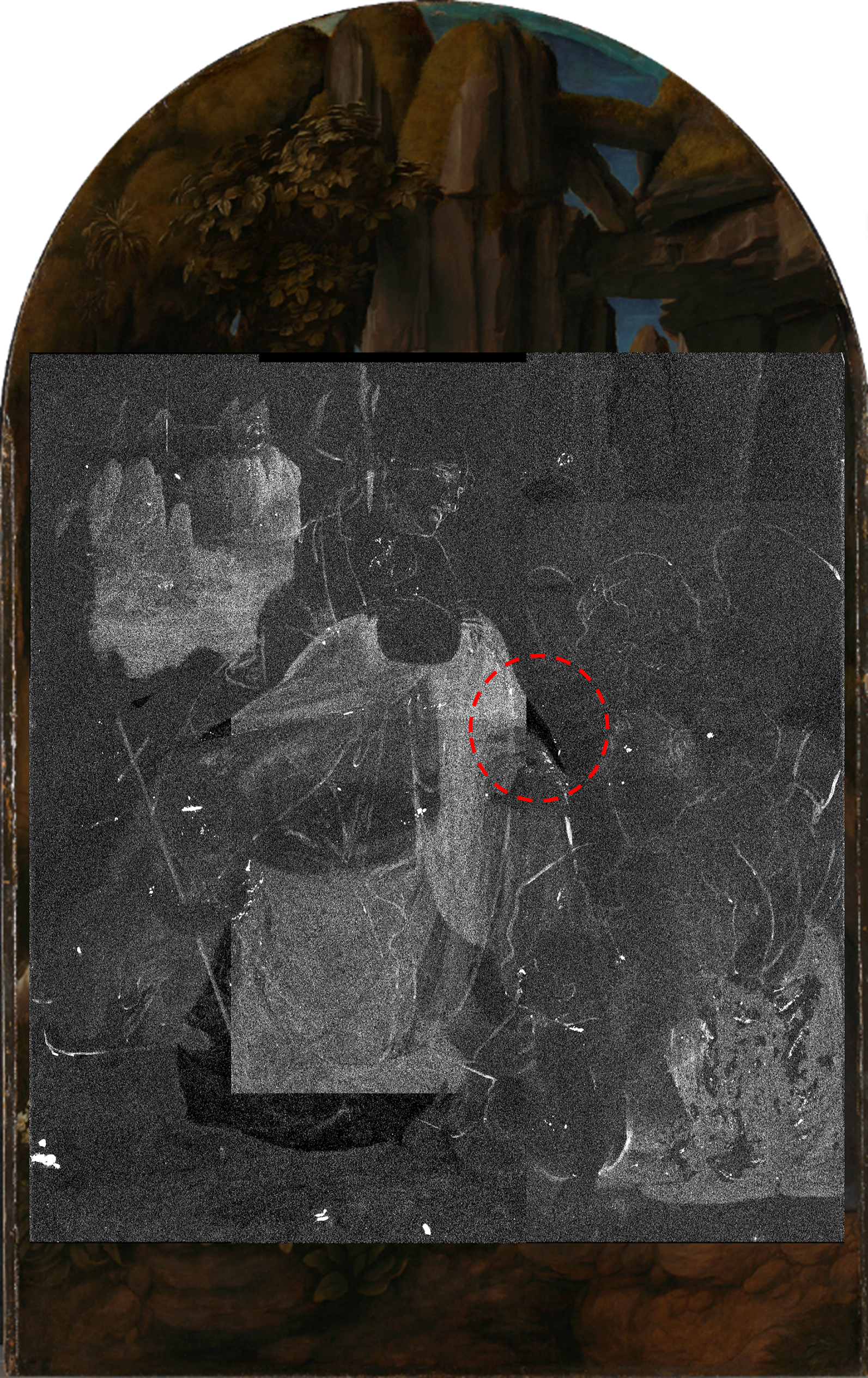}
        \vspace{-0.5cm}
        \caption{Bruker M6 software}
        \vspace{-0.1cm}
        \label{Fig:Zn_combine_Bruker}
    \end{subfigure}
    \hfill
    \begin{subfigure}{0.32\linewidth}
        \centering
        \includegraphics[width=1\linewidth]{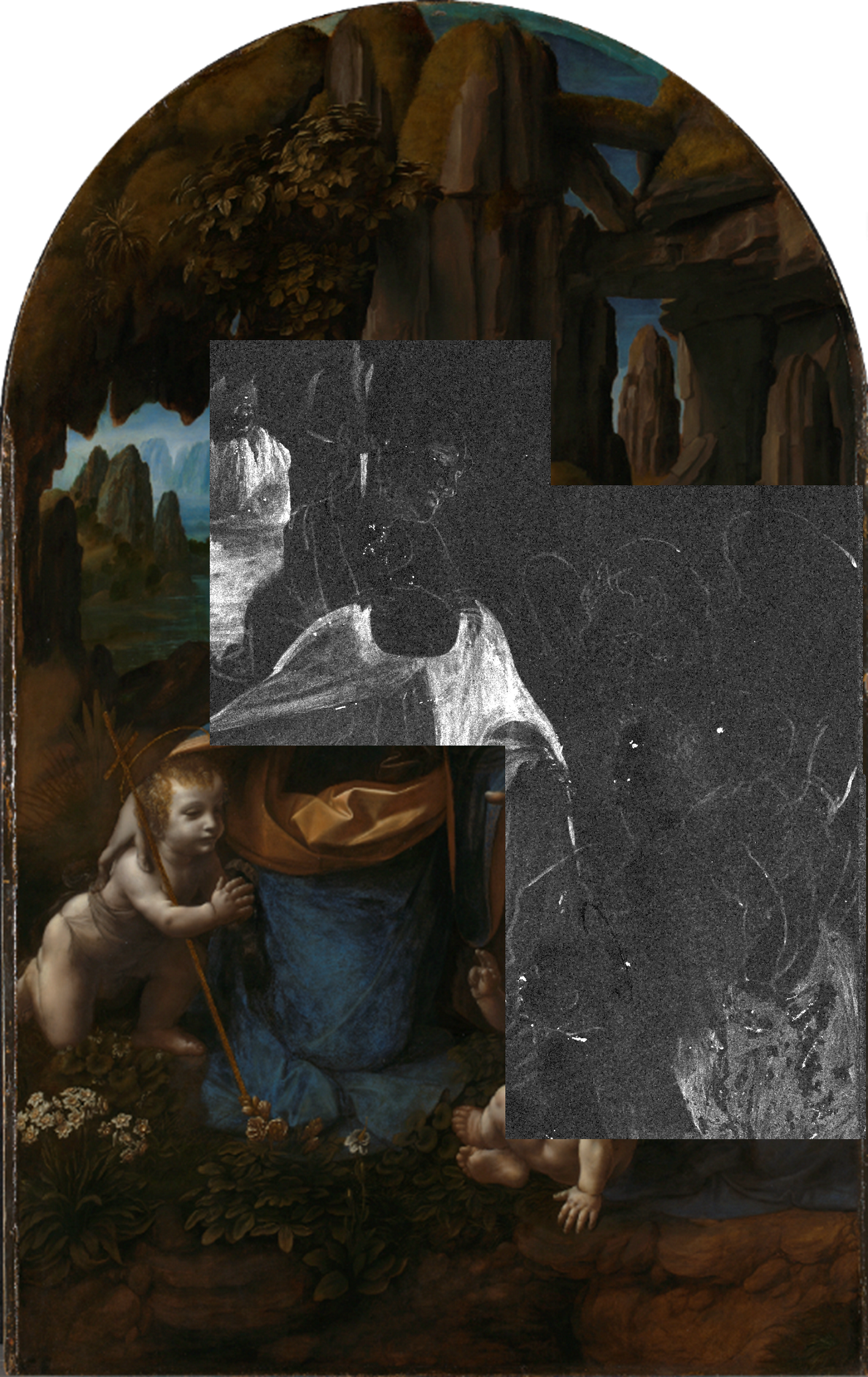}
        \vspace{-0.5cm}
        \caption{FAD method with ILF-ADMM scheme}
        \vspace{-0.1cm}
        \label{Fig:Zn_combine_ADMM}
    \end{subfigure}
    \hfill
    \begin{subfigure}{0.32\linewidth}
        \centering
        \includegraphics[width=1\linewidth]{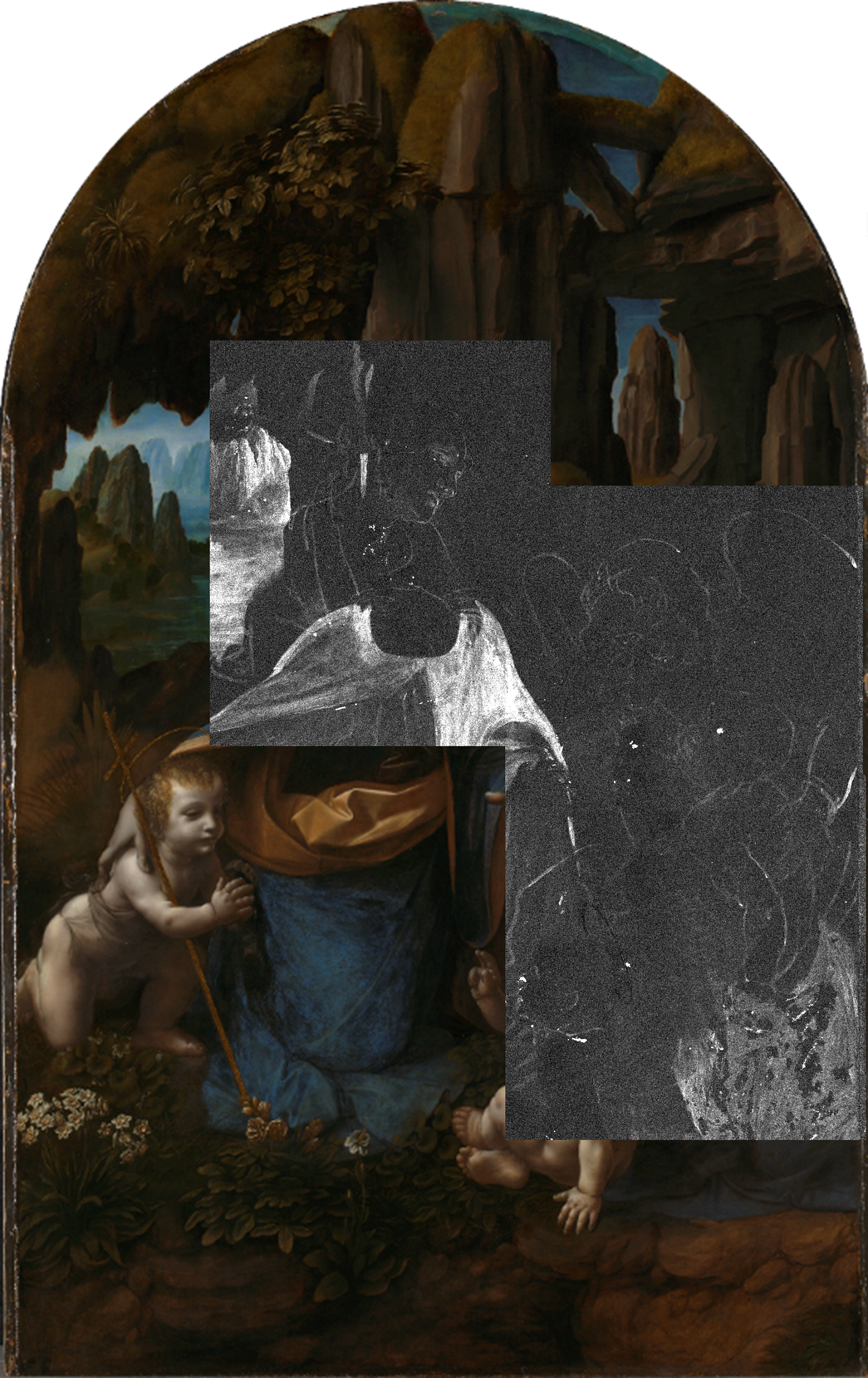}
        \vspace{-0.5cm}
        \caption{FAD method with FISTA scheme}
        \vspace{-0.1cm}
        \label{Fig:Zn_combine_FISTA}
    \end{subfigure}
    \hspace*{\fill}
\caption{Combined Zn $\text{K}_{\alpha}$ map of \emph{`The Virgin of the Rocks'} produced with different methods, revealing underdrawing related to an abandoned earlier composition. (a) deconvoluted with the Bruker M6 software, (b) deconvoluted with the proposed FAD method with the ILF-ADMM optimisation scheme (Algorithm \ref{Alg:ILF_ADMM}), (c) deconvoluted with the proposed FAD method with FISTA-inspired optimisation scheme (Algorithm \ref{Alg:FISTA}). The region highlighted in red shows the mismatches when combining nearby patches.}
\label{fig:NG1093_combine_Zn_FISTA}
\end{figure*}

To verify the performance of the proposed FAD method with the two optimisation schemes, we apply it to the MA-XRF datacubes collected from two easel paintings in the National Gallery collection, \emph{Sunflowers} (NG3863) by Vincent van Gogh \cite{NG_Technical_37, hendriks2019van} and \emph{The Virgin with the Infant Saint John the Baptist adoring the Christ Child accompanied by an Angel (`The Virgin of the Rocks')} (NG1093) created by Leonardo da Vinci \cite{NG_Technical_32,spring2021leonardo}, shown in Fig. \ref{fig:sunflowers} and Fig. \ref{fig:leonardo_oil_painting}, respectively. To acquire the MA-XRF datacubes, the highlighted regions in each painting were scanned with a Bruker M6 JETSTREAM MA-XRF instrument, consisting of a 30 \si{\watt} rhodium (Rh) anode X-ray tube with polycapillary optics operated at 50 \si{\kilo\volt} and 450-600 \si{\micro\ampere} and a 60 \si{\milli\meter^2} silicon (Si) drift detector with a threshold of 275 kcps. The deconvoluted element distribution maps generated after 50 iterations (epochs) of the proposed FAD method are also compared with the ones generated with both the AFRID method \cite{yan2021prony} and the Bruker M6 software, which is widely used for MA-XRF deconvolution in the cultural heritage sector.

\subsection{Automatic Deconvolution}
A common drawback of the existing methods \cite{van1977computer,vekemans1995comparison,vekemans1994analysis,sole2007multiplatform,alfeld2015strategies,ryan1993dynamic,ryan1995new,alfeld2013mobile,Conover2015} for MA-XRF deconvolution is that they require users to input a list of chemical elements which are expected to be present in the scanned target. This requirement of manual input adds complexity and a degree of variability to the use of these methods. However, as with the AFRID method, the FAD method proposed here is also able to deconvolute the MA-XRF datacube of a painting and produce element distribution maps without any user selection of chemical elements. We first focus on the MA-XRF datacube that was collected from the painting \emph{Sunflowers}, highlighted in Fig. \ref{fig:sunflowers}. The selection of processed results shown in Fig. \ref{fig:NG3863 elemental maps} demonstrate that, although the proposed method does not require any input of chemical elements, it still produces accurate element maps for the scanned region of \emph{Sunflowers}, which have similar distributions to the ones that are generated by the Bruker M6 software after detailed examination of the data by expert users to find a suitable selection of chemical elements. This improvement significantly reduces the complexity for the user to deconvolute the MA-XRF datacube of an easel painting. 

\subsection{Nearby Element Pulse Separation}
Another common difficulty in deconvoluting MA-XRF datacubes collected from easel paintings is the separation of overlapping pulses in the spectra that are due to different chemical elements. Failure to do so will lead to the deconvoluted element maps having interference from other chemical elements. For instance, the MA-XRF datacubes collected from three regions (d10 in yellow, d11 in pink and d12 in green) of the painting \emph{`The Virgin of the Rocks'} by Leonardo da Vinci, highlighted in Fig. \ref{fig:leonardo_oil_painting}, have the problem of separating copper (Cu) and zinc (Zn), since the $\text{K}_{\beta}$ characteristic line of Cu with energy of about 8905 \si{\electronvolt} is very close to the $\text{K}_{\alpha}$ line of Zn with energy of about 8631 \si{\electronvolt}. This can be seen by directly plotting the region of interest (ROI) of Zn $\text{K}_{\alpha}$ for the d10 scanned area of the painting in Fig. \ref{fig:NG1093 Zn bruker}, where some interesting zinc-containing underdrawing lines can be found (traced and overlaid onto the painting in Fig. \ref{fig:NG1093 underdrawing}) but also significant interference from Cu is present and required to be reduced with further processing. With a list of chemical elements selected by experts after multiple attempts, the deconvoluted element distribution maps of Cu $\text{K}_{\beta}$ and Zn $\text{K}_{\alpha}$ generated by the inbuilt deconvolution algorithm in the Bruker M6 software are shown in Fig. \ref{fig:NG1093 Cu bruker} and Fig. \ref{fig:NG1093 Zn-Cu bruker} respectively. Although the interference of Cu is removed from the Zn $\text{K}_{\alpha}$ map, the algorithm also reduces the intensity of the Zn $\text{K}_{\alpha}$ signal, making the underdrawing lines less visible and even removing the Zn signal in the Virgin's drapery completely. To solve this problem and produce an element map that accurately reflects the distribution of Zn in the painting, shown in Fig. \ref{fig:NG1093 Zn-Cu manually bruker}, experts had to manually scale and subtract the Cu map (Fig. \ref{fig:NG1093 Cu bruker}) from the ROI Zn map (Fig. \ref{fig:NG1093 Zn bruker}). Moreover, the selection of chemical elements input to the Bruker M6 software has a great impact on the deconvoluted element maps. For example, if the element nickel (Ni) is added in the selection, the new deconvoluted Zn $\text{K}_{\alpha}$ map (Fig. \ref{fig:NG1093 Zn_Ni bruker}) changes significantly and the zinc signal in the Virgin's drapery can be recovered, even though no measurable signal is found in the deconvoluted Ni $\text{K}_{\alpha}$ map.

The proposed FAD method is also applied on the same MA-XRF datacube. The element distribution maps deconvoluted with the FISTA-inspired optimisation scheme (Algorithm \ref{Alg:FISTA}) are displayed in Fig. \ref{fig:NG1093 d10 FISTA} and the ones deconvoluted with the ILF-ADMM optimisation scheme (Algorithm \ref{Alg:ILF_ADMM}) are shown in Fig. \ref{fig:NG1093 d10 ADMM}. Our Cu $\text{K}_{\alpha}$ maps shown in Fig. \ref{fig:NG1093_d10_Cu_K_alpha_FISTA} and \ref{fig:NG1093_d10_Cu_K_alpha_ADMM} are nearly identical to the one produced by the Bruker M6 software. However, our Zn $\text{K}_{\alpha}$ maps in Fig. \ref{fig:NG1093_d10_Zn_K_alpha_FISTA} and \ref{fig:NG1093_d10_Zn_K_alpha_ADMM}, where the interference from Cu is barely present, are very similar to the idealised one that is generated by experts after manual intervention (Fig. \ref{fig:NG1093 Zn-Cu manually bruker}). This proves that the proposed FAD method is able to separate Cu and Zn very well and produce accurate element distribution maps. More surprisingly, our proposed method can also produce the element map for the Zn $\text{K}_{\beta}$ line showing the underdrawing lines in Fig. \ref{fig:NG1093_d10_Zn_K_beta_FISTA} and \ref{fig:NG1093_d10_Zn_K_beta_ADMM}, which could not be generated with the Bruker M6 software. The reason is that the intensity of the Zn $\text{K}_{\beta}$ signal is very weak in the MA-XRF datacube, only around 2 photon counts in the areas of underdrawing. This success of deconvoluting Zn $\text{K}_{\beta}$ signal is also due to the proposed physical constraint expressed in (\ref{eq:adjustment}). With the physical constraint, the intensities of the signals for different lines of different elements are checked and re-adjusted to reasonable values during each optimisation loop according to the intensity ratios between different lines of the same element. To validate the physical constraint, we also show the element distribution maps of Cu and Zn generated using the proposed FAD method with the two optimisation schemes but without the physical constraint for the comparison. The Cu $\text{K}_{\alpha}$ maps (Fig. \ref{fig:NG1093_d10_Cu_K_alpha_FISTA_no_check} and \ref{fig:NG1093_d10_Cu_K_alpha_ADMM_no_check}) are similar to the ones with the physical constraint. However, the Zn $\text{K}_{\alpha}$ maps (Fig. \ref{fig:NG1093_d10_Zn_K_alpha_FISTA_no_check} and \ref{fig:NG1093_d10_Zn_K_alpha_ADMM_no_check}) and $\text{K}_{\beta}$ maps (Fig. \ref{fig:NG1093_d10_Zn_K_beta_FISTA_no_check} and \ref{fig:NG1093_d10_Zn_K_beta_ADMM_no_check}) are significantly different from the ones with the physical constraint. Not only has the Zn $\text{K}_{\alpha}$ signal in some regions been removed excessively, but most of underdrawing lines are lost in the Zn $\text{K}_{\beta}$ maps without the physical constraint. From the comparison, it can be concluded that the Zn signals have been better deconvoluted with the physical constraint than without it. The success of deconvoluting the Zn signals and producing the Zn element maps also highlights the excellent ability of our method to detect weak signals from noisy spectra. Furthermore, the deconvoluted element distribution maps produced with AFRID method \cite{yan2021prony} are also shown in Fig. \ref{fig:NG1093 d10 FRI}. Compared to the element distribution maps produced with AFRID method, the ones generated using the proposed FAD method are smoother and less noisy. This is because instead of applying element pulse detection on every pixel like the AFRID method, the proposed FAD method only detects the element pulses from the average and maximum spectra of the datacube, which reduces inconsistency in the deconvoluted element distribution maps. Moreover, the total-variation (TV) regularisation also imposes the spatial smoothness to the deconvoluted maps, which was not considered in the AFRID method. However, the pixel-wise pulse detection in AFRID method also has the advantage of generating a confidence map with values from 0 to 1 for each element line (Fig \ref{fig:NG1093_d10_Cu_K_alpha_FRI_confidence}-\ref{fig:NG1093_d10_Zn_K_beta_FRI_confidence}), indicating how likely it is that element is present in areas of the painting, which provides additional useful information for heritage scientists about the painting.
Finally the combined deconvoluted Zn $\text{K}_{\alpha}$ maps produced with the Bruker M6 software and the proposed FAD method with the ILF-ADMM and FISTA-inspired optimisation schemes are shown in Fig. \ref{fig:NG1093_combine_Zn_FISTA}. It can be seen that there are mismatches between the deconvoluted Zn maps from nearby patches in Fig. \ref{Fig:Zn_combine_Bruker} showing the instability of the Bruker M6 software. However, the Zn maps produced with our proposed method (Fig. \ref{Fig:Zn_combine_ADMM} and Fig. \ref{Fig:Zn_combine_FISTA}) can be seamlessly combined, which indicates the reliability and consistency of our method.

\begin{figure*}[t]
	\centering
	\begin{subfigure}[t]{0.245\linewidth}
		\centering
		\includegraphics[width=1\linewidth]{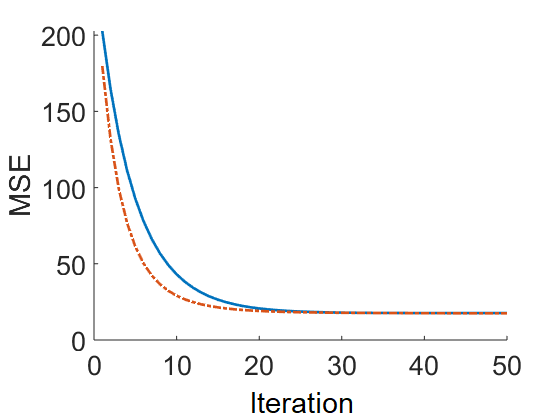}
		\caption{NG1093 d10}
		\label{fig:optimisation_loss_NG1093_d10}
	\end{subfigure}
	\begin{subfigure}[t]{0.245\linewidth}
		\centering
		\includegraphics[width=1\linewidth]{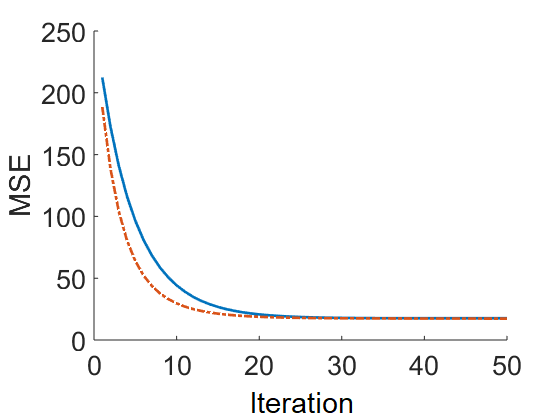}
		\caption{NG1093 d11}
		\label{fig:optimisation_loss_NG1093_d11}
	\end{subfigure}
	\begin{subfigure}[t]{0.245\linewidth}
		\centering
		\includegraphics[width=1\linewidth]{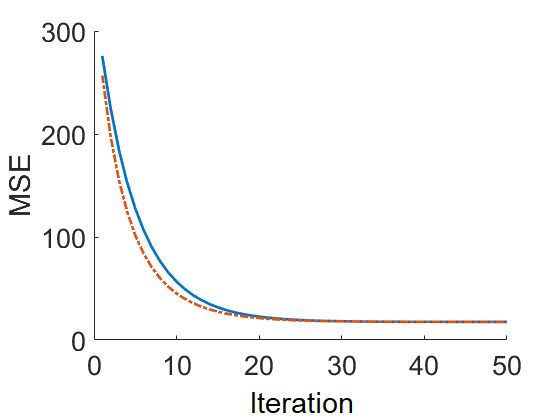}
		\caption{NG1093 d12}
		\label{fig:optimisation_loss_NG1093_d12}
	\end{subfigure}
	\begin{subfigure}[t]{0.245\linewidth}
		\centering
		\includegraphics[width=1\linewidth]{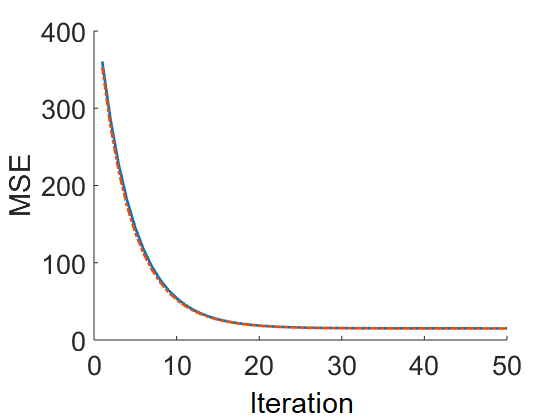}
		\caption{NG3863}
		\label{fig:optimisation_loss_NG3863}
	\end{subfigure}
	\caption{The MSE curves of the proposed FAD method optimised with the FISTA-inspired optimisation scheme (Algorithm \ref{Alg:FISTA}) for deconvoluting different MA-XRF datacubes with the physical constraint (blue solid lines) and without the physical constraint (red dash lines) expressed in (\ref{eq:adjustment}).}
	\vspace{0.3cm}
	\label{fig:optimisation_loss}
\end{figure*}

\begin{table*}[t]
	\centering
	\caption{Time consumed for deconvoluting different datacubes with different methods, measured in seconds, on a PC with AMD Ryzen 9 5900X CPU and 64 GB memory in MATLAB R2021b with parallel computing (parfor-loop) where possible.}
	\begin{tabular}{c | c | c | c | c}
		\hline
		\multirow{2}{*}{Datacube} & \multirow{2}{*}{Pixel number}  & \multicolumn{3}{c}{Time consumed in seconds}  \\ \cline{3-5}
		& & AFRID method \cite{yan2021prony} & FAD with ILF-ADMM & FAD with FISTA  \\
		\hline
		NG1093 d10 &  $1389\times1264$  & 21815 & 1015 & 170 \\
		NG1093 d11 &  $1401\times1417$  & 24251 & 2947 & 186 \\
		NG1093 d12 &  $1325\times1575$  & 24231 & 2804 & 194 \\
		NG3863 &  $1081\times851$  & 8212 & 457 & 64 \\
		\hline
	\end{tabular}
	\label{Table 1}
\end{table*}

\subsection{Convergence}
It is natural to wonder whether adding the physical constraint in (\ref{eq:adjustment}) affects the convergence of the optimisation method. To answer this question, we plot the mean square error (MSE) between the product of the pulse matrix and the estimated element distribution map matrix ($\mathbf{S}\mathbf{\hat{A}}$), and corresponding MA-XRF datacube ($\mathbf{Y}$) for each iteration, in Fig \ref{fig:optimisation_loss}. It can be seen that in all cases of optimisation with the physical constraint, the MSE gradually decreases with more iterations. Moreover, although in most cases the optimisations without the physical constraint converge more quickly, the ones with the physical constraint can still reach similar MSE values after several iterations. All this suggests that the proposed FAD method converges.

\subsection{Processing Speed}
As a main contribution of the proposed FAD method, we demonstrate its fast processing speed. Since other existing methods all require a number of manual steps and operations by the user when deconvoluting the MA-XRF datacube, the processing times required vary greatly and are not considered in the comparison. Therefore, we only compare the speeds of the proposed FAD method optimised with either the FISTA-inspired scheme (Algorithm \ref{Alg:FISTA}) and the ILF-ADMM-based scheme (Algorithm \ref{Alg:ILF_ADMM}) to the AFRID method proposed in \cite{yan2021prony}. All methods were implemented on a PC with AMD Ryzen 9 9500X CPU and 64 GB memory in MATLAB R2021b with parallel computing (parfor-loop) where possible. From Table \ref{Table 1}, it can be seen that the AFRID method takes the longest. This is because the AFRID method operates pixel by pixel. However, the proposed FAD method manages to process the datacube as a whole, which significantly reduces the time consumed. The FAD method with an ILF-ADMM optimisation scheme is more than 12 times faster than the AFRID method, but still not the best. The proposed FAD method with a FISTA-inspired optimisation scheme can deconvolute the XRF datacube at the highest speed, within minutes, and is approximately 10 times faster than the FAD method with an ILF-ADMM optimisation scheme and 124 times faster than the AFRID method. This is because our FISTA-inspired optimisation scheme requires fewer auxiliary variables than the optimisation with ILF-ADMM. Having fewer auxiliary variables in the optimisation loop of Algorithm \ref{Alg:FISTA} leads to fewer matrix multiplication operations and less memory usage, which not only makes the implementation much easier but also improves the processing speed.

\section{Conclusion}
\label{sec:conclusion}
In this paper, we have proposed a fast automatic deconvolution (FAD) method to process the MA-XRF datacubes of easel paintings and to produce the distribution maps of the chemical elements that are present in the paintings. To verify the performance of the proposed FAD method, we applied it on MA-XRF datacubes collected from two easel paintings at the National Gallery, London. The results show that the proposed FAD method can accurately deconvolute the MA-XRF datacubes collected from easel paintings without the requirement of manual input from users. Furthermore, by considering the spatial dependency, our deconvoluted element maps are of better quality and less noisy. More importantly, adding the proposed physical constraint to the optimisations with both ILF-ADMM-based and FISTA-inspired schemes shows demonstrable improvements in the deconvoluted element maps. Finally, we have shown that processing the datacube globally with the FAD method using either of the two optimisation schemes significantly reduces the running time, particularly when using the FISTA-inspired one.

\appendices

\section{Lookup Table of Element Characteristic X-Ray Line Energies}
\label{apdx:A}
\begin{table}[H]
	\centering
	\caption{Energies of the characteristic X-ray lines of 34 chemical elements commonly present in spectra acquired from easel paintings such as those considered in this study (in \si{\electronvolt}). Those lines outside the detection range of the MA-XRF device have been removed. Table taken and modified from \cite{thompson2009x}.}
	\includegraphics[width=0.95\linewidth]{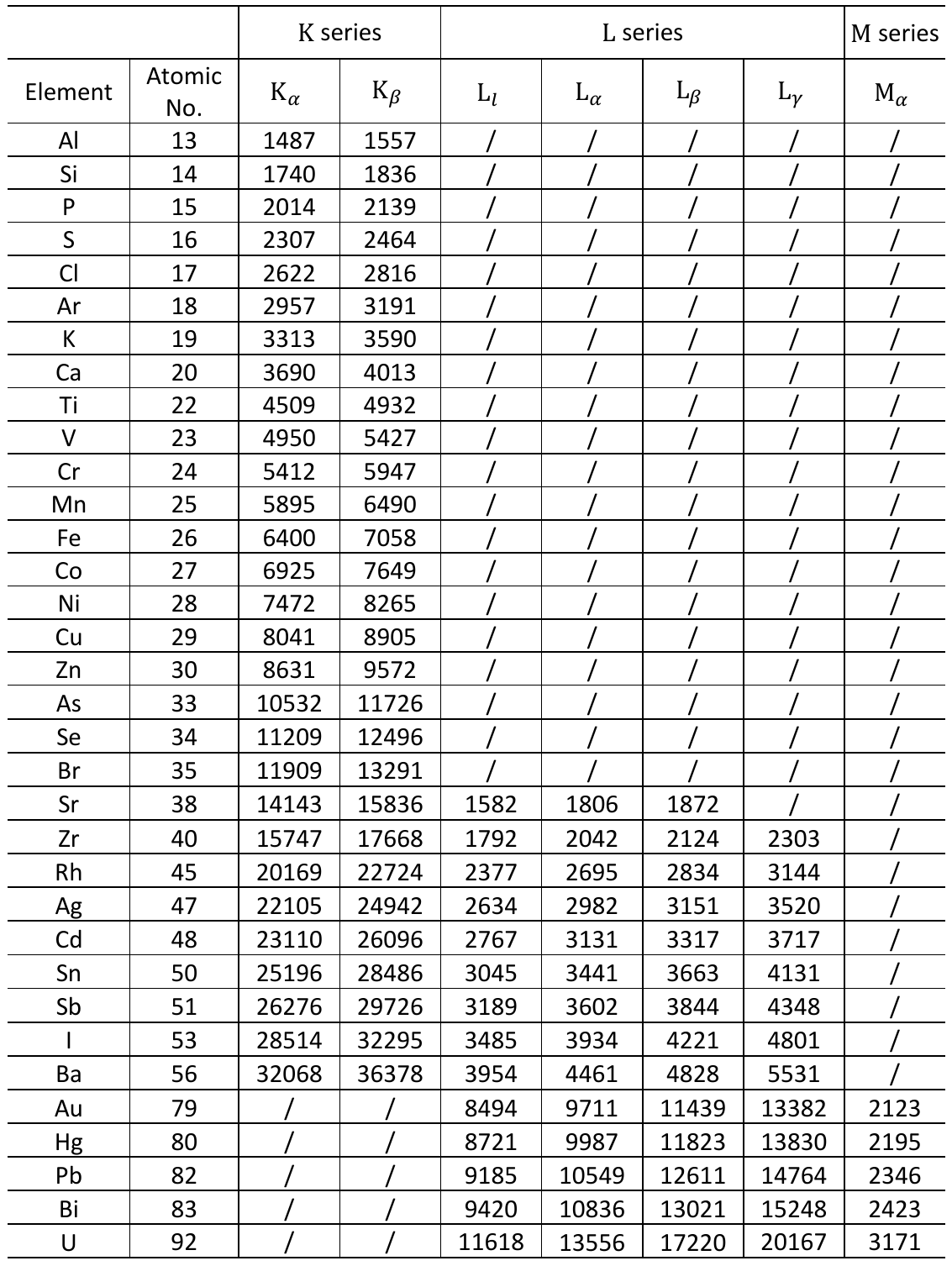}
	\label{tab:element_lins}
\end{table}

\ifCLASSOPTIONcaptionsoff
  \newpage
\fi

\bibliographystyle{IEEEtran}
\bibliography{reference}

\begin{thebibliography}{10}
\providecommand{\url}[1]{#1}
\csname url@samestyle\endcsname
\providecommand{\newblock}{\relax}
\providecommand{\bibinfo}[2]{#2}
\providecommand{\BIBentrySTDinterwordspacing}{\spaceskip=0pt\relax}
\providecommand{\BIBentryALTinterwordstretchfactor}{4}
\providecommand{\BIBentryALTinterwordspacing}{\spaceskip=\fontdimen2\font plus
\BIBentryALTinterwordstretchfactor\fontdimen3\font minus
  \fontdimen4\font\relax}
\providecommand{\BIBforeignlanguage}[2]{{%
\expandafter\ifx\csname l@#1\endcsname\relax
\typeout{** WARNING: IEEEtran.bst: No hyphenation pattern has been}%
\typeout{** loaded for the language `#1'. Using the pattern for}%
\typeout{** the default language instead.}%
\else
\language=\csname l@#1\endcsname
\fi
#2}}
\providecommand{\BIBdecl}{\relax}
\BIBdecl

\bibitem{huang2016computational}
X.~Huang, E.~Uffelman, O.~Cossairt, M.~Walton, and A.~K. Katsaggelos,
  ``{Computational imaging for cultural heritage: Recent developments in
  spectral imaging, 3-D surface measurement, image relighting, and X-ray
  mapping},'' \emph{IEEE Signal Processing Magazine}, vol.~33, no.~5, pp.
  130--138, 2016.

\bibitem{pizurica2015digital}
A.~Pizurica, L.~Platisa, T.~Ruzic, B.~Cornelis, A.~Dooms, M.~Martens,
  H.~Dubois, B.~Devolder, M.~De~Mey, and I.~Daubechies, ``{Digital image
  processing of the Ghent Altarpiece: Supporting the painting's study and
  conservation treatment},'' \emph{IEEE Signal Processing Magazine}, vol.~32,
  no.~4, pp. 112--122, 2015.

\bibitem{huang2020multimodal}
S.~Huang, B.~Cornelis, B.~Devolder, M.~Martens, and A.~Pizurica, ``{Multimodal
  Target Detection by Sparse Coding: Application to Paint Loss Detection in
  Paintings},'' \emph{IEEE Transactions on Image Processing}, vol.~29, pp.
  7681--7696, 2020.

\bibitem{dai2017spatial}
Q.~Dai, E.~Pouyet, O.~Cossairt, M.~Walton, and A.~K. Katsaggelos,
  ``{Spatial-Spectral Representation for X-Ray Fluorescence Image
  Super-Resolution},'' \emph{IEEE Transactions on Computational Imaging},
  vol.~3, no.~3, pp. 432--444, 2017.

\bibitem{dai2019adaptive}
Q.~Dai, H.~Chopp, E.~Pouyet, O.~Cossairt, M.~Walton, and A.~Katsaggelos,
  ``{Adaptive Image Sampling using Deep Learning and its Application on X-Ray
  Fluorescence Image Reconstruction},'' \emph{IEEE Transactions on Multimedia},
  2019.

\bibitem{sabetsarvestani2019artificial}
Z.~Sabetsarvestani, B.~Sober, C.~Higgitt, I.~Daubechies, and M.~R.~D.
  Rodrigues, ``{Artificial intelligence for art investigation: Meeting the
  challenge of separating x-ray images of the \textit{Ghent Altarpiece}},''
  \emph{Science Advances}, vol.~5, no.~8, p. eaaw7416, 2019.

\bibitem{8950399}
Z.~Sabetsarvestani, F.~Renna, F.~Kiraly, and M.~Rodrigues, ``{Source Separation
  With Side Information Based on Gaussian Mixture Models With Application in
  Art Investigation},'' \emph{IEEE Transactions on Signal Processing}, vol.~68,
  pp. 558--572, 2020.

\bibitem{pu2020connected}
W.~Pu, B.~Sober, N.~Daly, C.~Higgitt, I.~Daubechies, and M.~R. Rodrigues, ``A
  connected auto-encoders based approach for image separation with side
  information: with applications to art investigation,'' in \emph{ICASSP
  2020-2020 IEEE International Conference on Acoustics, Speech and Signal
  Processing (ICASSP)}.\hskip 1em plus 0.5em minus 0.4em\relax IEEE, 2020, pp.
  2213--2217.

\bibitem{pu2021learning}
W.~Pu, J.~Huang, B.~Sober, N.~Daly, C.~Higgitt, P.~L. Dragotti, I.~Daubechies,
  and M.~R. Rodrigues, ``A learning based approach to separate mixed x-ray
  images associated with artwork with concealed designs,'' in \emph{2021 29th
  European Signal Processing Conference (EUSIPCO)}.\hskip 1em plus 0.5em minus
  0.4em\relax IEEE, 2021, pp. 1491--1495.

\bibitem{pu2022mixed}
W.~Pu, J.-J. Huang, B.~Sober, N.~Daly, C.~Higgitt, I.~Daubechies, P.~L.
  Dragotti, and M.~R. Rodrigues, ``Mixed x-ray image separation for artworks
  with concealed designs,'' \emph{IEEE Transactions on Image Processing},
  vol.~31, pp. 4458--4473, 2022.

\bibitem{van2001handbook}
R.~Van~Grieken and A.~Markowicz, \emph{{Handbook of X-ray Spectrometry}}.\hskip
  1em plus 0.5em minus 0.4em\relax Marcal Dekker, New York, 2001.

\bibitem{beckhoff2007handbook}
B.~Beckhoff, B.~Kanngie{\ss}er, N.~Langhoff, R.~Wedell, and H.~Wolff,
  \emph{{Handbook of practical X-ray fluorescence analysis}}.\hskip 1em plus
  0.5em minus 0.4em\relax Springer Science \& Business Media, 2007.

\bibitem{wilkinson1971breit}
D.~H. Wilkinson, ``{Breit-Wigners viewed through gaussians},'' \emph{Nuclear
  Instruments and Methods}, vol.~95, no.~2, pp. 259--264, 1971.

\bibitem{van1977computer}
P.~Van~Espen, H.~Nullens, and F.~Adams, ``{A computer analysis of X-ray
  fluorescence spectra},'' \emph{Nuclear Instruments and Methods}, vol. 142,
  no. 1-2, pp. 243--250, 1977.

\bibitem{vekemans1995comparison}
B.~Vekemans, K.~Janssens, L.~Vincze, F.~Adams, and P.~Van~Espen, ``{Comparison
  of several background compensation methods useful for evaluation of
  energy-dispersive X-ray fluorescence spectra},'' \emph{Spectrochimica Acta
  Part B: Atomic Spectroscopy}, vol.~50, no.~2, pp. 149--169, 1995.

\bibitem{vekemans1994analysis}
------, ``{Analysis of X-ray spectra by iterative least squares (AXIL): New
  developments},'' \emph{X-Ray Spectrometry}, vol.~23, no.~6, pp. 278--285,
  1994.

\bibitem{sole2007multiplatform}
V.~A. Sol{\'e}, E.~Papillon, M.~Cotte, P.~Walter, and J.~Susini, ``{A
  multiplatform code for the analysis of energy-dispersive X-ray fluorescence
  spectra},'' \emph{Spectrochimica Acta Part B: Atomic Spectroscopy}, vol.~62,
  no.~1, pp. 63--68, 2007.

\bibitem{alfeld2015strategies}
M.~Alfeld and K.~Janssens, ``{Strategies for processing mega-pixel X-ray
  fluorescence hyperspectral data: a case study on a version of Caravaggio's
  painting Supper at Emmaus},'' \emph{Journal of analytical atomic
  spectrometry}, vol.~30, no.~3, pp. 777--789, 2015.

\bibitem{Conover2015}
D.~M. Conover, ``{Fusion of Reflectance and X-ray Fluorescence Imaging
  Spectroscopy Data for the Improved Identification of Artists' Materials},''
  Ph.D. dissertation, The George Washington University, 2015.

\bibitem{alfeld2013mobile}
M.~Alfeld, J.~V. Pedroso, M.~van Eikema~Hommes, G.~Van~der Snickt, G.~Tauber,
  J.~Blaas, M.~Haschke, K.~Erler, J.~Dik, and K.~Janssens, ``{A mobile
  instrument for in situ scanning macro-XRF investigation of historical
  paintings},'' \emph{Journal of Analytical Atomic Spectrometry}, vol.~28,
  no.~5, pp. 760--767, 2013.

\bibitem{Roald2017MAXRF}
R.~Tagle, M.~Bügler, F.~Reinhardt, and U.~Waldschläger, ``Processing of maxrf
  data with the m6 software,'' in \emph{ICXOM24: 24th International Congress on
  X-ray Optics and Microanalysis}, 2017.

\bibitem{yan2020revealing}
S.~Yan, J.-J. Huang, N.~Daly, C.~Higgitt, and P.~L. Dragotti, ``{Revealing
  Hidden Drawings in Leonardo's `the Virgin of the Rocks' from Macro X-Ray
  Fluorescence Scanning Data through Element Line Localisation},'' in
  \emph{ICASSP 2020-2020 IEEE International Conference on Acoustics, Speech and
  Signal Processing (ICASSP)}.\hskip 1em plus 0.5em minus 0.4em\relax IEEE,
  2020, pp. 1444--1448.

\bibitem{yan2021prony}
------, ``{When de Prony Met Leonardo: An Automatic Algorithm for Chemical
  Element Extraction From Macro X-Ray Fluorescence Data},'' \emph{IEEE
  Transactions on Computational Imaging}, vol.~7, pp. 908--924, 2021.

\bibitem{vetterli2002sampling}
M.~Vetterli, P.~Marziliano, and T.~Blu, ``{Sampling signals with finite rate of
  innovation},'' \emph{IEEE transactions on Signal Processing}, vol.~50, no.~6,
  pp. 1417--1428, 2002.

\bibitem{uriguen2013fri}
J.~A. Urig{\"u}en, T.~Blu, and P.~L. Dragotti, ``{FRI sampling with arbitrary
  kernels},'' \emph{IEEE Transactions on Signal Processing}, vol.~61, no.~21,
  pp. 5310--5323, 2013.

\bibitem{markowicz1993handbook}
A.~A. Markowicz and R.~E. Van~Grieken, \emph{{Handbook of x-ray spectrometry:
  methods and techniques}}.\hskip 1em plus 0.5em minus 0.4em\relax Marcel
  Dekker, Incorporated, 1993.

\bibitem{goldstein2017scanning}
J.~I. Goldstein, D.~E. Newbury, J.~R. Michael, N.~W. Ritchie, J.~H.~J. Scott,
  and D.~C. Joy, \emph{{Scanning electron microscopy and X-ray
  microanalysis}}.\hskip 1em plus 0.5em minus 0.4em\relax Springer, 2017.

\bibitem{Prony1795}
G.~R. de~Prony, ``{Essai {\'e}xperimental et analytique: sur les lois de la
  dilatabilit{\'e} de fluides {\'e}lastique et sur celles de la force expansive
  de la vapeur de l’alkool,a diff{\'e}rentes temp{\'e}ratures},''
  \emph{Journal de l’{\'E}cole Polytechnique}, vol.~1, no.~22, pp. 24--76,
  1795.

\bibitem{hua1990matrix}
Y.~Hua and T.~K. Sarkar, ``{Matrix pencil method for estimating parameters of
  exponentially damped/undamped sinusoids in noise},'' \emph{IEEE Transactions
  on Acoustics, Speech, and Signal Processing}, vol.~38, no.~5, pp. 814--824,
  1990.

\bibitem{rudin1992nonlinear}
L.~I. Rudin, S.~Osher, and E.~Fatemi, ``Nonlinear total variation based noise
  removal algorithms,'' \emph{Physica D: nonlinear phenomena}, vol.~60, no.
  1-4, pp. 259--268, 1992.

\bibitem{babacan2008variational}
S.~D. Babacan, R.~Molina, and A.~K. Katsaggelos, ``Variational bayesian blind
  deconvolution using a total variation prior,'' \emph{IEEE Transactions on
  Image Processing}, vol.~18, no.~1, pp. 12--26, 2008.

\bibitem{figueiredo2010restoration}
M.~A. Figueiredo and J.~M. Bioucas-Dias, ``Restoration of poissonian images
  using alternating direction optimization,'' \emph{IEEE transactions on Image
  Processing}, vol.~19, no.~12, pp. 3133--3145, 2010.

\bibitem{donati2019inner}
L.~Donati, E.~Soubies, and M.~Unser, ``{Inner-loop-free admm for cryo-em},'' in
  \emph{2019 IEEE 16th International Symposium on Biomedical Imaging (ISBI
  2019)}.\hskip 1em plus 0.5em minus 0.4em\relax IEEE, 2019, pp. 307--311.

\bibitem{boyd2011distributed}
S.~Boyd, N.~Parikh, and E.~Chu, \emph{{Distributed optimization and statistical
  learning via the alternating direction method of multipliers}}.\hskip 1em
  plus 0.5em minus 0.4em\relax Now Publishers Inc, 2011.

\bibitem{beck2009fast}
A.~Beck and M.~Teboulle, ``{A fast iterative shrinkage-thresholding algorithm
  for linear inverse problems},'' \emph{SIAM journal on imaging sciences},
  vol.~2, no.~1, pp. 183--202, 2009.

\bibitem{NG_Technical_37}
A.~Roy and E.~Hendriks, ``{Van Gogh's ``Sunflowers'' in London and
  Amsterdam},'' \emph{National Gallery Technical Bulletin}, vol.~37, pp.
  60--77, 2016.

\bibitem{hendriks2019van}
C.~Higgitt, G.~Macaro, and M.~Spring, ``{`Methods, Materials and Condition of
  the London Sunflowers' in Van Gogh's Sunflowers Illuminated: Art Meets
  Science},'' \emph{Van Gogh Museum Studies}, pp. 49--83, 2019.

\bibitem{NG_Technical_32}
L.~Keith, A.~Roy, R.~Morrison, and P.~Schade, ``{Leonardo da Vinci's "Virgin of
  the Rocks": Treatment, Technique and Display},'' \emph{National Gallery
  Technical Bulletin}, vol.~32, pp. 32--56, 2011.

\bibitem{spring2021leonardo}
M.~Spring, M.~M. Di~Crescenzo, C.~Higgitt, and R.~Billinge, ``Leonardo's virgin
  of the rocks in the national gallery london; new discoveries from macro x-ray
  fluorescence scanning and reflectance imaging spectroscopy,'' \emph{Nat.
  Gallery Tech. Bull.}, vol.~41, pp. 68--117, 2021.

\bibitem{ryan1993dynamic}
C.~Ryan and D.~Jamieson, ``{Dynamic analysis: on-line quantitative PIXE
  microanalysis and its use in overlap-resolved elemental mapping},''
  \emph{Nuclear Instruments and Methods in Physics Research Section B: Beam
  Interactions with Materials and Atoms}, vol.~77, no. 1-4, pp. 203--214, 1993.

\bibitem{ryan1995new}
C.~Ryan, D.~Jamieson, C.~Churms, and J.~Pilcher, ``{A new method for on-line
  true-elemental imaging using PIXE and the proton microprobe},'' \emph{Nuclear
  Instruments and Methods in Physics Research Section B: Beam Interactions with
  Materials and Atoms}, vol. 104, no. 1-4, pp. 157--165, 1995.

\bibitem{thompson2009x}
A.~Thompson, D.~Attwood, E.~Gullikson, M.~Howells, J.~Kortright, A.~Robinson
  \emph{et~al.}, ``{X-ray data booklet (2009)},'' 2009.

\end{thebibliography}

\end{document}